\documentclass[a4paper,11pt]{article}
\usepackage{jheppub}

\usepackage{slashed}
\usepackage{amsmath}
\usepackage{amssymb}

\def\sss{\scriptscriptstyle}

\newcommand{\Tr}[1]{\textrm{Tr}\left[#1\right]}

\def \bq {\mathbf{q}}

\def \nc {N_c}
\def \ca {C_{\sss A}}
\def \nf {N_f}

\def \dr {d_{\sss R}}
\def \da {d_{\sss A}}

\def \bk {\mathbf{k}}

\def \bp {\mathbf{p}}

\newcommand{\Tint}[1]{{\hbox{$\sum$}\!\!\!\!\!\!\!\int\,}_{\!\!\!\!\raise-0.9ex\hbox{$\scriptstyle{#1}$}}}

\def \cc{{\mathcal C}}
\def \Aa{{\mathcal A}}

\def\siml{{\ \lower-1.2pt\vbox{\hbox{\rlap{$<$}\lower6pt\vbox{\hbox{$\sim$}}}}\ }}
\def\simg{{\ \lower-1.2pt\vbox{\hbox{\rlap{$>$}\lower6pt\vbox{\hbox{$\sim$}}}}\ }}

\def \bq {\mathbf{q}}

\def \crr {C_{\sss R}}

\def \bk {\mathbf{k}}

\def \bp {\mathbf{p}}

\def \als {\alpha_{\mathrm{s}}}

\def \m2   {\mu^{2 \epsilon}}

\newcommand{\order}[1]{\mathcal{O}\left(#1\right)}

\def\siml{{\ \lower-1.2pt\vbox{\hbox{\rlap{$<$}\lower6pt\vbox{\hbox{$\sim$}}}}\ }}
\def\simg{{\ \lower-1.2pt\vbox{\hbox{\rlap{$>$}\lower6pt\vbox{\hbox{$\sim$}}}}\ }}

\def\nn {\nonumber}

\def\sgn{\mathrm{sgn}}

\def\pp {p_\perp}
\def\xp {x_\perp}

\def\qp {q_\perp}
\def\bpp {\mathbf{p}_\perp}
\def\bqp {\mathbf{q}_\perp}
\def\bxp {\mathbf{x}_\perp}

\def\mm {m_\infty^2}
\def\md {m_{\sss D}}
\def\dgk{\frac{d\Gamma_\gamma}{d^3k}\bigg\vert}
\def\ddgk{\frac{d \delta \Gamma_\gamma}{d^3k}}
\def\ddgkv{\frac{d \delta \Gamma_\gamma}{d^3k}\bigg\vert}
\def\2to2{{2\leftrightarrow 2}}
\def\soft{{\mathrm{soft}}}
\def\hard{{\mathrm{hard}}}
\def\coll{{\mathrm{coll}}}
\def\sc{{\mathrm{semi-coll}}}
\def\LO{{\mathrm{LO}}}
\def\NLO{{\mathrm{NLO}}}
\def\dm{\delta m}

\def\ra{{$r,a$}\,}
\def\dZg{\delta Z_{g}}
\def\ZgLO{Z_g^{\rm{LO}}}
\def\Isc{\frac{\hat{q}(\delta E)}{g^2 \crr}}

\def\alphas{\alpha_{\rm s}}
\def\Eq#1{Eq.~(\ref{#1})}
\def\Fig#1{Fig.~\ref{#1}}
\def\Sect#1{Section \ref{#1}}
\def\k{{\bf{k}}}
\def\f{{\bf{f}}}
\def\q{{\bf{q}}}
\def\b{{\bf{b}}}
\def\p{{\bf{p}}}
\def\x{{\bf{x}}}
\def\OO{{\mathcal{O}}}

\def\nfd{n_{\!\sss F}}
\def\nbe{n_{\!\sss B}}

\title{Next-to-Leading Order Thermal Photon Production 
  in a Weakly Coupled Quark-Gluon Plasma}
\author[1]{Jacopo Ghiglieri,}
\author[2]{Juhee Hong,}
\author[1]{Aleksi Kurkela,}
\author[1]{Egang Lu,}
\author[1]{Guy D. Moore,}
\author[3]{and Derek Teaney}

\affiliation[1]{McGill University, Department of Physics,\\3600 rue University, Montreal QC H3A 2T8, Canada}
\affiliation[2]{WCI Center for Fusion Theory, National Fusion Research Institute, \\ Daejeon 305-806, Korea} 
\affiliation[3]{Department of Physics and Astronomy, Stony Brook University,\\
Stony Brook, New York 11794-3800, United States}
\emailAdd{jacopo.ghiglieri@physics.mcgill.ca}
\emailAdd{jhong7@nfri.re.kr}
\emailAdd{aleksi.kurkela@mcgill.ca}
\emailAdd{legang@physics.mcgill.ca}
\emailAdd{guymoore@physics.mcgill.ca}
\emailAdd{derek.teaney@stonybrook.edu}

\abstract{
We compute the next-to-leading order $\OO(g)$ correction to the thermal photon
production rate in a QCD plasma. The NLO contributions can be expressed in
terms of gauge invariant condensates on the light cone, which are amenable to
novel sum rules and Euclidean techniques. We expect these technologies to be
generalizable to other NLO calculations. For the phenomenologically interesting
value of $\als=0.3$, the NLO correction represents a 20\% increase and has a
functional form similar to the LO result.
}


\begin{document}

\maketitle

\flushbottom

\section{Introduction}
\label{sec_intro}

Photon production has long been considered a key ``hard probe'' for
studying the formation and evolution of the quark-gluon plasma in heavy
ion collisions.  A chief advantage is that the coupling of the
plasma to photons is weak, which means that the re-absorption rate of
photons is expected to be negligible.  Once formed, a photon will escape
to the detector, carrying direct information about its formation process
unmodified by hadronization or other late time physics.

Experimentally, there are now detailed data on real photon production at
RHIC \cite{Adare:2008ab,Adare:2011zr,Afanasiev:2012dg} and 
the LHC \cite{Lee:2012cd,delaCruz:2012ru,Milov:2012pd,Steinberg:2012tv}.
Photons arising
from meson decays following hadronization 
are subtracted from the data experimentally, and the remaining sample of direct photons 
arises from several (hopefully distinct) processes.  There are
``prompt'' photons produced in the scattering of the partons from the
colliding nuclei.  The production rate here should be calculable using
perturbative QCD \cite{Gordon:1993qc}.  There are also photons associated
with the fragmentation of jets and with jet-medium interactions
\cite{Fries:2002kt,Zakharov:2004bi}, and photons produced by the
interaction of excitations of the nearly thermal Quark-Gluon Plasma,
which appears to be produced in the collision.  The thermal and
jet-medium photons are the most interesting (to us) because they
represent a signal specifically of the plasma and its evolution.

On the theoretical side, the calculation of the photon production rate
from the quark-gluon plasma has mostly been carried out within the
context of the perturbative or weak-coupling expansion.  The photon
emission rate from the plasma was computed to leading order in the
{\sl logarithm} of the strong coupling in 1991, when Kapusta {\it et al}
and Baier {\it et al} computed the rate of Compton and pair annihilation
processes \cite{Kapusta:1991qp,Baier:1991em}.  It was later pointed out
that this calculation is not complete at leading order, as
bremsstrahlung processes arise at the same power of $\alphas$
\cite{Aurenche:1998nw}.  The complete treatment of these processes was
completed in 2001, when Arnold, Moore and Yaffe performed a
leading-order calculation of the photon production rate from an
equilibrium plasma \cite{Arnold:2001ba,Arnold:2001ms}.  All these
calculations are for the production of photons from a thermal medium,
but they are rather easily adapted to include jet-medium interaction
photons as well \cite{Zakharov:2004bi}.

A leading order calculation begs many questions.  By itself it gives no
information on its own reliability; we do not know how quickly the
perturbative expansion will converge, or what will be the sign of the
next correction.  There is some concern that a leading order calculation
will not be very reliable for the photon production rate.  First of all,
the coupling is rather large at the modest temperatures achieved in
heavy ion collisions.  Second, the convergence of other perturbative
expansions in the context of finite temperature QCD is not very
comforting.  For instance, the pressure of the Quark-Gluon
Plasma has been computed to very high order in the perturbative
expansion \cite{Arnold:1994eb,Arnold:1994ps,Braaten:1995jr,%
Kajantie:2002wa}, with the result that the convergence of
the series is rather poor.  The pressure is a thermodynamical quantity,
as in fact are almost all the quantities computed beyond leading order
in the coupling.
Among dynamical transport quantities, such
as the photon rate and the rather closely related shear viscosity, heavy
quark energy loss, and heavy quark diffusion rate, only one quantity is
known beyond leading order -- the heavy quark diffusion rate
\cite{CaronHuot:2008uh}.  In this case the next-to-leading order
corrections prove to be very large.  But this case also may not be very
representative, since it involves rather different physics than photon
production or shear viscosity.  Heavy quark diffusion involves particles
which are nearly at rest and interact only via spacelike longitudinal
gluons.  But the other transport coefficients involve species moving at
almost the speed of light, exchanging transverse and longitudinal gluons
at finite frequencies.  They also involve light quarks and their hard
thermal loops in a much more direct way than the heavy quark calculation
did.

We therefore think it would be extremely useful to compute the photon
production rate at next-to-leading order in the coupling.  As we have
emphasized, this may be of phenomenological interest.  And it is most
definitely of theoretical interest, since it extends our understanding
of the convergence properties of the perturbative expansion for
dynamical quantities in the Quark-Gluon Plasma.  It also allows us to
develop the theoretical tools and understanding which will be needed for
other quantities, such as the shear viscosity, in the context of a very
cleanly defined calculation.
In the remainder of this paper we will present precisely this
next-to-leading order calculation of the photon production rate from an
equilibrium quark-gluon plasma.

The reason that there are $\OO(g)$ NLO corrections is because of the
complex self-interactions of soft highly occupied gauge fields.  As such, all of the NLO
corrections arise from the interaction with soft gluons.  Since
we are concerned with hard photons with $k \geq T$, the photons must be
produced by hard quarks moving at essentially the speed of light.
But such quarks only ``see'' the soft fields in an eikonalized way,
feeling soft-sector correlations at lightlike separated points.
And bosonic soft correlators at spacelike or
lightlike separated points can be determined from correlators of the
Euclidean theory; in fact, at leading and next-to-leading order, they
{\sl are} the correlators of the 3-D Euclidean theory, EQCD \cite{CaronHuot:2008ni}.  Better still, the
hard particles are only sensitive, at NLO, to two effects from the soft
physics: a shift in the dispersion relation, and transverse momentum
exchange.  And both of these properties are already known at NLO
\cite{CaronHuot:2008uw,CaronHuot:2008ni}.

The goal of our paper is to derive and explain these facts within the
context of the calculation of the NLO photon production rate.
We begin with an overview of the calculation in Section
\ref{sec_overview}.  We start by reviewing the leading-order
calculation, which arises from two distinct kinematic regions, one
running from ``hard'' to ``soft'' fermionic momenta and one involving
``collinear'' fermionic momenta.  Next we show
that the first corrections to the
photon production rate arise at $\OO(g)$, not at $\OO(g^2)$, and that
these corrections arise in several kinematic regions; the collinear
region, the ``semi-collinear'' region which lies between the hard and
collinear regions, and 
the ``soft'' infrared region.  We handle the collinear region
in Section \ref{sec_coll}, the soft region in Section \ref{sec_soft},
and the ``semi-collinear'' region in Section \ref{sec_semicollin}.
We present our results in Section \ref{sec_results}.

Because the hard modes experience the soft modes at lightlike
separations, we can use analyticity and Euclidean methods very
successfully in the calculation.  This was impossible
in the heavy quark diffusion calculation \cite{CaronHuot:2008uh}, and
means that the photon production calculation is actually simpler than
the heavy quark diffusion calculation.  This technical
development makes it more likely that other transport coefficients, such
as shear viscosity, can be computed beyond leading order with a
reasonable amount of effort.

On the phenomenological side, we find that there are rather large
corrections to the photon production rate from the different kinematic
regions, but that they are of both sign and surprisingly similar
magnitude.  In practice the partial NLO corrections nearly cancel in the
full result for the
phenomenologically interesting energy range of several times the
temperature.  As far as we can tell this is an accident.  However, the
individual (canceling) contributions are also not as large, relative to
the leading-order result, as in the heavy quark case.  So it appears
plausible that NLO corrections are in general not as severe as was
feared.

\section{Overview of the calculation}
\label{sec_overview}
The photon production rate is given at leading order in $\alpha$ by
\begin{equation}
	\label{defrate}
   (2\pi)^3\frac{d\Gamma_\gamma}{d^3k}
        =\frac{1}{2k}\sum_a\epsilon^\mu_a(k)\epsilon^{\nu\,*}_a(k)W^<_{\mu\nu}(K)\,, 
\end{equation}
where $\epsilon^\mu_a(k)$ are a basis of transverse polarization vectors and 
$W^<_{\mu\nu}(K)$ is the backward Wightman correlator of the
electromagnetic current
\begin{equation}
	\label{defwightman}
	W^<_{\mu\nu}(K)\equiv\int d^4X e^{-i K\cdot X}\left\langle J_\mu(0)J_\nu(X)\right\rangle.
\end{equation}
Here and throughout the paper capital letters stand for four-vectors, lowercase 
italic letters for the modulus of the spatial three-vectors and the metric signature 
is $({-}{+}{+}{+})$, so that $P^2=p^2-p^2_0$. $K=(k,\bk)=(k,0,0,k)$ is the lightlike momentum 
of the photon, which we choose to be oriented along the $z$ axis. We furthermore assume 
$k\simg T$, which is the validity region of the LO and NLO calculations.
We will work perturbatively in the strong coupling $g$, meaning that we
treat the scale $gT$ (the soft scale) as parametrically smaller than the
scale $T$ (the hard scale).

Throughout the paper we will often use light-cone coordinates, which we
define as $p^{-}\equiv p^0- p^z$ and $p^+ \equiv \frac{p^0 + p^z}{2}$.
This normalization is nonstandard, but we find it convenient because
$dp^0 dp^z = dp^+ dp^-$, and because we will frequently encounter cases
in which $p^-=0$, in which case $p^z = p^0 = p^+$ with our conventions.
The transverse coordinates are written as $\bpp$, with modulus $\pp$.

We finally remark that for convenience we will mostly work in the Keldysh, or \ra, 
basis of the real-time formalism for the computation
of \Eq{defwightman}. The two elements of this basis are defined as $\phi_r\equiv(\phi_1+\phi_2)/2$,
 $\phi_a\equiv\phi_1-\phi_2$, $\phi$ being a generic field and the subscripts 1 and 2 
labeling the time-ordered and anti-time-ordered branches of the Schwinger-Keldysh contour respectively.
The propagator is a $2\times2$ matrix, where one entry is always zero and only one entry depends on the 
thermal distribution, \emph{i.e.},
\begin{equation}
	\label{raprop}
	D=\left(\begin{array}{cc} D_{rr}&D_{ra}\\D_{ar}&D_{aa}\end{array}\right)
	=\left(\begin{array}{ccc} \left(\frac12\pm n(p^0)\right)(D_R-D_A)&&D_{R}\\D_{A}&&0\end{array}\right),
\end{equation}
where $D_R$ and $D_A$ are the retarded and advanced propagators, the plus (minus) sign refers 
to bosons (fermions). $n(p^0)$ is the corresponding thermal distribution, either 
$\nbe(p^0)=(\exp(p^0/T)-1)^{-1}$ for bosons or $\nfd(p^0)=(\exp(p^0/T)+1)^{-1}$ for fermions. 
We also define the spectral function as the difference of the retarded and 
advanced propagators, $\rho \equiv D_R - D_A$. 
We will denote the gluon propagator by $G$ and the quark one $S$.

We will adopt strict Coulomb gauge throughout.  The treatment of soft
momenta in propagators and vertices requires the use of Hard Thermal
Loop (HTL) resummation \cite{Braaten:1989mz}.  For convenience we list
the Coulomb gauge retarded HTL resummed propagators for fermions and
gluons in App.~\ref{app_props}. We will discuss the power-counting rules
of the HTL theory with fermions in \ra basis in Section~\ref{sec_soft}.

\subsection{Leading-order calculation}

\label{sub_overview_LO}
\begin{figure}[ht]
	\begin{center}
	\begin{minipage}{0.125\textwidth}
	\mbox{$\dgk_\mathrm{\hard}=$}
	\end{minipage}
	\begin{minipage}{0.85\textwidth}
	\includegraphics[width=12.85cm]{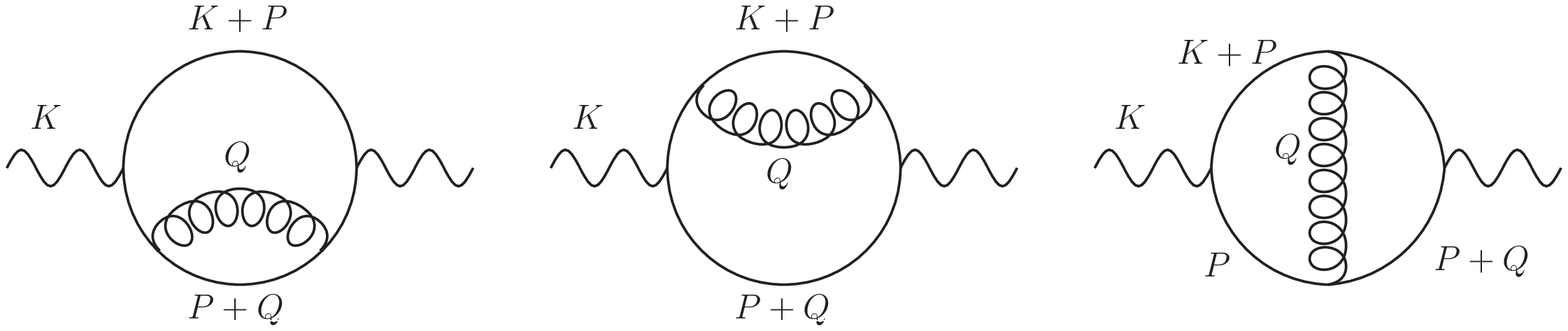}
	\end{minipage}
	\end{center}
	\caption{Two-loop diagrams necessary for the evaluation of the $\2to2$ region. 
	The wavy lines are photons, curly lines are gluons and plain lines are quarks. 
	No assumption is made yet on the scaling (hard, soft, collinear...) of the internal lines. 
	The momentum assignments shown here will be used throughout the paper.}
	\label{fig_diagrams}
\end{figure}

At leading order (and also at NLO) in $g$ the photon production rate arises from 
diagrams where the photon attaches to a single connected quark loop with a number 
of gluon lines. We denote the momenta flowing in the quark lines that attach to one of the 
photon vertices $P$ and $K+P$ -- see for example \Fig{fig_diagrams}. Then the leading order rate arises from three kinematic regimes:
\begin{enumerate}
\item 
One of the quarks attaching to the photon is
on-shell, $(K+P)^2\sim g^2 T^2$, and the other one is far off-shell, $P^2\sim T^2$. This is the \emph{hard} $\2to2$
\emph{region}.
\item 
The photon attaches to one hard on-shell, $(P+K)^2\sim g^2T^2$, fermionic line and one soft fermionic line with $P^2\sim g^2T^2$ and 
$P\cdot u \sim gT$, where $u$ is the rest-frame of the medium.
This is
the \emph{soft $\2to2$ region}, and it is the soft limit of the hard $\2to2$ kinematic region. 
\item 
The photon line attaches to two
fermionic lines which are hard ($P\cdot u \sim T $ and $(K+P)\cdot u \sim T $),  nearly collinear ($K\cdot P\sim g^2 T^2$), and nearly on-shell ($P^2\sim g^2 T^2$ and $(P+K)^2\sim g^2 T^2$); this is the \emph{collinear region}.
\end{enumerate}
The total leading order rate is the sum of these three kinematic regions:
\begin{equation}
\dgk_\mathrm{LO}=\dgk_\hard+\dgk_\soft+\dgk_\coll \, .
\end{equation}
The physics of each of these regions will be summarized in the remainder
of this section.

\begin{figure}[ht]
	\begin{center}
		\includegraphics[width=12cm]{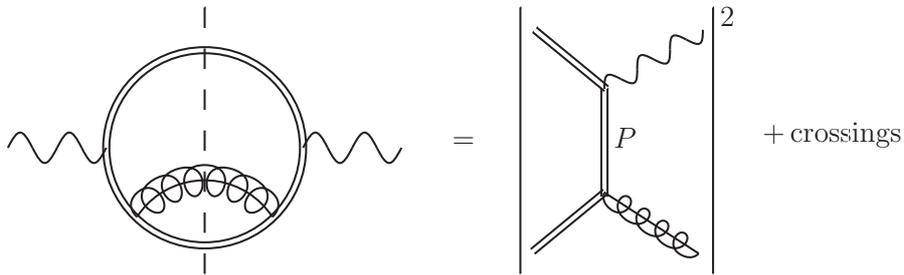}
	\end{center}
	\caption{The two-loop diagram on the left corresponds to the square of the amplitude 
	of the diagram on the right and on the squares of its crossing. The interference terms 
	arise from the two-loop diagram where the gluon is exchanged between the two fermionic 
	lines. Double internal lines stand for hard particles, \emph{i.e.}, particles whose momentum is 
	$\OO(T)$ in at least one component.}
	\label{fig_hard_cuts}
\end{figure}

\begin{figure}[ht]
\begin{center}
\includegraphics[width=10cm]{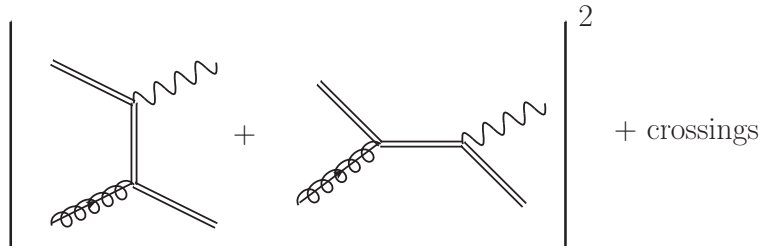}
\end{center}
\caption{The squared matrix element for $\2to2$  processes. The corresponding annihilation
diagrams are obtained by crossing and are not shown.}
\label{twototwo}
\end{figure}

Diagrammatically, the \emph{hard $\2to2$ region}  consists of the simple two-loop diagrams obtained by adding 
a single gluon to the one-loop diagram for $W^<$, which is kinematically forbidden
 for a real photon. These diagrams are shown in Fig.~\ref{fig_diagrams}; 
 their cuts (when all lines are hard) correspond to the $\2to2$ processes
 $qg\to q\gamma$ (Compton) and $q\overline q\to g \gamma$
 (annihilation).
In Fig.~\ref{fig_hard_cuts} 
we show an example diagram and its corresponding cut.%
\footnote{In our graphical notation 
the double line represents particles
whose momentum is hard, \emph{i.e.}, $\OO(T)$ or larger, in at least one component. This thus includes
not only hard particles, but also collinear and semi-collinear ones.}
Fig.~\ref{twototwo} shows the processes contributing to the sum of
the cut diagrams.

\begin{figure}[ht]
  \begin{center}
   \begin{minipage}{0.13\textwidth}
   $\dgk_\mathrm{soft}=$\\ \vspace{0.5cm}
   \end{minipage}
   \begin{minipage}{0.3\textwidth}
   \includegraphics[width=4cm]{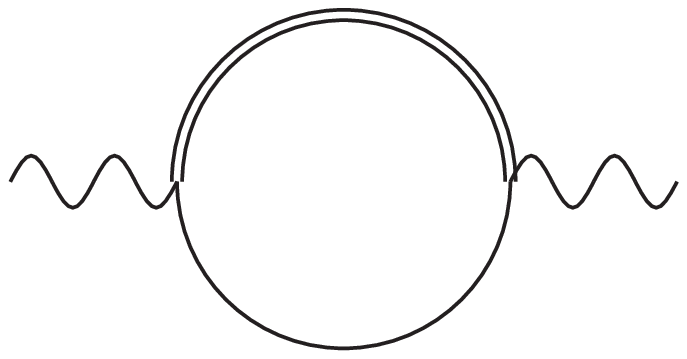}\\\vspace{0.1cm}
   \end{minipage}
   \includegraphics[width=\textwidth]{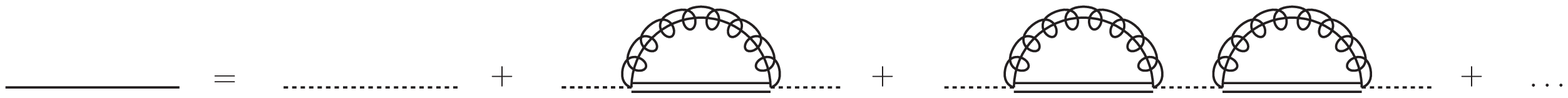}
  \end{center}
  \caption{The diagram contributing to the soft region at leading order. The double lines are the hard lines whereas 
  the dotted single line represent bare soft propagators. The single plain line is the HTL-resummed soft propagator. The momentum assignments are given in \Fig{fig_diagrams}. }
  \label{fig_lo_soft}
\end{figure}

The two first (self-energy) diagrams in  Fig.~\ref{fig_diagrams} receive 
parametrically equal contributions from all logarithmic momentum scales (up to $T$) of the intermediate
virtual quark with momentum $P$. Thus, the naive computation of the diagrams of Fig.~\ref{fig_hard_cuts}
leads to a logarithmic infrared divergence.
However, this IR divergence is regulated by the physics of screening and collective plasma excitations, and indeed
when the momentum $P$ becomes
soft, $P \sim g T$, the self-energy insertion
is not anymore a $g$-suppressed perturbation in the dispersion relation of the intermediate quark
 and needs to be resummed.
 This is the soft region. In this kinematic region, the third diagram in \Fig{fig_diagrams} gives a subleading contribution and thus
the diagrams that contribute to the leading order result are those shown
in Fig.~\ref{fig_lo_soft}. These diagrams can be most conveniently resummed in Hard Thermal Loop -effective
theory, and in particular the leading order HTL-resummed diagram is  displayed in Fig.~\ref{fig_lo_soft}.
The cuts of the diagram correspond to  \emph{conversion} processes, where a soft fermion exchange with the medium converts a hard quark with momentum $K+\OO(gT)$ into a photon with momentum $K$.
The computation of the resummed diagram will be discussed in detail in Sec.~\ref{sec_soft}, where a sum rule for its 
evaluation will be introduced.

The soft and the $\hard$ regions are smoothly connected and when $gT\ll P \ll T$,
each set of diagrams gives an equivalent and correct description. In a practical calculation
one introduces a momentum cutoff for the integrals $gT\ll \mu_\perp^{\textrm{LO}}\ll T$ dividing the two regions.
Both contributions then individually depend on $\mu_\perp^{\textrm{LO}}$, but this dependence
cancels exactly in the sum of the two terms yielding a logarithm of the temperature over the 
asymptotic mass $m_\infty\sim gT$ of the quark. 

\begin{figure}[ht]
	\begin{center}
		\includegraphics[width=7cm]{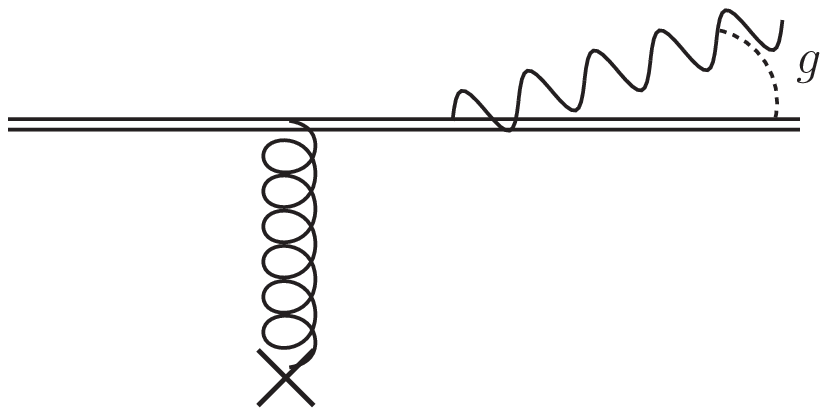}
		\includegraphics[width=6.7cm]{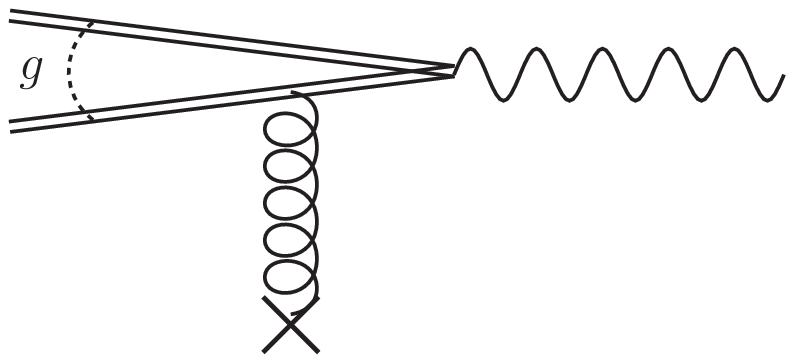}
	\end{center}
	\caption{Collinear diagrams. In the first case, called the \emph{bremsstrahlung} 
	diagram, the angle between the emitted photon and the outgoing emitting fermion 
	is of order $g$. In the second case, called the \emph{pair annihilation} diagram, 
	it is the angle between the annihilating quark and antiquark that is of order $g$. 
	The diagrams where the  gluon is attached to the other fermionic line
	are not shown.  
	In both graphs the gluon is soft and is scattering off the hard 
   quarks and gluons  of the 
	plasma as indicated by the crosses, \emph{i.e.} it is an HTL gluon in the Landau cut.
}
	\label{fig_collinear}
\end{figure}

 In \cite{Aurenche:1998nw,Aurenche:1999tq,Aurenche:2000gf} it was pointed 
out that there is another phase space region that contributes at LO
 besides the $\hard$ and soft regions: the 
\emph{collinear region}. 
In this  region
\begin{itemize}
\item the momenta of the quark lines are hard, nearly on shell and collinear to each other, 
\emph{i.e.},  $p^+ \sim T$, $\pp \sim g T$, and $p^- \sim g^2 T$ such that $P^2\sim g^2 T$,
and
\item the momentum $Q$ of the gluon is spacelike and soft with $ q^+ \sim \qp \sim g T$ and $q^{-}\sim g^2T$, 
so that the kinematics of the quarks are unaffected by the gluon.
\end{itemize}
These constraints force the angles between the quarks and the photon to be
small $\sim g$, and  therefore the cuts of  the third diagram in Fig.~\ref{fig_diagrams}
correspond to the \emph{bremsstrahlung} and \emph{pair annihilation} processes shown in 
Fig.~\ref{fig_collinear}.%
In this case the intermediate virtual quark is almost on shell and
thus has a long lifetime of order $\sim 1/g^2T$ which is parametrically of the same order as the small
angle scattering rate in the plasma $1/\Gamma \sim 1/g^2T$. Hence during the formation
of the photon, there is an $\mathcal{O}(1)$ probability that the intermediate quark 
will undergo one or several additional soft scatterings with the constituents, and 
processes involving multiple soft scatterings are not suppressed by powers of $g$. 
These multiple scatterings lead to destructive interference, that is known as the \emph{Landau-Pomeranchuk-Migdal} (LPM)
effect that leads to an $\mathcal{O}(1)$ suppression of the collinear rate.

\begin{figure}[ht]
	\begin{center}
	\begin{minipage}{0.1\textwidth}
		$\dgk_\mathrm{coll}$=
		\end{minipage}
		\begin{minipage}{0.875\textwidth}
		\includegraphics[width=13.2cm]{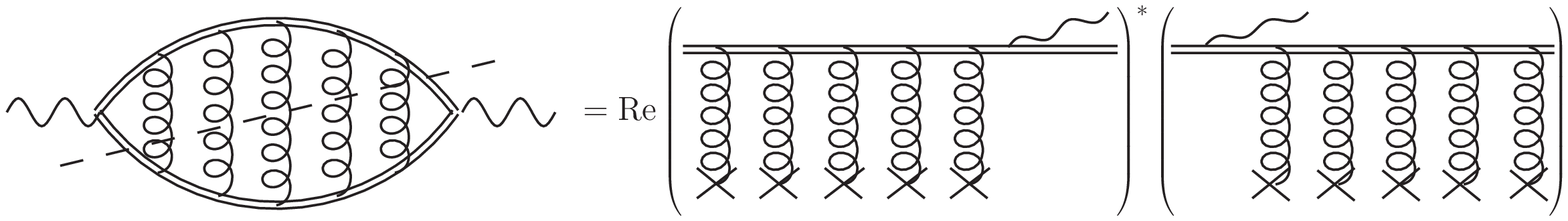}
		\end{minipage}
	\end{center}
	\caption{The uncrossed ladder diagrams that need to be resummed to account 
	for the LPM effect in the collinear region. The cut shown here corresponds to 
	the interference term on the right-hand side. The rungs on the l.h.s. are HTL 
	gluons in the Landau cut. On the r.h.s., the crosses at the 
	lower end of the gluons represent the hard scattering centers, either 
	gluons or fermions.}
	\label{fig_lpm}
\end{figure}

In terms of the two-point function these processes correspond to diagrams 
with the two nearly collinear fermion lines connected with arbitrary number of
soft spacelike gluons with same kinematics as $Q$. In \cite{Arnold:2001ba,Arnold:2001ms} Arnold, Moore and Yaffe (AMY)
showed that it is only the ladder-type diagrams shown in Fig.~\ref{fig_lpm} that
contribute to a leading order calculation; the factors of $g$ arising from additional
vertices are canceled by near on-shell propagators and large statistical
factors arising from the gluonic propagators. The near on-shellness of the quark lines makes
the diagrams sensitive to the thermal mass $\mm\sim g^2T^2$ and the thermal width $\Gamma\sim g^2T$ 
of the quark lines, which need to be consistently resummed.
Furthermore AMY showed how these diagrams can
be resummed in terms of a Schr\"odinger equation type differential equation,
and they obtained the complete leading-order results in \cite{Arnold:2001ms}.
In Sec.~\ref{sec_coll} we will discuss in detail this equation in the context of the treatment of 
its NLO corrections.

The leading order result arising from 
the $\hard$, soft and collinear regions can be summarized  as \cite{Arnold:2001ms}
\begin{eqnarray}
\dgk_{\hard} &=& \frac{\mathcal{A}(k)}{(2\pi)^3}\left[ \ln\left(\frac{T}{\mu_\perp^{\textrm{LO}}}\right) + C_{\hard}\left( \frac{k}{T}\right) \right]\\
\dgk_{\soft} &=& \frac{\mathcal{A}(k)}{(2\pi)^3}\left[ \ln\left(\frac{\mu_\perp^{\textrm{LO}}}{m_\infty}\right) \right] \\
\dgk_{\coll} &=& \frac{\mathcal{A}(k)}{(2\pi)^3}\left[ C_\coll^{\mathrm{LO}}\left(\frac{k}{T},\kappa\right) \right]
\end{eqnarray}
or
\begin{equation}
	\label{totallo}
	\frac{d\Gamma_\gamma}
	{d^3k}\bigg\vert_\mathrm{LO}
        = \frac{\mathcal{A}(k)}{(2\pi)^3}\left[\ln\left(\frac{T}{m_\infty}\right)
	+C_\mathrm{\hard}\left(\frac{k}{T}\right)
        +C_\mathrm{coll}^{\mathrm{LO}}\left(\frac{k}{T},\kappa\right)
      \right],
\end{equation}
where $\Aa(k)$ is the leading-log coefficient of the photon production
rate
\begin{equation}
	\label{A_of_k}
	\Aa(k) =  2 \alpha_{\rm EM} \frac{\mm}{ k} \nfd(k)
	  \sum_s \dr q_s^2
\quad \left( = \frac{4 \alpha_{\rm EM} \nfd(k) g^2 T^2}{3 k}  \, 
\mbox{for QCD with $uds$ quarks} \right) .
\end{equation}
Here $\dr$ is the dimension of the quark's representation 
($\dr=\nc$ in the fundamental representation of $SU(\nc)$), $q_s$ is its
abelian charge, and the sum runs over the number of light fermions flavors, $\nf$.
The parameter $\kappa$  is the square of the ratio of $\mm = g^2T^2\crr/4$
to the Debye mass $\md^2=g^2T^2(\ca + \nf T_R)/3$ at leading order in $g$. $\kappa$ encodes the dependence
on the number of colors and light flavors in the plasma, \emph{i.e.}
\begin{equation}
	\label{defkappa}
	\kappa\equiv\frac{\mm}{\md^2}=\frac{3\crr}{4(\ca+T_R\nf)}
\quad
\left( = \frac{2}{6+\nf} \mbox{ for QCD} \right),
\end{equation}
where $\crr$ and $\ca$ are the quadratic Casimirs of the representations 
of the quarks and gluons respectively, and $T_R$ is the index of the representation of the quarks.\footnote{
$\crr=(\nc^2-1)/2\nc$ and $T_R=1/2$ for quarks in the fundamental representation, and $T_A=\ca=\nc$ for the adjoint representation of $SU(\nc)$.}
The definition of $\mm$ is discussed in Appendix \ref{app_mass}.
 
The
functions $C_\mathrm{\hard}(k/T)$ and $C_\mathrm{coll}^\mathrm{LO}(k/T,\kappa)$
describe the momentum dependence of the $\hard$ and collinear regions
and have to be obtained numerically, the former by integrating the matrix elements
for the hard processes folded over the thermal distributions and the latter by solving
the integral equation for collinear processes. For further convenience we list
the parametrization of these functions as given in \cite{Arnold:2001ms} for a $\nc=3$ QCD
plasma with $\nf$ flavors:
\begin{eqnarray}
	\label{c2to2fit}
	C_\mathrm{\hard}\left(\frac{k}{T}\right)&\approx&\frac12
	\ln\left(\frac{2k}{T}\right)+ 0.041\frac{T}{k} -0.3615 + 1.01 e^{-1.35 k/T},\quad 0.2<\frac{k}{T} , \\
\nn	C_\mathrm{coll}^{\mathrm{LO}}\left(\frac{k}{T},\kappa\right)
	&\approx&
	\frac{1}{\sqrt{\kappa}}\left[\frac{0.316\ln(12.18+T/k)}{\left(k/T\right)^{3/2}}
	+\frac{0.0768k/T}{\sqrt{1+k/(16.27 T)}}\right],\quad 0.2<\frac{k}{T}<50.\\
	&&\label{ccolllofit}
\end{eqnarray}

\begin{figure}[ht]
	\begin{center}
		\includegraphics[width=13cm]{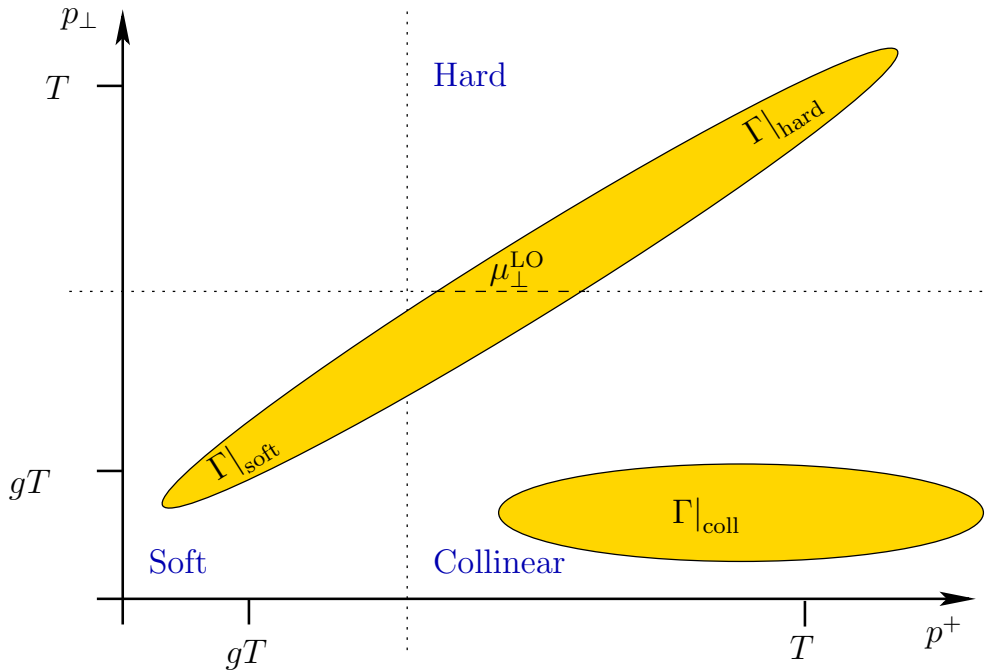}
	\end{center}
   \caption{Momentum integration regions in the $(p^+,\pp)$ plane contributing to the leading-order calculation.
The $\mu_\perp^\LO$ label indicates a LO cancellation of UV/IR log 
divergences between the soft and hard regions respectively. }
	\label{fig_lomap}
\end{figure}

We conclude this overview of the leading-order calculation by noting that 
the momentum integration regions that contribute here are best identified by their 
scaling in terms of $P$. 
In Fig.~\ref{fig_lomap} we map these momentum regions 
in the $(p^+,\pp)$ plane. 
The scaling of $p^-$ can be obtained from momentum 
conservation.

\subsection{Next-to-leading order corrections}
\label{sub_overview_NLO}
At next to leading order, the full result is a sum of the leading order rate and 
its $\OO(g)$ correction
\begin{eqnarray}
\dgk_{\LO+\NLO} &=& \dgk_{\LO} + \ddgk.
\end{eqnarray}
As in  the leading order calculation, the NLO rate arises from distinct kinematic 
regions and the NLO correction can be parametrized as
\begin{eqnarray}
\ddgk &=& \ddgkv_{\soft} + \ddgkv_{\coll} + \ddgkv_{\sc}.
\end{eqnarray}
The soft and collinear regions are the same kinematic regions as in the
leading-order calculation, while the \emph{semi-collinear} region is an 
additional kinematic region whose contribution starts at NLO.

In the hard region, corrections come about by adding an extra loop to the diagrams 
shown in Fig.~\ref{fig_diagrams}. As long as momenta stay hard, we need not worry 
about these corrections, which are suppressed by a factor of $g^2$ relative to the 
leading-order result.

However, both the soft propagator and the location of the quasiparticle pole 
at hard momenta have $\OO(g)$ corrections, and thus diagrams that are
sensitive to these quantities
may also receive $\OO(g)$ corrections.
Indeed, in the collinear case the individual rungs in the ladder resummation 
are soft gluons, and the collinear quarks are near the quasiparticle pole.
Consequently, the parameters of the collinear integral equation 
that record the soft gluon scattering rate and the quasiparticle masses, ${\mathcal C}(q_\perp)$ and $\mm$,  are modified at NLO. The structure of the  NLO
correction arising from the collinear region is then
\begin{equation}
\ddgkv_{\coll} = \ddgkv_{\dm} + \ddgkv_{\delta C},
\end{equation}
where the first term is due to a $\OO(g)$ shift in $\mm$, and 
the second term arises from a one-loop $\OO(g)$ correction  to the 
soft scattering kernel, ${\mathcal C}(q_\perp)$.
In \Sect{sec_coll} we will discuss these corrections in detail, as 
well as two $\order{g}$ subregions where the collinear phase space starts to overlap with the soft and semi-collinear regions, requiring subtractions.

\begin{figure}[ht]
	\begin{center}
		\includegraphics[width=15cm]{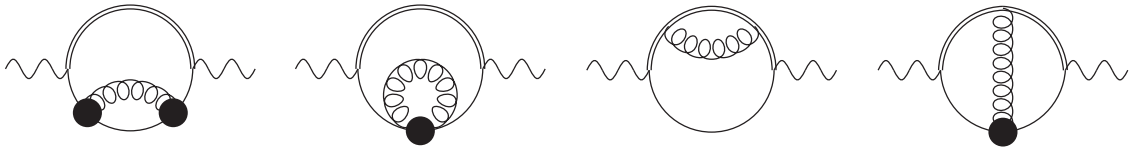}
	\end{center}
	\caption{Diagrams contributing to the NLO fully soft rate. The black blobs 
	are bare+HTL vertices, plain lines and gluons are soft. We call these four diagrams, from 
	left to right, the soft-soft self-energy, the tadpole, the hard-soft self-energy and 
the cat eye. The momentum assignments are given in \Fig{fig_diagrams}.}
	\label{fig_nlo_soft}
\end{figure}

In the soft region, the addition of an extra soft gluon to the diagram in 
Fig.~\ref{fig_lo_soft} results in the diagrams shown in Fig.~\ref{fig_nlo_soft}, 
which represent an $\order{g}$ correction to the conversion process.
In particular, wherever a gluon ends on a soft fermion line, all momenta flowing 
in that quark-gluon vertex are of order $gT$. This causes the bare and HTL vertices to 
be of the same order, requiring the inclusion of the HTL vertex, as shown in the first and last 
diagrams in Fig.~\ref{fig_nlo_soft}. Furthermore, the two-quark, two-gluon HTL gives rise 
to a new topology, the second diagram in that figure.

The complicated analytic structure of the HTL vertices and propagators, with their branch cuts and 
imaginary parts, as well as the non-trivial functional dependence on the momenta, would in 
principle make the calculation of the diagrams in Fig.~\ref{fig_nlo_soft} technically intricate 
and only amenable to a multi-dimensional numerical integration. However, in Sec.~\ref{sec_soft} we 
develop a set of sum rules using the analytic properties of these amplitudes, 
which are in turn related to causality. These sum rules, as we shall show,
simplify the calculation dramatically leading to an analytical result.

\begin{figure}[ht]
	\begin{center}
		\includegraphics[width=5cm]{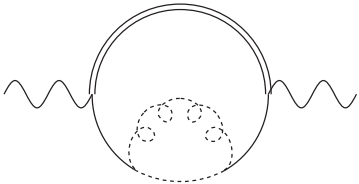}
	\end{center}
	\caption{Mistreated $\OO(g)$ part of the leading order soft calculation.
   The dotted lines indicate bare soft propagators.}
	\label{fig_nlo_soft_counterterm}
\end{figure}

When evaluating these soft diagrams we must correct the LO treatments 
of the soft region to avoid double counting. Note that the 
first diagram in Fig.~\ref{fig_nlo_soft} is the soft limit of the
HTL self-energy already included in the soft-LO calculation, see
Fig.~\ref{fig_lo_soft}.
 The HTL self-energy used at LO
includes an integral over the hard thermal loop momentum $Q$ which extends
down to zero, with $\OO(g)$ of the contribution arising from 
the soft region of integration where the HTL approximation is no longer valid.
Thus, while the first diagram in Fig.~\ref{fig_nlo_soft}
has already been included at LO, the LO treatment is incorrect at NLO.
To avoid double counting and to correct this mistreatment
we have to
subtract this soft-loop part of the HTL calculation.
We do this by subtracting a
counterterm diagram shown in Figure \ref{fig_nlo_soft_counterterm}.

Similarly, in the calculation of the leading order collinear rate, an $\OO(g)$ part of the $p^+$ integration
arises from the kinematic region where $p^+$ is soft and  overlaps with the soft region. For example, the LO treatment of
the collinear bremsstrahlung diagram shown \Fig{fig_collinear}
integrates over the momentum fraction of the final state quark, 
and is incorrect when this momentum fraction is $\OO(g)$.
This region is correctly dealt with by the soft graphs of \Fig{fig_nlo_soft}, 
and the mistreated LO collinear contribution must be subsequently subtracted. 
The structure of the soft region at NLO is 
then
\begin{eqnarray}
\label{eq_soft}
\ddgkv_{\soft} &=& \ddgkv_{\soft}^{\rm{diags.}} - \dgk_{\textrm{soft}}^{\textrm{subtr.}},
\end{eqnarray}
where the first term arises from the difference of the diagrams in Fig.~\ref{fig_nlo_soft} and Fig.~\ref{fig_nlo_soft_counterterm} and the second is
the soft part of the leading order collinear rate.
The complete treatment of the soft region, including the diagrams of \Fig{fig_nlo_soft} and all necessary subtractions, is given in \Sect{sec_soft}.

Finally, a third region contributes to the NLO rate, corresponding to the uncharted, wedge-shaped
area in Fig.~\ref{fig_lomap} between the $\hard$ and collinear regions
where $Q\sim gT$ is soft and
 the scaling of $P$ obeys
$p^+\sim T$, $p^-\sim gT$ and $\pp\sim \sqrt{g}T$,
so that $P^2 \sim gT^2 \sim P\cdot K$ --- this is the 
\emph{semi-collinear region}.
This region is closer to
the mass shell and closer to collinearity with $K$ than the hard region
(where $P^2\sim T^2 \sim P\cdot K$), but farther from the mass
shell and less collinear than the collinear region (where
$P^2 \sim g^2 T^2 \sim P\cdot K$).%
\footnote{%
    Naive power-counting arguments suggest that this region, with the
    exchange of a soft $Q\sim gT$ gluon, should actually be leading
    order.  But there is a cancellation between the diagrams shown in
    Fig.~\ref{fig_diagrams}.  This cancellation, pointed
    out in footnote 5 of Ref.~\cite{Arnold:2001ba}, renders this phase
    space region $\OO(g)$ with respect to the leading contributions.}
The physical processes in this region are characterized by the sign of $Q^2$. For timelike gluon momenta 
the semi-collinear contribution can be seen as a correction to the $\hard$ region 
from the emission/absorption of soft gluons on their massive, quasiparticle plasmon poles. 
In the spacelike region, on the other hand, one can interpret this contribution as a less 
collinear emission, the angle being now $\sqrt{g}$ instead of $g$. This in turn implies 
a formation time of the order of $1/(gT)$, much shorter than the inverse scattering rate. 
The LPM effect is then not relevant in this region.

The leading order treatment of the collinear region also mistreats the
semi-collinear region by an $\OO(g)$ amount, and the part of the leading 
order collinear rate arising from the semi-collinear region again needs to be
subtracted. Also, the leading order hard calculation receives an $\OO(g)$
mistreated contribution from the semi-collinear regions so that  
the structure of the semi-collinear correction becomes
\begin{eqnarray}
\label{eq_semicoll}
\ddgkv_{\sc} &=& \ddgkv_{\sc}^{\rm{diags.}} - \dgk_{\sc}^{\textrm{coll. subtr.}} - \dgk_{\sc}^{\textrm{hard subtr.}}.
\end{eqnarray}
The complete treatment of the semi-collinear region is given in \Sect{sec_semicollin}.

\begin{figure}[ht]	
	\begin{center}
		\includegraphics[width=13cm]{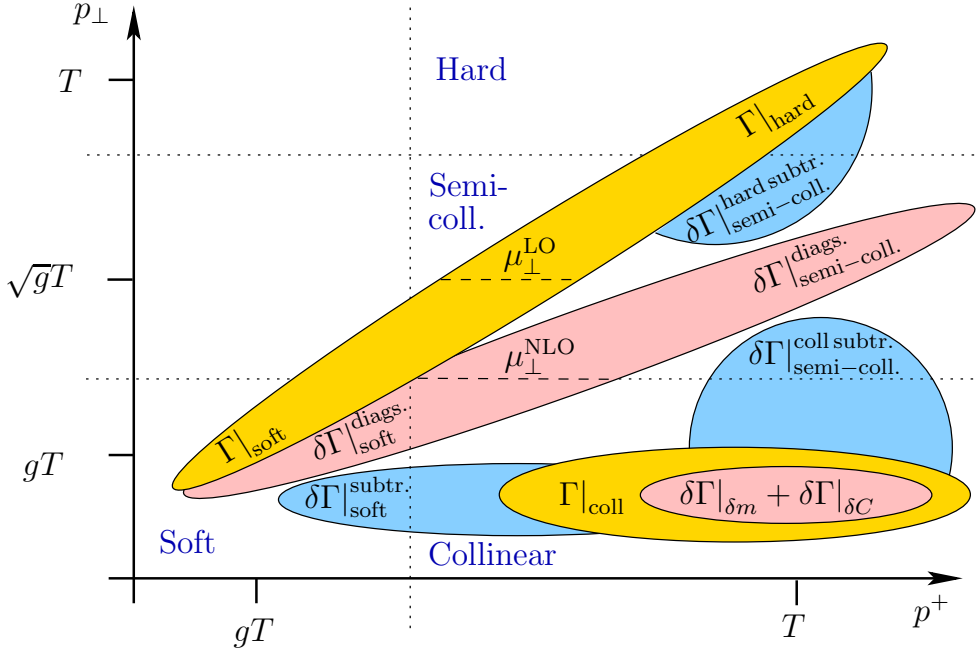}
	\end{center}
   \caption{Momentum integration regions in the $(p^+,\pp)$ plane contributing to the  LO and NLO calculations. 
   The yellow and pink bands 
   indicate the LO and NLO integration regions respectively. 
   The blue 
bands indicate an overlap region where the LO contribution
must be subtracted to avoid double counting.
The hard region does not contribute at NLO. Finally, 
the $\mu_\perp^\NLO$ label anticipates a cancellation of UV/IR log 
divergences between the soft and semi-collinear regions, while the
$\mu_\perp^{\LO}$ label indicates a similar LO cancellation between the soft and hard regions.
}
	\label{fig_nlomap}
\end{figure}

We conclude this section by mapping the  NLO integration regions in the $(p^+,\pp)$ plane in \Fig{fig_nlomap}, in analogy  the LO map in \Fig{fig_lomap}.

\section{The collinear region}
\label{sec_coll}
The evaluation of the collinear region at 
leading order requires the resummation of an infinite number of soft gluon exchanges through
an integral equation.
Such an equation was derived by Arnold, Moore, and Yaffe \cite{Arnold:2001ba,Arnold:2001ms} 
and gives rise to a LO contribution to the photon production rate of
\begin{eqnarray}
	 \dgk_{\coll} & = & \frac{\Aa(k)}{(2\pi)^3}
\int_{-\infty}^\infty dp^+ \left[ \frac{(p^+)^2 + (p^+{+}k)^2}{(p^+)^2 (p^+{+}k)^2} \right]\:
   \frac{\nfd(k{+}p^+) [1-\nfd(p^+)]}{\nfd(k)}
\nonumber \\
\label{fpproblem}
 & & \times \frac{1}{g^2 \crr T^2} \int \frac{d^2\pp}{(2\pi)^2}
  \:{\rm Re}\: 2\bpp \cdot \f(\bpp,p^+,k) \,,
\\
\label{qspace}
 2\bpp & = & i\delta E \; \f(\bpp) +
 \int \frac{d^2\qp}{(2\pi)^2} \: \cc(\qp)
  \Big[ \f(\bpp) - \f(\bpp {+} \q_\perp) \Big] \,,
\\
\label{deltaE}
\delta E & = & \frac{k(\pp^2 + \mm)}{2p^+(k{+}p^+)} \,.
\end{eqnarray}
Here $\delta E=E_{\bp+\bk}-E_\bp-k$ is the eikonalized energy difference between 
having a quark of energy corresponding to a momentum of $k+p$ and having a quark of 
energy corresponding to momentum $p$ with a photon of energy $k$. 
$\cc(\qp)$ is the differential soft scattering rate, which at leading order reads
\cite{Arnold:2001ba,Aurenche:2002pd}
\begin{equation}
	\label{C_LO}
	 \cc(\qp) =g^2\crr\int\frac{dq^0dq_z}{(2\pi)^2}2\pi\delta(q^0-q_z) G^{rr}_{\mu\nu}(Q)v_k^\mu v_k^\nu=
	          g^2\crr T \frac{\md^2}{\qp^2(\qp^2{+}\md^2)} \,,
\end{equation}
where $v_k^\mu\equiv(1,0,0,1)$ and $G_{\mu \nu}^{rr}$ is the cut HTL gluon propagator.

The physical interpretation of this expression is as follows.  Photon
production involves a current operator insertion in the amplitude,
followed by time evolution, and then a current insertion in the
conjugate amplitude.  At times between the current insertions, the
density matrix contains an off-diagonal term with a quark with momentum
$p^+,\bpp$ and a photon of momentum $k$ in the amplitude, but a
quark of momentum $(p^+{+}k),\bpp$ in the conjugate amplitude.  The size
for this entry in the density matrix is $\f(\bpp)$ and \Eq{qspace}
is the time-integrated evolution equation for this amplitude;
$\delta E$ represents the phase accumulation because the state in the
amplitude and conjugate amplitude do not have the same energy, while the
$\cc(\qp)$ term describes the effect of scattering processes on the
evolution of the density matrix.  The second line in \Eq{fpproblem}
describes the overlap of the current operator on this density matrix
element.  The term in square brackets in the first line of \Eq{fpproblem} is
the DGLAP splitting kernel.

The NLO corrections to this leading order calculation arise from an $\OO(g)$
correction to thermal mass and to the differential soft scattering rate.
These contributions, like the LO contribution, need to be solved for
numerically. The most
convenient way to do so is to Fourier-transform \Eq{qspace}
to impact parameter space, where the integral equation turns into a 
differential equation with mixed boundary conditions. This will be 
discussed in subsection \ref{sub_coll_NLO}.
First we find the behavior of \Eq{fpproblem}, \Eq{qspace} at small $p^+$
and at large $\pp$, which we need as counterterms in
Eqs.~(\ref{eq_soft}) and (\ref{eq_semicoll}).  We
can find these analytically by perturbing \Eq{qspace} in $\delta E$.

\subsection{Leading order subtractions: $\Gamma_\gamma\,\vline\,_{\sc}^{{\rm coll.}\,\,{\rm subtr.}}$ and $\Gamma_\gamma\,\vline\, _{\soft}^{\rm{subtr.}}$}
In the leading order calculation of the collinear rate in \Eq{fpproblem}, the 
integral over $p^+$ extends artificially all the way down to
$p^+\lesssim gT$; and
the $\pp$ integral extends up to $\pp \gg gT$, contrary to the
definition of the
collinear kinematic region. In these regions, the integrals start to 
probe the soft and semi-collinear regions, respectively.
There, the assumptions  made in arriving
at \Eq{fpproblem} and \Eq{qspace} start to break down, so that a more
detailed analysis is needed.  To find the leading order contribution we
simply extend the $p^+$ integral to zero and the $\pp$ integral to
infinity; then when we handle the $p^+\sim gT$ and $\pp^2 \sim gT^2$
regions more carefully, we will subtract the (incorrect) amount already
included in this way in the leading-order calculation, via the
subtractions of Eqs.~(\ref{eq_soft}) and (\ref{eq_semicoll}).  So let us
evaluate \Eq{fpproblem}, \Eq{qspace} in these regions.

In the soft and semi-collinear regions,
one finds that $\delta E \sim gT$, and
therefore in these regions the first $\delta E$ term in \Eq{qspace} is larger than the
second $\cc(\qp)$ term.  Hence, the evaluation of
\Eq{fpproblem} and \Eq{qspace} can be simplified by
working to first order in $\cc(\qp)$.
Physically  this is because, if the opening angle becomes large
$\OO(\sqrt{g})$ or if an external quark becomes soft and therefore
scatters at a large angle, then the formation time becomes
$\sim 1/gT$, parametrically shorter than the mean time between
scatterings $\sim 1/g^2 T$.  Therefore the LPM effect becomes subleading
and the rate is determined by the single scattering rate.

In order to solve \Eq{fpproblem} by perturbing in $\delta E^{-1}$
write $\f = \f_1 + \f_2 + \ldots$ with $\f_1 \propto (\delta E)^{-1}$,
$\f_2 \propto (\delta E)^{-2}$ {\it etc}.\ and evaluate iteratively:
\begin{equation}
2\bpp = i\delta E \, \f_1(\bpp) \quad \Rightarrow \quad
\f_1 = \frac{2\bpp}{i\delta E(\bpp)},
     \end{equation}
which gives zero in \Eq{fpproblem} because it is imaginary. 
Substituting $\f_1$ into the collision term to determine $\f_2$, we get
\begin{eqnarray}
0 & = & i\delta E(\bpp) \f_2(\bpp) + \int \frac{d^2\qp}{(2\pi)^2}
  \cc(\qp) \left[ \f_1(\bpp) - \f_1(\bpp + \q_\perp)
 \right] \,, \nonumber \\
\f_2(\bpp) & = &  \frac{2}{\delta E(\bpp)}
  \int \frac{d^2\qp}{(2\pi)^2} \cc(\qp)
 \left( \frac{\bpp}{\delta E(\bpp)}
       -\frac{\bpp + \q_\perp}{\delta E(\bpp+\bqp)}
 \right) .
\end{eqnarray}
Since this term is real, it contributes to \Eq{fpproblem}.  The next
term $\f_3$ is again imaginary, and $\f_4$ is $\OO(g)$ in the regions of
interest and therefore negligible.  Integrating over $\pp$ as in
\Eq{fpproblem} and symmetrizing the resulting expression with respect to
$\pp,(\pp{+}\qp)$, we find%
\footnote{\label{foot_shift}This symmetrization is natural, indeed
  necessary, from the point of view of the momentum labeling which we
  introduced on the diagrams shown in Fig.~\ref{fig_diagrams}. Namely,
  the $1/(\delta E(\bpp))^2$ term corresponds to the first diagram there
  (self-energy on the lower line), the $1/(\delta E(\bpp{+}\bqp))^2$
  term is the next diagram (self-energy on the top line), and the
  cross-terms are the cat eye diagram.  In terms of
  Fig.~\ref{fig_collinear}, for the bremsstrahlung diagram the
  $1/(\delta E(\bpp))^2$ term is the diagram shown, the
  $1/(\delta E(\bpp{+}\bqp))^2$ term is the square of the diagram where
  the gluon attaches after the photon, and the cross-term is the
  interference term between these diagrams.  Not performing the symmetrization
  corresponds to evaluating one of the diagrams with different momentum
  assignments on the external lines.}
\begin{equation}
	\label{shiftbethe}
	\int \frac{d^2 \pp}{(2\pi)^2} 2\,\bpp\cdot\f_2(\bpp)  =  2
	  \int \frac{d^2\qp d^2\pp}{(2\pi)^4} \cc(\qp)
	 \left( \frac{\bpp}{\delta E(\bpp)}
	       -\frac{\bpp + \q_\perp}{\delta E(\bpp+\bqp)}
	 \right)^2 .
\end{equation}
We now turn to the application of this equation to the specifics of the 
soft and semi-collinear limits.

\subsubsection{The soft fermion, collinear contribution}
\label{sub_coll_soft}
We now consider \Eq{fpproblem} in the region where either $p^+$ or $p^++k^+$ is small.
We will show that the integrand goes over to a constant, so
$\OO(g)$ of the contribution to $\Gamma_\gamma$ arises from the region
where $p^+$ is $\OO(g)$.  This $p^+$-independent behavior will turn into
a linearly divergent subtractive counterterm when we evaluate the soft
region.

First consider \Eq{fpproblem} in the regime where $p^+$ is formally
$\OO(T)$ but soft, $p^+ \ll T$ and $p^+ \ll k$.  In this case
we can approximate \Eq{fpproblem} as
\begin{equation}
\label{smallp}
  \dgk_{\textrm{soft}}^{\textrm{subtr.}}
=\frac{\Aa(k)}{(2\pi)^3} \int dp^+ \frac{1}{(p^+)^2} \frac{1}{2}
\; \frac{1}{g^2 \crr T^2} \int \frac{d^2\pp}{(2\pi)^2}
{\rm Re}\: 2\bpp \cdot \f(\bpp,p^+) \,.
\end{equation}
Upon plugging the soft-$p^+$ limit of $\delta E$, \emph{i.e.},
\begin{equation}
\delta E = \frac{\pp^2 + \mm}{2p^+} \,,
\end{equation}
into \Eq{shiftbethe} we obtain
\begin{equation}
\int \frac{d^2 \pp}{(2\pi)^2} 2\,\bpp\cdot \f_2(\bpp)  =  8(p^+)^2 
	  \int \frac{d^2\qp d^2 \pp}{(2\pi)^4} \cc(\qp)
	 \left( \frac{\bpp}{\pp^2 {+} \mm}
	       -\frac{\bpp {+} \q_\perp}{(\bpp{+}\q_\perp)^2 + \mm}
	 \right)^2 .
\end{equation}

In terms of $p^+$ scaling, we see that \Eq{qspace} gives $(p^+)^2$ times a
$p^+$-independent function.  This cancels the $(p^+)^{-2}$ in the integrand in
\Eq{smallp}, so indeed the integrand in \Eq{smallp} is independent of
$p^+$ at small $p^+$.  Since $p^+ \sim gT$ represents $\OO(g)$ of the phase
space of $p^+$ values available, this region therefore represents an
$\OO(g)$ fraction of the photon production rate, as claimed.

The region where $p^+{+}k$ is soft gives an identical contribution.
Inserting $2\f_2$ into \Eq{smallp} we then get
\begin{eqnarray}
\nn \dgk_{\textrm{soft}}^{\textrm{subtr.}}
 &=& \frac{\Aa(k)}{(2\pi)^3} \int_{-\mu^+}^{+\mu^+} dp^+\: \frac{8}{T}\!
  \int \frac{d^2\pp d^2\qp}{(2\pi)^4}
  \frac{\md^2}{\qp^2(\qp^2+\md^2)}\\
&&\hspace{3.2cm}\times
 \left( \frac{\bpp}{\pp^2+\mm}
      -\frac{\bpp+\q_\perp}{(\bpp{+}\q_\perp)^2+\mm}
 \right)^2,
\label{divergent_bit}
\end{eqnarray}
where we introduced a regulator
$gT\ll\mu^+\ll T$ for the linear divergence. 

\subsubsection{The semi-collinear fermion, collinear contribution}
\label{sub_coll_semicollin}

The semi-collinear region represents another $\OO(g)$ contribution to the integral in 
\Eq{fpproblem}. As in the previous case, the approximations that lead
to that equation are no longer valid when $P$ becomes semi-collinear ($\pp\to \sqrt{g}T$,
$p^-\sim gT$). This limit is then incorrectly described by \Eq{fpproblem} and,
as in the previous subsection, we need to derive its limit in order to subtract it from
the semi-collinear region, where this momentum scaling will be correctly treated.

We can again use \Eq{shiftbethe}, but now there is an additional
simplification; $\pp^2 \gg \mm$ and $|\pp|\gg |\qp|$.  Therefore we can
drop $\mm$ and work to lowest order in $\qp$, which is
\begin{eqnarray}
\int \frac{d^2 \pp}{(2\pi)^2} 2\,\bpp\cdot\f_2(\bpp)
 &=& 2 \int \frac{d^2 \pp d^2 \qp}{(2\pi)^4} \cc(\qp)
  \frac{\qp^2}{(\delta E(\pp))^2}  \nonumber \\
  &=& 2\int \frac{d^2\pp}{(2\pi)^2} \; \frac{4(p^+)^2(p^++k)^2}{k^2\pp^4}
 \int \frac{d^2\qp}{(2\pi)^2}\, \qp^2\,\cc(\qp). 
\end{eqnarray}
When plugged in \Eq{fpproblem}, this yields
\begin{eqnarray}
	 \ddgkv_{\sc}^{\textrm{coll subtr.}} & = &2 \frac{\Aa(k)}{(2\pi)^3}
\int dp^+ \left[ \frac{(p^+)^2 + (p^++k)^2}{(p^+)^2 (p^++k)^2} \right]
   \frac{\nfd(k+p^+) [1-\nfd(p^+)]}{\nfd(k)}
\nonumber \\
\label{colltosemi}
 & & \times \frac{1}{g^2 \crr T^2} \int \frac{d^2\pp}{(2\pi)^2}
  \,\frac{4(p^+)^2(p^++k)^2}{k^2\pp^4}\int \frac{d^2\qp}{(2\pi)^2}\, \qp^2\,\cc(\qp) \,.
\end{eqnarray}
The $\pp$ integration is power $\pp$ divergent and the $\qp$ integral is
log UV divergent.  This is not surprising, since this expression was
obtained based on $\qp\ll\pp$.

\subsection{NLO corrections to the collinear regime: $\delta\Gamma_\gamma\,\vline\,_{\delta m}$ and $\delta \Gamma_\gamma\,\vline\,_{\delta C}$}
\label{sub_coll_NLO}
Even at leading order  \Eq{qspace} has to be solved numerically in order to get 
the collinear contribution. The most convenient way to do so
is by Fourier transforming $\bpp$ and $\q_\perp$ into impact-parameter
variables, as first proposed in \cite{Aurenche:2002wq}.  The advantages are, 
first, that the convolution over the collision kernel $\cc(\qp)$ 
becomes a product, turning an integral equation into a differential equation; 
second, that the source on the left-hand side becomes a boundary condition at $\b=0$; and third,
that the desired final integral, \Eq{fpproblem}, becomes a boundary
value of the ODE solution.  Specifically, defining
\begin{equation}
\f(\b) = \int \frac{d^2\qp}{(2\pi)^2} e^{i\b \cdot \q_\perp}\f(\bqp)\,,
\end{equation}
we have
\begin{equation}
\label{b_want}
{\rm Re}\:\int \frac{d^2\pp}{(2\pi)^2} 2 \bpp \cdot \f(\bpp)
 = {\rm Im}( 2 \nabla_{b} \cdot \f(b))\,,
\end{equation}
and \Eq{qspace} becomes
\begin{equation}
\label{bspace}
-2i \nabla \delta^2(\b) = \frac{i k}{2p^+(k+p^+)} 
\Big( \mm - \nabla_b^2 \Big) \f(\b) + \cc'(b) \f(\b) \,,
\end{equation}
with%
\footnote{Note that $\cc'(b)$ is not the Fourier transform of
$\cc(\qp)$, but rather the difference between the Fourier
transform at zero $\b$ and at finite $\b$, which is better behaved
(in particular, not sensitive to the divergent total
cross-section).  Alternately, we can redefine
$\cc(\qp)$ to have a negative delta function at $\qp=0$, normalized so
that its integral $\int d^2 \qp \cc(\qp)$ vanishes, in which case
\Eq{qspace} does not need the first term in square brackets, and $\cc'(b)$
is minus the Fourier transform of $\cc(\qp)$.}
\begin{equation}
\label{eq:C(b)}
\cc'(b) \equiv \int \frac{d^2\qp}{(2\pi)^2}
     \Big( 1 - e^{i\b \cdot \q_\perp} \Big) \cc(\qp)\,.
\end{equation}

In the collinear regime the $\OO(g)$ corrections enter then in two places: 
both the effective thermal mass squared $\mm$ and the collision kernel 
$\cc(\qp)$ get $\OO(g)$ corrections which modify \Eq{bspace},
\begin{eqnarray}
m^2_{\infty,\rm LO+NLO}& =& \mm +\delta \mm, \\
\cc'_{\rm LO+NLO}(b) &=& \cc'(b) + \delta \cc'(b).
\end{eqnarray}
The computation of the NLO thermal mass from \cite{CaronHuot:2008uw} is rederived
in Appendix \ref{app_mass}. The NLO collision kernel is computed in
\cite{CaronHuot:2008ni} in momentum space; in Appendix \ref{sub_nlo_cq}
we perform the Fourier transformation into impact parameter space.

  \Eq{bspace} is then solved perturbatively, by treating $\f(\b)$
formally as an expansion in powers of $\delta \mm, \delta \cc$;
$\f(\b) = \f_0(\b)+\f_1(\b)+\ldots$, and expanding to first order.
The zero-order expression is just \Eq{bspace}, while at the linear order
the expression reads
\begin{equation}
\label{bspace3}
0 = \left( \frac{i k}{2p^+(k+p^+)} \Big(  \mm - \nabla_b^2 \Big)
+ \cc'(b) \right) \f_1(\b)
+
\left( \frac{i k \, \delta \mm}{2p^+(k+p^+)} 
      + \delta\cc'(b) \right) \f_0(\b) \,,
\end{equation}
where the leading order solution $\f_0(\b)$ acts as a source term in the
differential equation for $\f_1(\b)$.

When evaluating \Eq{bspace} one must deal with mixed boundary
conditions; the function $\f(\b)$ must decay to zero as
$\b\rightarrow \infty$ (one boundary condition), and it must
yield the correct normalization
\begin{equation}
\label{boundaryatzero}
\nabla_b^2 \f_0(\b) = \frac{4 p^+ (k{+}p^+)}{k} \nabla \delta^2(\b)
\end{equation}
at zero.  This is done by evolving the differential equation starting at
large $\b$, with starting data which ensure that the solution will decay
away as $\b\rightarrow \infty$ but with an arbitrary normalization.
One then solves the differential equation going in towards the
origin, generally resulting in a mis-scaled solution.  But this solution
can be multiplied by a complex constant so that \Eq{boundaryatzero} is
satisfied.  Similarly, when solving \Eq{bspace3} for $\f_1(\b)$, the
boundary condition that $\f_1(\b)$ should decay at large $\b$ is not
enough to fix the solution completely; so one generically gets a
solution which is a mixture of
the solution to \Eq{bspace3} with correct boundary condition
$\lim_{\b\rightarrow 0} \nabla^2 \f_1(\b) = 0$, plus a multiple of the
homogeneous solution (that is, the solution of \Eq{bspace3} at $\f_0=0$)
with the wrong boundary condition at zero.  But the
homogeneous solution is proportional to $\f_0$, which is known; so it can
be subtracted to obtain the solution with correct boundary condition.

We solved \Eq{bspace3} as a function of $k/T$. 
The results can be parametrized as follows:
\begin{eqnarray}
\ddgkv_{\coll} &=& \ddgkv_{\dm} + \ddgkv_{\delta C}, \label{totalcoll}\\
\ddgkv_{\delta m} &=& \frac{\Aa(k)}{(2\pi)^3} \Big[\frac{\delta\mm}{\mm} C_\mathrm{coll}^{\delta m}\left(\frac{k}{T},\kappa\right) \Big]\,, \\
\ddgkv_{\delta C } &=& \frac{\Aa(k)}{(2\pi)^3} \Big[ \frac{g^2\ca T}{\md} C_\mathrm{coll}^{\delta\cc}\left(\frac{k}{T},\kappa\right) \Big]\,.
\end{eqnarray}
where
\begin{equation}
	\label{defdmm}
\frac{\delta\mm}{\mm}=-\frac{2\md}{\pi T}.
\end{equation}
The fitting functions read
\begin{eqnarray}
C^{\delta m}_{\rm coll}(x) &=& \left(\frac{2/9}{\kappa}\right)^{0.25}\Big(-0.3664/x - 0.08478 - 0.0799\,\log(x)  \nn 
\\ &&+0.0315\, x - 0.0050\, x\log(x) - 0.0681 \,(\log(x))^2\Big),\\
C^{\delta C}_{\rm coll}(x) &=& (-0.7207 - 0.8236 \delta\kappa +3.986 \delta \kappa^2 )/x \nn \\
&&+ (0.7056+ 0.0998 \delta\kappa - 1.186 \delta\kappa^2) \nn \\
&&+ (-0.8309 - 0.2610 \delta\kappa + 2.247 \delta\kappa^2) \log(x)\nn  \\
&& +(0.12305 - 0.0108 \delta\kappa - 0.2871 \delta\kappa^2)\,x\nn \\
&& +(-0.01777 + 0.00148 \delta\kappa + 0.0434 \delta\kappa^2)x\log(x)\nn \\
&& +(0.2804 - 0.0369 \delta\kappa  - 0.2375 \delta\kappa ^2)\log(x)^2\nn  \\
&& +(-0.0702 + 0.00440  \delta\kappa + 0.1149  \delta\kappa^2)\log(x)^3\,,
\end{eqnarray}
with $\delta \kappa \equiv \kappa-2/9$. The fitting functions have a relative error smaller than 2\%
for $N_f=3$ QCD ($\kappa=2/9$) in the momentum range $0.5< x < 70$. In the range
$0.15 < \kappa < 0.35$ the relative error is less than 5\%.

\begin{figure}
	\begin{center}
		\includegraphics[width=0.6\textwidth]{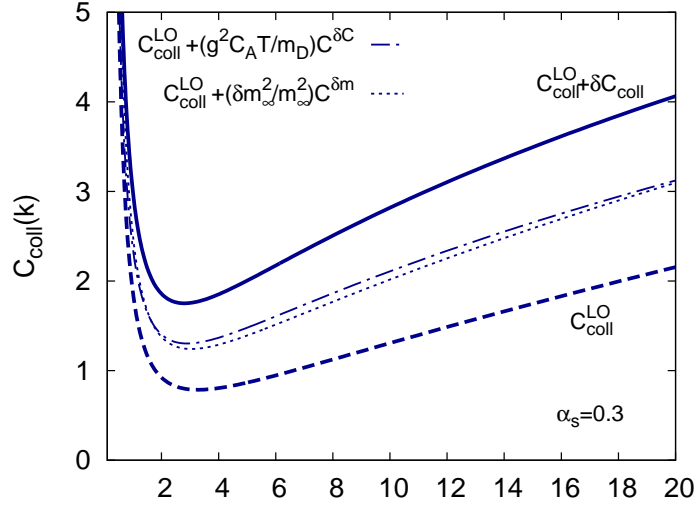}
	\end{center}
   \caption{
      Two NLO functions,
   $(\delta\mm/\mm) C_\mathrm{coll}^{\delta m}(k/T)$ and 
   $(g^2\ca T/\md)C_\mathrm{coll}^{\delta\cc}(k/T)$, 
   which parameterize the changes in the  
   the collinear emission rate  due to the NLO quasi-particle masses and collision kernel respectively  -- see \Eq{totalcoll}.
   The full LO+NLO collinear emission function is a sum
       these two corrections and the 
    leading order result, $C_{\rm coll}^{\rm LO} + \delta C_{\rm coll}$.  The curves are for  $\nc=\nf=3$ and $\alpha_s=0.30$ }
\label{plot_collinear}
\end{figure}
We will present most of the numerical results, for different values of
the parameters such as the coupling, in the exposition of the final
results in Sec.~\ref{sec_results}.
Here we just show in Fig.~\ref{plot_collinear} the size of
the mass correction, $[\delta\mm/\mm] C_\mathrm{coll}^{\delta m}(k/T,\kappa)$,  and  the collision kernel correction, 
$[g^2\ca T/\md] C_\mathrm{coll}^{\delta\cc}(k/T,\kappa)$, relative
to the the LO collinear result, $C_\mathrm{coll}^\mathrm{LO}(k/T)$.
The NLO correction $\delta C_\mathrm{coll}(k/T)$ is an $\OO(100\%)$ correction
for most of the considered range.

\section{The soft region}
\label{sec_soft}

To introduce the NLO calculation, we begin by reproducing the
soft-momentum part of the leading-order calculation introducing the
notation. We perform the leading order calculation using novel sum
rule technology, which admits a generalization to the NLO calculation.

\subsection{Leading-order evaluation and introduction to the fermionic sum rules}

\label{sub_soft_LO}

\begin{figure}
	\begin{center}
		\includegraphics[width=14cm]{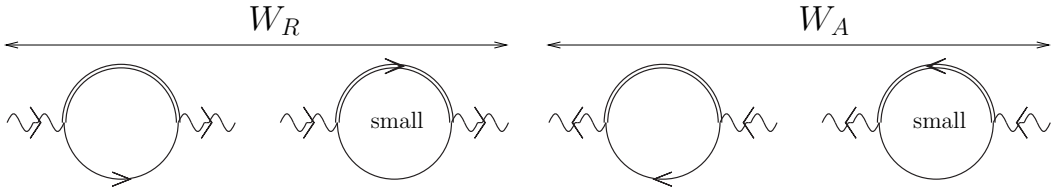}
	\end{center}
\caption{\label{fig:ra} Leading-order diagrams in the \ra basis.  
   The lines with arrows indicate retarded $ra$ propagators with the 
   arrow flow from $a$ to $r$, while lines without arrows indicate
   $rr$ propagators -- see the text for further discussion.
  The first
  two diagrams contribute to $W_R$, with either the bottom or the top
  propagator retarded; the last two contribute to $W_A$.  The diagrams
  with cut soft lines (second and fourth diagrams) are
  suppressed by the small statistical function on the cut soft line.}
\end{figure}

The most straightforward approach to the leading-order calculation is to
evaluate the trace of the Wightman correlator $W^<(K) \equiv g^{\mu\nu}
W^{<}_{\mu\nu}(K)$ in \Eq{defrate} directly in the $1,2$ basis.
However, it turns out that the NLO calculation is much simpler to perform in the
\ra basis, so we will instead use the \ra basis also at leading
order. The object that is most conveniently calculated in the \ra basis is 
the retarded (advanced) correlator, which is related to the backward Wightman
function via the KMS relation
\begin{equation}
	\label{kmswightman}
W^<(K) = 2 \nbe(k) \:{\rm Im}\: i W_R(K) = \nbe(k) (W_R(K) - W_A(K)) \,.
\end{equation}

At leading order, $W_R(K)$ and $W_A(K)$ each arise from two \ra assignments of 
the one loop diagram, shown in Figure
\ref{fig:ra}.  
Our graphical notation for the \ra assignments follows the one in \cite{CaronHuot:2007nw}: we
draw outgoing arrows for $a$ fields at vertices and incoming ones for $r$ fields. 
For the $rr$ propagator, rather than drawing two arrows pointing in opposite directions, we omit to draw 
them; this should cause no confusion.
The double lines refer to hard $(K+P)\sim T$
propagators whereas the single lines refer to soft $P\sim gT$ HTL propagators.
The different \ra assignments of resummed HTL propagators are easily
obtained from the retarded ones listed in App.~\ref{app_props} by using \Eq{raprop}.

The cut ($rr$) soft line carries a factor of $-\nfd(p^0)+1/2 \ll 1$ and is
therefore suppressed;%
\footnote{It is also odd in $p^0$ and will be even more suppressed when
  averaging over $p^+ \rightarrow -p^+$.}
we may therefore drop the two diagrams containing cut soft lines.
Summing the other diagrams gives the difference between retarded and
advanced propagators on the soft line.  This difference is
the spectral function $\rho\equiv S_R-S_A$.  Also the cut line can be
expressed in terms of the spectral function;
$S_{rr}(K+P) = (\frac{1}{2} - \nfd(k^0+p^0))\rho(K+P)$.
We approximate this statistical function as
$\frac{1}{2} - \nfd(k^0+p^0) \simeq \frac{1}{2} - \nfd(k)$ and use
the identity
\begin{equation}
\nbe(k) \left( 1 - 2\nfd(k) \right)
= \nfd(k) \,.
\end{equation}
Bringing everything together,
\begin{equation}
W^<(K)= \sum_s \frac{q^2_se^2 \dr \nfd(k)}{2} \int\frac{d^4P}{(2\pi)^4} 
   \Tr{\gamma_\mu \rho(K+P)\gamma^\mu \rho(P)}.
	\label{Wtrace}
\end{equation}
If we had evaluated $W^<$ directly without going to the \ra basis we
would have written this down immediately.

Now we evaluate \Eq{Wtrace} expanding in $P\sim gT \ll K \sim T$. This expansion will enforce eikonality on the hard line, which is   
essential for the sum rules described below. It is convenient to write each
propagator in terms of its components of positive and negative
chirality-to-helicity ratio:
\begin{equation}
\rho(P) =  h^+_P \rho^+(P) + h^-_P \rho^-(P) \, , \qquad
h_P^{\pm} \equiv \frac{\gamma^0 \mp \hat{p}\cdot \vec\gamma}{2} \,,
\end{equation}
with $\hat{p}=\p/p$.
For the hard line,  we  use an eikonal approximation
\begin{equation}
\rho^+(P{+}K) \simeq 2\pi \delta(v_k\cdot P) = 2\pi \delta(p^-) \,  ,
\qquad \rho^-(P{+}K) =2\pi \delta(p^0+k+|\p{+}\k|) \simeq 0\, , 
\end{equation}
and thus the hard line is a function of $p^-$ only.
Using this delta function simplifies the traces, which we expand in
small $P$:
\begin{equation}
	\label{leadingtrace}
	\Tr{\gamma_\mu h^+_{\bk+\bp}\gamma^\mu h^{\pm}_\bp}=2\left(1\mp\frac{p^+}{p}\right)
	\mp 2\frac{\pp^2}{pk}\pm p^+ \frac{3\pp^2}{pk^2}+\order{\frac{p^3}{k^3}}.
\end{equation}
We insert the leading-order piece of this trace into \Eq{Wtrace}, finding
\begin{equation}
\label{leadingW}
 W^{<}(K) = 2\sum_s q^2_se^2 \dr \nfd(k)
  \int\frac{dp^+ d^2 \pp}{(2\pi)^3}
  \left[\left(1{-}\frac{p^+}{p}\right) \rho^+(p^+\!,\pp) 
         +\left(1{+}\frac{p^+}{p}\right)\rho^-(p^+\!,\pp) \right]\! .
\end{equation}
The factor of two in front takes care of the kinematic region where $K+P$ is soft and
$P$ is hard, which gives the same result after a shift of integration
variables.

\begin{figure}
	\begin{center}
		\includegraphics[width=9cm]{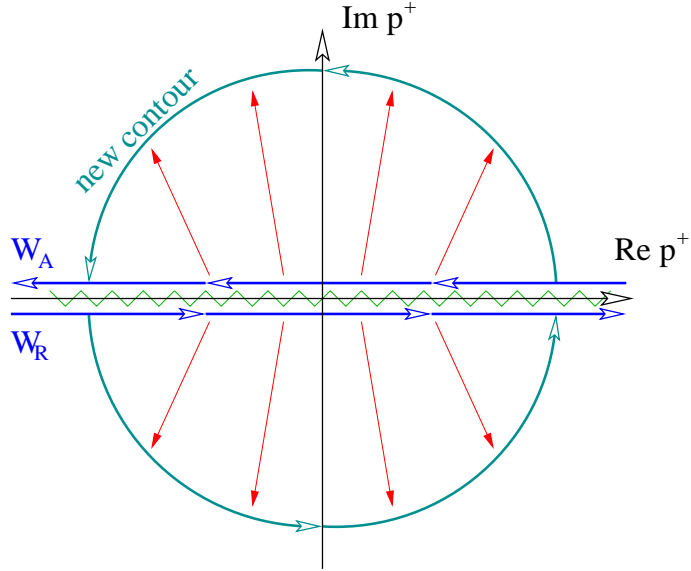}
	\end{center}
\caption{\label{contour}  Integration contour in the complex $p^+$
  integration, and the deformation we use to render $p^+ \gg gT$. $W_R$ 
runs below the real axis and $W_A$ above. This happens because the letters
$R$ and $A$ refer to the causal prescription with respect to $K$ and our momentum
assignments imply that when $W$ is retarded in $K$ it is advanced in $p^+$.}
\end{figure}

Next consider the $p^+$ integration in \Eq{leadingW}.
We will perform the integral using analyticity methods
similar to those discussed in App.~\ref{condensates}.
The key is that \Eq{leadingW} involves
$\rho(p^-,p^+,\pp) = S_R(p^-,p^+,\pp)-S_A(p^-,p^+,\pp)$.
Due to causality, the retarded and advanced functions
are analytic  in any timelike or null momentum variable 
in the upper and lower half planes respectively,
generalizing the familiar analyticity  properties of these functions in $p^0$ \cite{CaronHuot:2008ni}.
In
particular, $S_R(p^-,p^+,\pp)$ is analytic in the upper half of the
complex $p^+$ plane, while holding $p^-$ and $\pp$ fixed. 
We are therefore free to deform the $p^+$ integration contour: 
 instead of 
 integrating just above and below the real axis for $S_R$ and $S_A$ respectively, we integrate along an arc at large $p^+$   where  $gT \ll p^+ \ll T$ -- see  Figure \ref{contour}.
Along these arcs the integrand has a remarkably simple behavior, which can be obtained
by expanding the HTL propagator listed in App.~\ref{app_props}, yielding
\begin{equation}
	\label{upperarc}
\left(1-\frac{p^+}{p}\right)S^+(p^+,p)+\left(1+\frac{p^+}{p}\right)S^-(p^+,p)
\bigg\vert_{\vert p^+\vert\to+\infty}
  =\frac{i}{p^+}\frac{\mm}{p_\perp^2+\mm}+\order{\frac{1}{(p^+)^2}}.
\end{equation}
Integrating along the arcs at positive and negative $\mathrm{Im}(p^+)$ for the retarded and advanced 
contributions one then obtains
\begin{equation}
	\label{finalw}
		W^{<}(K)=2\sum_s q^2_se^2 \dr \nfd(k) 
  \int\frac{d^2\pp}{(2\pi)^2} \frac{\mm}{p_\perp^2+\mm}\,.
\end{equation}
A numerical integration of \Eq{leadingW} agrees perfectly with this expression.
This result was also recently obtained in \cite{Besak:2012qm}. 

To summarize, the reason why it is possible to deform the $p^+$ integration can
be understood diagrammatically from \Fig{fig:ra}. When evaluating $W_R$ and $W_A$ (the first and third diagrams in \Fig{fig:ra}), the soft
fermionic lines are either fully retarded or fully advanced, since   soft $rr$
fermionic propagators are suppressed. This is seen from the 
flow of 
arrows on soft fermionic lines, which indicates the \ra assignments.  Thus, since the cut hard line is eikonal and is 
only a function of $p^-$ and not $p^+$,  the $p^+$ integration is over either a
fully  retarded or fully advanced function and can be deformed 
away from the real axis.

It is worth noting that, while each of the two components $(1\mp p^+/p)S^\pm(p^+,p)$ 
presents poles separately at $p^+=\pm i\pp$ and branch cuts in $(-i\infty,-i\pp)$, $(i\pp,i\infty)$, 
their sum is analytic for $\mathrm{Im}(p^+)\ne 0$. This must be so as our sum rule is
based on causality; but the chirality-to-helicity decomposition is not
Lorentz covariant, so the individual terms need not respect causality.
This can be seen from the properties
\begin{equation}
	\label{splusminusprops}
	S^+(-P)=S^{-\,*}(P)\,,\qquad S^-(-P)=S^{+\,*}(P)\,,
\end{equation}
 from which one can also see that the sum of the two components
is covariant and does respect causality.

To constrain the integral to the soft region only, the $p^+$ integral
should be cut off at a finite momentum scale $\mu^+ \gg gT$. However,
if we cut off the integral at a finite $\mu^+ \gg gT$, the
$\OO((p^+)^{-n})$ subleading corrections to \Eq{upperarc} give rise to
$(gT/\mu^+)^{n-1}$ suppressed corrections.  But these will be canceled
by corrections which will arise when we perform the calculation of the
region above $\mu^+$, since the total result should be $\mu^+$
independent.  Therefore we need not compute them.

The $d^2\pp$ integration in \Eq{finalw} should be cut off 
at some large momentum $gT\ll \mu^\mathrm{LO}_\perp \ll T$, where it should match 
with the contribution from the hard region. The explicit expression reads
\begin{equation}
	\label{losoftrate}
	\dgk_\soft
	=\frac{\mathcal{A}(k)}{2(2\pi)^3}\ln\left(\frac{(\mu^\mathrm{LO}_\perp)^2}{\mm}+1\right)
	\approx\frac{\mathcal{A}(k)}{(2\pi)^3}\ln\frac{\mu^\mathrm{LO}_\perp}{m_\infty},
\end{equation}
which agrees with the original calculations of the LO soft region in 
\cite{Kapusta:1991qp,Baier:1991em}.\footnote{The calculation of \cite{Baier:1991em} 
used a different regularization, cutting off the $d^3p/(2\pi)^3$ integral at $p=\mu$. For any UV
log-divergent function in three dimensions that for $p> \mu$ is approximated
by its asymptotic behavior $1/p^3$, the difference between our cylindrical regularization
( $\pp<\mu$, $-\infty<p^+<\infty$) and their spherical one is  $(1-\ln(2))/(2\pi^2)$. Indeed, by inspecting Eq.~(17)
in \cite{Baier:1991em} and fixing the overall normalization one sees that their result is 
$\mathcal{A}(k)(\ln(\mu/m_\infty)+\ln(2)-1)$ (their numerical term on the second line
is $-0.31\approx \ln(2)-1$). }

The manipulations made in arriving from \Eq{Wtrace} to \Eq{leadingW} are valid up to NNLO corrections,
and we do not need revisit them in the NLO computation.
While the approximations we have made to statistical functions have
$\OO(g)$ corrections, they give rise to odd integrands in $p^+$ and would hence give a
vanishing integral when plugged in \Eq{leadingW} since the term in
square brackets is even as given by \Eq{splusminusprops}.
In a similar way one can show that the order-$g$ correction from the
trace, {\it i.e.}, the 
order $\pp^2/(pk)$ term in \Eq{leadingtrace}, results in an odd
integration. Finally, we can consider
the $\OO(g)$ correction to the dispersion relation of the 
hard line, which changes $\delta(p^-)$ to
$\delta(p^- -(\pp^2+\mm)/(2k))$.  But the difference
between these delta functions again yields
an odd integrand at the $\OO(g)$ level. 
\Eq{leadingW} is then free of $\OO(g)$ corrections.

However, the soft HTL fermion propagator is
resummed in the hard self-energy insertions as shown in
Figure~\ref{fig_lo_soft}. Whenever the momentum flowing
inside these internal loops becomes $\OO(gT)$ soft, the approximations
made in the computation of the HTL propagator fail. The diagrams
where exactly one of the internal loops becomes soft represent a
mistreated relative $\OO(g)$ contribution which needs to be subtracted
in the NLO calculation. We will return to this contribution in Sec.~\ref{sec_soft_subtr}.

\subsection{The structure of the soft NLO corrections; a quick derivation}
\label{sub_soft_NLO}

We saw in Sec.~\ref{sub_overview_NLO} that the leading-order diagram of
Fig.~\ref{fig_lo_soft} receives order-$g$ corrections from four
diagrams, shown in Fig.~\ref{fig_nlo_soft}.  These correspond to four
different ways to add a soft gluon to a LO diagram.
In these diagrams HTL corrections appear on all gluon propagators, all soft fermionic
propagators, and on vertices in three of the four diagrams.
We will compute each diagram in detail in the following subsections.
But here we will present a not-quite-rigorous argument which establishes
what the sum of the diagrams must yield.

Consider the sum of the four diagrams for $W_R$ shown in Figures \ref{fig_soft},
\ref{fig_hard}, and \ref{fig_cateye}, paying particular attention
to the \ra arrow flow.
These diagrams correspond to those of
\Fig{fig_nlo_soft}, but the causality (or \ra) structure  is clarified by the arrows. Indeed,
examining  these diagrams, we see  that the soft
fermionic line is either fully retarded or fully advanced,  as indicated by
uni-directional arrow flow along soft fermionic lines. As in the leading order
case, this is a consequence of the fact that soft fermionic $rr$ propagators
 are suppressed.  The cut fermionic lines (those without arrows) are hard, and
are only a function of $p^-$ and not $p^+$ as is typical of an eikonal
approximation.  Thus,  the $p^+$ integration is over a fully retarded or fully
advanced function, and  we are again free to deform the $p^+$ contour as in
Fig.~\ref{contour}. After this deformation  $p^+$ is everywhere large (albeit complex) relative to $p^-, p_\perp,Q$, and we are free to expand the
integrand at large $p^+$. 

The leading contribution in this expansion should be $(p^+)^0$ (arising,
for instance, from the soft propagator width, which first arises at NLO
and should give rise to precisely such a $(p^+)^0$ contribution); and the
next order should be $(p^+)^{-1}$.  Higher orders are suppressed and need
not be considered.  The $(p^+)^0$ term gives rise to a pure linear
divergence $\int d^2 p_\perp d^4 Q \int dp^+ (p^+)^0 F(Q,p_\perp)$.  A
linearly divergent $p^+$ behavior in the $p^+\sim gT$ region corresponds
to a leading-order behavior for $p^+\sim T$, so the linear divergence
{\sl must} appear as a $p^+$-independent small-$p^+$ limit of a
hard-$p^+$, leading-order contribution.  There is precisely one such
contribution, namely, the small $p^+$ limiting behavior of the collinear
region found in \Eq{divergent_bit} of Subsec.~\ref{sub_coll_soft}.
Therefore the $\OO(g)$, $(p^+)^0$ behavior must be precisely
\Eq{divergent_bit}, which was already included in the treatment of the
collinear region and so should be subtracted to avoid double counting.

Next we consider the subleading $(p^+)^{-1}$ behavior.  In the last
subsection we saw that such behavior arose already at the leading order,
and that its physical interpretation was as an asymptotic thermal mass.
While it is not obvious, it is at least not surprising that the
subleading contribution should be precisely a shift to \Eq{finalw} in
which $\mm$ is replaced by $\mm+\delta \mm$, as defined in
\Eq{minfNLO1}.  If we make this replacement and then expand to linear
order in $\delta \mm$, we find
\begin{eqnarray}
\frac{\mm + \delta \mm}{\pp^2 + \mm + \delta \mm}
& = & \frac{\mm}{\pp^2+\mm} 
 + \delta \mm \left( \frac{1}{\pp^2+\mm} - \frac{\mm}{(\pp^2+\mm)^2} \right)
\nonumber \\
& = &  \frac{\mm}{\pp^2+\mm} + \delta \mm \; \frac{\pp^2}{(\pp^2+\mm)^2} \,.
\label{soft_guess}
\end{eqnarray}
The first term corresponds to the leading-order result and should be
subtracted off; the second term is a true NLO correction.  That is, we
expect that the soft contribution at NLO should be
\begin{equation}
(2\pi)^3\ddgkv_{\soft}  = \sum_s q_s^2 e^2 \dr \frac{\nfd(k)}{k}
\int \frac{d^2 \pp}{(2\pi)^2} \; \delta \mm \:
\frac{\pp^2}{(\pp^2+\mm)^2} \,.
\label{soft_quick}
\end{equation}
Note that,
like the leading-order term, this will also give rise to a logarithmic
large-$\pp$ divergence, which must be balanced by some logarithmic
behavior at larger $\pp$.  In this case the corresponding logarithmic
behavior will be found in the semi-collinear region.

In conclusion, we expect three contributions from the soft region; the
leading-order contribution which should be subtracted off, the infrared
limit of the collinear contribution which should also be subtracted, and
\Eq{soft_quick}.  The argument supporting this result
is not rigorous, so we need to proceed with the actual evaluation of
each diagram, making full use of the $p^+$ contour deformation
technique.  Since $p^+$ can always be taken as large, we will actually
not need the vertex HTL's at all, and each diagram will become an
expansion in $p^+$ as described above.
We then sum the diagrams to get a gauge invariant total,
whereupon we can perform the $\int d^4 Q$ integral to find that
we indeed get exactly the behavior described above.

We believe that the very simple form of the large $p^+$ expanded result
can be made rigorous and understood physically in terms of eikonalized
dipole propagation in the medium, and that this method can then be
extended to other problems such as gluon radiation; we plan to return to
this topic elsewhere.

\subsection{Soft diagrams}
We now turn to the diagram-by-diagram evaluation of the soft diagrams.
The purpose of the
next three subsections is to support with a concrete calculation the
arguments of the last subsection, yielding in the end the same result
presented there.  We will concentrate on computing $W_R$; $W_A$ is
trivially related, corresponding to a contour in the other half-plane.

The diagrams that contribute to the soft region at NLO are 
those in Fig.~\ref{fig_nlo_soft}. We parametrize the different contributions 
of the soft NLO diagrams by
\begin{equation}
(2\pi)^3 \ddgkv_{\rm soft}^{\rm diags.} =\frac{1}{2 k}\left( W^{<}_s + W^{<}_h  + W^{<}_c - W^{<}_{\rm{subtr.}}\right)
\end{equation}
where $W^{<}_s$ includes contributions arising
from the two first diagrams of Fig.~\ref{fig_nlo_soft}, \emph{i.e.}, from 
diagrams where the additional soft propagator gives a self-energy
correction to the soft fermion. $W^{<}_h$ 
arises from the diagram where the hard fermion receives a self-energy
correction and $W^{<}_c$ gives the contribution of the 
remaining ``cat eye'' diagram. The last term $W^{<}_{\rm subtr.}$ arises
from a mistreated kinematical region in the leading order calculation
where the momentum in one of the hard loops in the resummed HTL 
propagator becomes soft, shown in Fig.~\ref{fig_nlo_soft_counterterm}.

For each diagram there are several \ra assignments of the 
propagators and vertices that may
contribute to the diagram. In practice though, most of these
are suppressed by powers of $g$ and
give subleading contributions. The power-counting in the pure glue theory
has been worked out in \cite{CaronHuot:2007nw}. Here we extend the 
power counting to a theory with fermions. 
Because $G_R \propto 1/p^2$ while $S_R \propto 1/p$, the soft retarded gauge
and fermionic propagators scale as $1/(gT)^2$ and $1/gT$ respectively.
The symmetrized $rr$ gauge propagator is proportional to
$\nbe(p^0)+1/2 \simeq T/p^0 \sim 1/g$ and is therefore enhanced by an
extra factor of $1/g$; contrarily, the fermionic $rr$ propagator is
proportional to $-\nfd(p^0)+1/2 \simeq p^0/4T \sim g$ and so is suppressed by
an extra factor of $g$.  Therefore in practice one always needs diagrams
with the maximum number of soft $rr$ gauge boson propagators, but no
soft fermionic $rr$ propagators.  These rules are summarized in
Table~\ref{tab_pc}.
\begin{table}[ht]
	\begin{center}
	\begin{tabular}{|c|c|}
	\hline
	Propagator & Scaling\\
	\hline
	Soft, retarded gluon &$1/(g^2T^2)$\\
	Soft, $rr$ gluon &$1/(g^3T^2)$\\
	Soft, retarded fermion& $1/(gT)$\\
	Soft, $rr$ fermion& $1/T$\\
	\hline
	\end{tabular}
	\end{center}
	\caption{Power counting for the soft propagators.}
	\label{tab_pc}
\end{table}

All bare vertices have an odd
number of $a$ indices; 
those with one $a$ index are the same as the
zero-temperature ones, while those with three $a$ indices carry a factor
of $1/4$.  The three-$a$ vertices remove Bose stimulation factors by reducing 
the number of available $rr$ propagators and hence lead to subleading
corrections in the current context, and we will not encounter them. 

 Fig.~\ref{fig_nlo_soft} suggests that we would need
the explicit forms of the HTL quark-gluon vertex
and the two-quark, two-gluon vertex.
However, while the HTL vertices are the same
order as the bare vertices when all momenta are $\sim gT$, if a momentum
$p$ entering an HTL vertex becomes large $p\gg gT$, the vertex becomes
suppressed, relative to the bare vertex, by $(gT/p)^2$.  We will see
that sum rules allow all diagrams (or combinations of them) to be evaluated in terms of their
large-momentum behavior; therefore the total effect of HTL vertices will
cancel exactly at NLO and we will not need their detailed form in the
calculation. Therefore we defer their treatment to
App.~\ref{app_htl}.

\subsubsection{The soft-soft self-energy diagrams: $W_s$}

\begin{figure}[ht]
	\begin{center}
		\includegraphics[width=5cm]{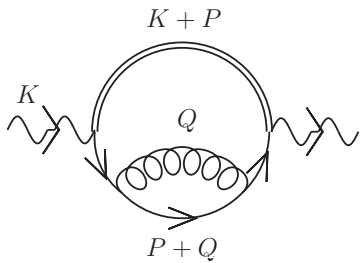} 
	\end{center}
	\caption{The retarded diagram for the soft self-energy insertion. The arrows indicate \ra
	flow. Internal lines without arrows are understood to be $rr$  propagators. The direction of 
	fermion flow is clockwise and fermion momenta are always oriented along fermion flow. Wherever 
	the arrow of \ra flow is parallel (antiparallel) to fermion flow this gives rise to a 
	retarded (advanced) propagator.}
	\label{fig_soft}
\end{figure}

We start with $W^{<}_s(K)$, the contribution to the retarded correlator from the self-energy insertion 
on the soft line. We observe that deviations from eikonality on the hard line are now NNLO 
at the largest, being suppressed by $P/K\sim g$, thus effectively reducing the hard line 
to an integral over $x^+$ of an eikonal Wilson line in the same direction
multiplied by $\slashed{v}_k$ and the appropriate statistical 
factor, \emph{i.e.}, $(\nfd(k)-1/2)\slashed{v}_k\delta(iv_k\cdot D)$.

As in the leading order case, we evaluate the retarded correlator and apply the KMS
relation of \Eq{kmswightman} to obtain the backward Wightman correlator. Therefore, 
the left external line is of type $r$ and the right one is of type $a$. There are several
possible choices of \ra assignments for the internal lines, but
given the power-counting rules in Table~\ref{tab_pc}, only one represents 
an order-$g$ correction to the leading-order result. It is the one where the gluon is $rr$, 
thus receiving a $1/g$ Bose enhancement, and the hard fermion is $rr$ too. This \ra 
assignment and its corresponding momenta are shown in Fig.~\ref{fig_soft}.

Enforcing eikonality on the hard line,
$S_{rr}(K+P)=-\slashed{v}_k(1/2-\nfd(k))2\pi\delta(2v_k\cdot P)$,
the retarded amplitude reads%
\footnote{
We remark that there is an extra subtlety for fermions, since $S_R(-P)=-S_A(P)$. To assign the right 
prescription to propagators one should consistently assign fermion momenta parallel to 
fermion flow: wherever the arrow of \ra flow is parallel (antiparallel) to momentum/fermion flow
this gives rise to a retarded (advanced) propagator.
}
\begin{eqnarray}
\hspace{-0.5cm} \nn W^{R}_s(K)&=&-2 e^2\sum_s q^2_s e^2\dr g^2 \crr \int\frac{d^4P}{(2\pi)^4}
\int\frac{d^4Q}{(2\pi)^4}\left(\frac12-\nfd(k)\right)2\pi\delta(v_k\cdot P) G^{rr}_{\mu\nu}(Q)\\
			\label{defws}
			&&\hspace{5.5cm}\times\Tr{\slashed{v}_kS_A(P)\gamma^\mu S_A(P+Q)\gamma^\nu S_A(P)},
\end{eqnarray} 
where the factor of 2 accounts again for the possibility of having either of the two lines soft. 
Let us recall that in the soft approximation $G^{rr}_{\mu\nu}(Q)=T/q^0\rho_{\mu\nu}(Q)$ and
that the retarded HTL propagators in Coulomb gauge are given in App.~\ref{app_props}.

Performing the $p^-$ integration in \Eq{defws} over the $\delta$-function and defining 
for conciseness 
$\mathcal{B}(k)\equiv e^2\sum_s q^2_s e^2\dr g^2 \crr(1/2-\nfd(k))$ we have
\begin{equation}
	\label{wspz}
	 W^{R}_s(K)=-2 \mathcal{B}(k) \int\frac{dp^+d^2\pp}{(2\pi)^3}\int\frac{d^4Q}{(2\pi)^4}
	G^{rr}_{\mu\nu}(Q)\mathrm{Tr}\bigg[\slashed{v}_kS_A(P)\gamma^\mu S_A(P+Q)\gamma^\nu S_A(P)\bigg]_{p^-=0}.
\end{equation}
This expression is a fully advanced function of $p^+$. Using the analyticity arguments
introduced in the previous section, we can again deform the contour away from the real axis 
in the lower half-plane without encountering poles or branch cuts. Let us call $\cc_A$ 
the arc going from $-\mu^+ -i\epsilon$ 
to $+\mu^+ -i\epsilon$ at $\vert p^+\vert=\mu^+\gg gT$, $\mathrm{Im}(p^+)<0$. The integrand 
simplifies dramatically along this integration contour: the result of the trace and the 
propagators can be expanded for large $\vert p^+\vert$. As we argued, we need only the 
terms up to order $1/p^+$. We then have
\begin{eqnarray}
	\nn W_s^{R}(K)&=&-2 \mathcal{B}(k) \int\frac{d^2\pp}{(2\pi)^2}\int_{\cc_A}
	\frac{dp^+}{2\pi}\int\frac{d^4Q}{(2\pi)^4}\Bigg[\frac{i\pp^2  G^{++}_{rr}(Q)}
	{(p^+)^2\delta E_\bp^2(q^--i\epsilon)}\left(1+\frac{\delta E_{\bp+\bq}}{(q^--i\epsilon)}\right)\\
	\label{wsexpanded}
	&&\hspace{4cm} +\frac{i\pp^2 G_T^{rr}(Q)}{2(p^+)^3 \delta E^2_\bp}\left(-\frac{2q_z\left(1-\frac{q_z^2}{q^2}\right)}{(q^--i\epsilon)}+1+\frac{q_z^2}{q^2}\right)\Bigg],
\end{eqnarray}
where $G_T$ is defined as $G_{ij}(Q)=(\delta_{ij}-\hat{q}_i\hat{q}_j)G_T(Q)$
 (see \Eq{htltrans}), terms proportional to 
$\bpp\cdot\bqp$, that average to zero in the azimuthal integration, have been omitted, and
\begin{equation}
	\label{defdeltae}
	\delta E_\bp=\frac{\pp^2+\mm}{2p^+},\qquad \delta E_{\bp+\bq}=\frac{(\bpp+\bqp)^2+\mm}{2p^+}.
\end{equation}
We have furthermore used the fact that along $\cc_A$ the components of $S(P+Q)$ become
\begin{equation}
	\label{spqcontour}
	S_A^+(P+Q)\to\frac{i}{q^--\delta E_{\bp+\bq}-i\epsilon},\qquad S_A^-(P+Q)\to \frac{i}{2p^+},
\end{equation}
where in obtaining \Eq{wsexpanded} we have expanded the ``$+$'' component for small 
$\delta E_{\bp+\bq}$, since there is no other pole for $q^-=0$ on the opposite side
of the complex plane, which would cause a pinch singularity. This will no longer 
be true when evaluating the cat eye diagram. 

\subsubsection{Soft leading order subtraction: $W_{\rm subtr.}$}
\label{sec_soft_subtr}

Let us now turn to the subtracted counterterm of Figure \ref{fig_nlo_soft_counterterm}. 
The entire advanced HTL self-energy, which is given by the simple one-loop self-energy
graph taking one of the two bare propagators in the loop to be $rr$ and the other to be 
advanced, with loop momentum hard and external momentum soft, results in 
the well known $\Sigma_\mathrm{HTL}\sim g T$. The self-energy we have inserted
in Fig.~\ref{fig_soft} is however $\OO(g^2T)$ by construction. This implies that 
we have already implicitly subtracted all of the $\OO(gT)$ HTL self-energy and we
have to only worry about $\OO(g)$ regions in the calculation of the HTL
self-energy, where the approximations taken for its derivation fail. The only such region
is the limit where the gluon becomes soft, which clearly overlaps with the
phase space of the calculation we have just performed.
A certain care is then needed in subtracting
only this part of the HTL self-energy. To this end we take \Eq{wspz} and replace $G^{rr}(Q)$
with $G^{(0)\,rr}(Q)=T/q^0\rho^{(0)}(Q)$, the soft limit of the bare gluon propagator 
($\rho^{(0)}(Q)$ is the bare spectral density). 
For what concerns the fermion propagator, we replace $S_A(P+Q)$ with the bare one 
and, following with the HTL
approximation, we keep only $\slashed{Q}$ at the numerator, \emph{i.e.},
\begin{equation}
	\label{wspzsubtr}
	 W^R_{\rm subtr.}(K)=-2i \mathcal{B}(k) \int\frac{d^3p}{(2\pi)^3}\int\frac{d^4Q}{(2\pi)^4}\left.
	\frac{G^{(0)\,rr}_{\mu\nu}(Q)\mathrm{Tr}\big[\slashed{v}_kS_A(P)\gamma^\mu \slashed{Q}
	\gamma^\nu S_A(P)\big]}{P^2+Q^2+2P\cdot Q+i\epsilon(p^0+q^0)}\right\vert_{p^-=0}.
\end{equation}
The attentive reader might think that we could be neglecting other $\OO(g)$ regions,
since in general, for a soft particle, the one-loop self-energy is equal to the HTL
self-energy plus $\OO(g)$ corrections, such as those that would arise from the inclusion
of $\slashed{P}$ at the numerator. However the sum rule approach leads us to evaluate
the self-energy at $p^+\gg gT$ and very close to the light-cone, where the full
and HTL self-energies agree (as long as loop momenta remain hard, of course) and those $\OO(g)$
corrections vanish. Indeed, we have explicitly checked that the inclusion of $\slashed{P}$
in \Eq{wspzsubtr} leads to vanishing extra contributions, at least up to order $1/p^+$.

We can then evaluate \Eq{wspzsubtr} in analogy with the previous calculation, 
expanding along $\cc_A$. We obtain
\begin{equation}
	 W^{R}_{\rm subtr.}(K)=-2i \mathcal{B}(k) \int\frac{d^2\pp}{(2\pi)^2}\int_{\cc_A}\frac{dp^+}{2\pi}\int\frac{d^4Q}{(2\pi)^4}\frac{\pp^2  G^{(0)\,rr}_{T}(Q)}{(p^+)^3\delta E_\bp^2},
		\label{wsexpandedct}
	\end{equation}
where we have used the fact that the bare longitudinal spectral density vanishes
in Coulomb gauge.

\subsubsection{The hard-soft self-energy diagram: $W_{h}$}
\begin{figure}[ht]
	\begin{center}
		\includegraphics[width=5cm]{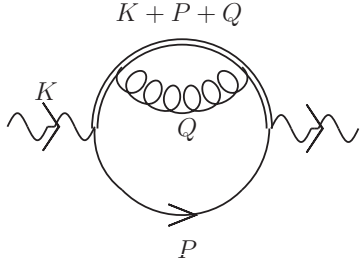}
	\end{center}
	\caption{The retarded diagram for the hard self-energy insertion. 
	Conventions are as in Fig.~\ref{fig_soft}. The effective one-loop propagator 
	for the hard line is understood to be $rr$.}
	\label{fig_hard}	
	\end{figure}
In this subsection we compute the contribution arising from the hard-soft self-energy diagram of Figure \ref{fig_hard}
which we denote $W_h$. 
The power counting requires again that the soft line be advanced 
and the hard line, considered as a one-loop propagator, be $rr$, as shown in Fig.~\ref{fig_hard}, 
together with the chosen momentum assignments.\footnote{\label{foot_shift_h}
The momentum assignments differ from those adopted
in Fig.~\ref{fig_diagrams} and in the semi-collinear calculation. They amount to a shift, which will
be undone in the end, when we will perform the subtraction of \Eq{divergent_bit}.}
The KMS relation gives 
$S_{rr}(K+P)=(1/2-\nfd(k+p))(S_R(K+P)-S_A(K+P))$, as in \Eq{raprop}. The one-loop
retarded and advanced propagators are obtained by inserting the retarded or advanced
self-energy in the corresponding propagator. 
When plugged in our diagram, the advanced term gives rise to a fully advanced loop and thus vanishes, 
since all poles are on the same side of the $p^-$ complex plane. 
The expression for $W^{R}_h(K)$  reads
\begin{equation}
	\label{defwh}
	W^R_h(K)=-2i \mathcal{B}(k)\int\frac{d^4P}{(2\pi)^4}\int\frac{d^4Q}{(2\pi)^4} \frac{ G^{++}_{rr}(Q)\Tr{\slashed{v}_kS_A(P)}}{(v_k\cdot P-i\epsilon)^2(v_k\cdot(P+Q)-i\epsilon)},
\end{equation}
where the Dirac and Lorentz structures of the hard line are again those of an eikonal Wilson line,
i.e. $v^\mu v_k^\nu\slashed{v}$. Furthermore, the Wilson line gives rise to the 
retarded eikonal propagators, which are functions of $p^-$ and $q^-$ only. We can
thus exploit their independence on $p^+$ to deform the contour along $\cc_A$
and expand $\Tr{\slashed{v}_k S_A(P)}$ along it, yielding
\begin{eqnarray}
\nn	W^{R}_h(K)&=&+4i \mathcal{B}(k)\int_{\cc_A}\frac{dp^+}{2\pi}\int\frac{dp^-}{2\pi}\int\frac{d^2\pp}{(2\pi)^2}\int\frac{d^4Q}{(2\pi)^4} \frac{ G^{++}_{rr}(Q)}{(v_k\cdot P-i\epsilon)^2(v_k\cdot(P+Q)-i\epsilon)}\\
	\label{whdeform}
	&&\hspace{5.8cm}\times\left(\frac{i\pp^2}{2(p^+)^2(p^--\delta E_\bp-i\epsilon)}+\frac{i}{p^+}\right),
\end{eqnarray}
where the $(p^--\delta E_\bp-i\epsilon)^{-1}$ in round brackets on the second line
 has not been expanded for small $\delta E_\bp$ due to the presence of a pinch
singularity with the double pole at $p^-=-i\epsilon$. We can then perform the 
$p^-$ integration by closing the contour above, picking the 
residue from the pole in the first term in round brackets
\begin{equation}
	W^{R}_h(K)=2i \mathcal{B}(k)\int_{\cc_A}\frac{dp^+}{2\pi}\int\frac{d^2\pp}{(2\pi)^2}\int\frac{d^4Q}{(2\pi)^4} \frac{ G^{++}_{rr}(Q)}{\delta E_\bp^2(q^-+\delta E_\bp+i\epsilon)}\frac{\pp^2}{(p^+)^2}.
	\label{whdeform2}
\end{equation}
We now observe that the $Q$ integration is free of pinch singularities when 
$\delta E_\bp$ goes to zero, so that we can safely expand the denominator. 
Furthermore the change $Q\to-Q$ (recall that $G_{rr}(Q)$ is even in $Q$) 
brings the resulting expression to be  identical to some of the terms in 
\Eq{wsexpanded}, \emph{i.e.}
\begin{equation}
W^{R}_h(K)=-2i \mathcal{B}(k)\int_{\cc_A}
  \frac{dp^+}{2\pi}\int\frac{d^2\pp}{(2\pi)^2}
  \int\frac{d^4Q}{(2\pi)^4} \frac{\pp^2 G^{++}_{rr}(Q)}
  {(p^+)^2\delta E_\bp^2(q^--i\epsilon)}
  \left(1+\frac{\delta E_\bp}{q^--i\epsilon}\right)\,.
	\label{whexpanded}
\end{equation}

\subsubsection{The cat eye diagram: $W_c$}
\begin{figure}[ht]
\begin{center}
	\includegraphics[width=5cm]{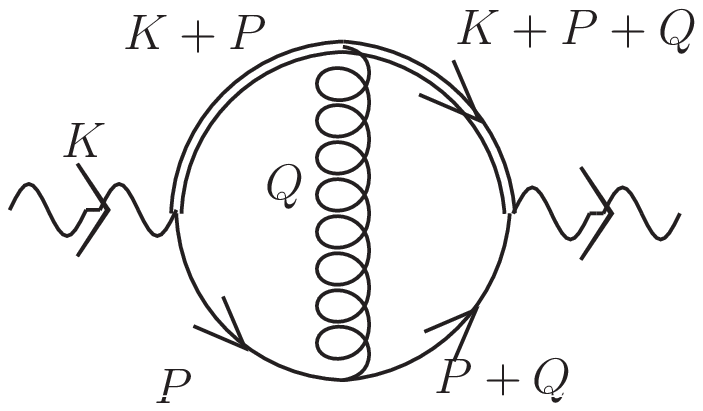}
	\includegraphics[width=5cm]{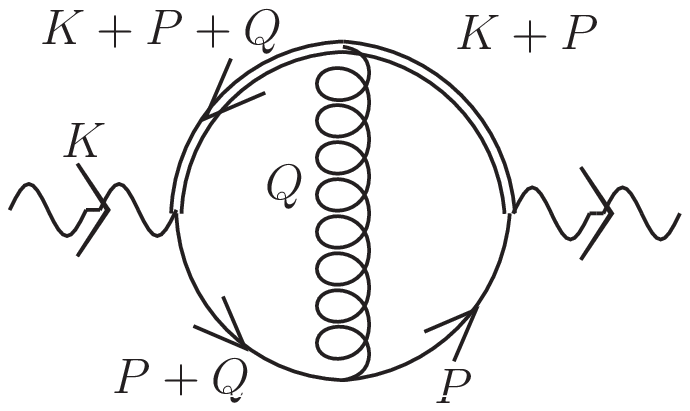}
\end{center}
\caption{The  assignments needed for the $\order{g}$ soft correction
 to the cat eye diagram. Conventions are as in Fig.~\ref{fig_soft}; 
in particular, internal lines without arrows are understood to be $rr$ propagators.	}
\label{fig_cateye}
\end{figure}

Next we consider the cat-eye diagram of Fig.~\ref{fig_cateye}.
We label its retarded amplitude $W_c^R(K)$. 
The two assignments contributing are shown 
in Fig.~\ref{fig_cateye}. For the assignment on the right we have operated a shift, 
so that the momentum flowing in the hard $rr$ propagator is always $K+P$.\footnote{
\label{foot_shift_c}This is analogous to what was done for the hard-soft self-energy.} 
The amplitude obtained by summing the two assignments 
in Fig.~\ref{fig_cateye} is then
\begin{eqnarray}
\nn		W_c^{R}(K)&=&2i \mathcal{B}(k)\int\frac{d^4P}{(2\pi)^4}\int\frac{d^4Q}{(2\pi)^4}\left(\frac{1}{v_k\cdot(P+Q)+i\epsilon}+\frac{1}{v_k\cdot(P+Q)-i\epsilon}\right) 2\pi\delta(v_k\cdot P)\\
&&\hspace{5cm}\times G_{\mu\nu}^{rr}(Q)v^\mu\Tr{\slashed{v}_kS_A(P+Q)\gamma^\nu S_A(P)},
\label{defwc}
\end{eqnarray}
where we have used the fact that $\Tr{\slashed{v}_kS_A(P+Q)\gamma^\nu S_A(P)}=\Tr{\slashed{v}_kS_A(P)\gamma^\nu S_A(P+Q)}$.
At this point we integrate over the $\delta$-function in $dp^-$, \emph{i.e.}, 
	\begin{eqnarray}
\nn		W^{R}_c(K)&=&2i \mathcal{B}(k)\int\frac{dp^+d^2\pp}{(2\pi)^3}\int\frac{d^4Q}{(2\pi)^4}\left(\frac{1}{v_k\cdot Q+i\epsilon}+\frac{1}{v_k\cdot Q-i\epsilon}\right) \\
&&\hspace{3.6cm}\times\delta G_{\mu\nu}^{rr}(Q)v^\mu\Tr{\slashed{v}_kS_A(P+Q)\gamma^\nu S_A(P)}\big\vert_{p^-=0}.
\label{wcpzero}
	\end{eqnarray}
We	observe again that the resulting function is fully advanced in $p^+$, 
allowing for a deformation along $\cc_A$. We furthermore notice that 
the retarded eikonal propagator $(v_k\cdot Q-i\epsilon)^{-1}$ introduces a pinch 
singularity in the $Q$ integration, since $S^+_A(P+Q)$ turns into the form given
in \Eq{spqcontour}. 
In order to make the pinch explicit we rewrite the terms in round brackets above as
	\begin{eqnarray}
\nn		W^{R}_c(K)&=&2i \mathcal{B}(k)\int\frac{dp^+d^2\pp}{(2\pi)^3}\int\frac{d^4Q}{(2\pi)^4}\left(\frac{2}{v_k\cdot Q+i\epsilon}+2i\pi\delta(v_k\cdot Q)\right) \\
&&\hspace{3.6cm}\times G_{\mu\nu}^{rr}(Q)v^\mu\Tr{\slashed{v}_kS_A(P+Q)\gamma^\nu S_A(P)}\big\vert_{p^-=0}.
\label{wcdelta}
	\end{eqnarray}
	Upon deforming the contour to $\cc_A$ we obtain
		\begin{eqnarray}
	\nn		W^{R}_c(K)&=&2i \mathcal{B}(k)\int\frac{d^2\pp}{(2\pi)^2}\int_{\cc_A}\frac{dp^+}{2\pi}\int\frac{dq^+d^2\qp}{(2\pi)^3}\left\{2\left[\int\frac{dq^-}{2\pi}\frac{\pp^2  G^{++}_{rr}(Q)}{(p^+)^2\delta E_p(q^--i\epsilon)^2}\right]\right.\\
\nn	&&\hspace{3.5cm}+i\frac{\pp^2+\bpp\cdot\bqp}{(p^+)^2\delta E_\bp\delta E_{\bp+\bq}} G^{++}_{rr}(q^+,\qp)\left(1-\frac{q^+}{p^+}\right)\\
&&\hspace{0.2cm}	\left.-i\frac{q^+(\qp^2+\bpp\cdot\bqp)G^{rr}_T(q^+,\qp)}{q^2\,(p^+)^3\delta E_\bp\delta E_{\bp+\bq}}\left(\bpp\cdot\bqp+\pp^2-\delta E_p p^+\right)\right\},
	\label{wcexpanded}
		\end{eqnarray}	
	where we have again dropped terms proportional to $\bpp\cdot\bqp$  in the trace
	wherever they would have vanished in the angular integration. We furthermore 
	observe that the terms proportional to $q^+/p^+$ on the second and third lines 
	vanish upon integration, since $G_{rr}(q^+,\qp)$ is an even function of $q^+$.

\subsubsection{Summary and result}
We can now sum Eqs.~\eqref{wsexpanded}, \eqref{whexpanded} and
\eqref{wcexpanded} and subtract the counterterm given by \Eq{wsexpandedct}
to obtain the NLO retarded amplitude
$W^R(K)=W^{R}_s+W^{R}_h+W^{R}_c-W^{R}_{\rm subtr.}$. Most of the terms
proportional to $(q^--i\epsilon)^{-2}$ cancel, yielding
\def\bigvbox{\vphantom{\frac{2q_z\left(1-\frac{q_z^2}{q^2}\right)}{(q^-)}}}
\begin{eqnarray}
\hspace{-0.5cm} W^R(K)&=&-2i \mathcal{B}(k) 
  \int\frac{d^2\pp}{(2\pi)^2}
  \int_{\cc_A}\frac{dp^+}{2\pi}
  \int	\frac{d^3q}{(2\pi)^3}
  \left\{-i\frac{\pp^2+\bpp\cdot\bqp}
  {(p^+)^2\delta E_\bp\delta E_{\bp+\bq}} G^{++}_{rr}(q^+,\qp)
 \bigvbox \right.\\ \nn &&\hspace{3.5cm}
  +\int\frac{dq^0}{2\pi}\frac{\pp^2}{(p^+)^2\delta E_\bp^2}
  \left[  \frac{ G^{++}_{rr}(Q)}{(q^--i\epsilon)}
  \left(2+\frac{\qp^2}{2p^+(q^--i\epsilon)}
 \bigvbox \right)\right.\\ & & \hspace{2.2cm}\left.\left. 
 +\frac{ G_T^{rr}(Q)}{2p^+}
 \left( -\frac{2q_z\left(1-\frac{q_z^2}{q^2}\right)}{(q^--i\epsilon)}
  +1+\frac{q_z^2}{q^2}\right)\right]
  -\frac{\pp^2 G^{(0)\,rr}_{T}(Q)}
 {(p^+)^3\delta E_\bp^2}\right\}, \nonumber
\label{wexpandedtotal1}
\end{eqnarray}
where the last term is the subtracted counterterm. Furthermore, 
in dealing with the contribution from \Eq{wsexpanded}, we have  
dropped the term proportional to $\bpp\cdot\bqp$ in 
$\delta E_{\bp+\bq}=\delta E_\bp+(\qp^2+2\bpp\cdot\bqp)/(2p^+)$. Exploiting the even 
nature of the gluon $rr$ propagator we can rewrite the first term in round brackets on 
the second line in terms of $\delta(q^-)$, yielding, after some rearrangements,
\begin{eqnarray}
 \nn  W^R(K)&=&-2i \mathcal{B}(k) 
 \int\frac{d^2\pp}{(2\pi)^2}\int_{\cc_A}
 \frac{dp^+}{2\pi}\int\frac{d^3q}{(2\pi)^3}
\left\{ \frac{iG^{++}_{rr}(q^+,\qp)}{(p^+)^2}
 \left(\frac{\pp^2}{\delta E_\bp^2}-\frac{\pp^2+\bpp\cdot\bqp}{\delta
     E_\bp\delta E_{\bp+\bq}}\right) \bigvbox \right.
 \\ \nn &&\hspace{-1.6cm} \left.
 +\int\frac{dq^0}{2\pi}\frac{\pp^2}{2(p^+)^3\delta E_\bp^2}
 \left[  \frac{\qp^2 G^{++}_{rr}(Q)}{(q^--i\epsilon)^2} 
 + G_T^{rr}(Q)\left(1+\frac{q_z^2}{q^2}
 -\frac{2q_z\left(1-\frac{q_z^2}{q^2}\right)}{(q^--i\epsilon)}\right)
 -2G^{(0)\,rr}_{T}(Q)\right]\!\!\right\}.\\
		&&	\label{wexpandedtotal}
\end{eqnarray}
The first line  is independent 
of $p^+$, since $\delta E_{\bp},\,\delta E_{\bp+\bq}\propto 1/p^+$.  We
already encountered the $dq_z$ integral in \Eq{C_LO} (see also
App.~\ref{app_cq}):
\begin{equation}
\label{agz}
\int\frac{dq_z}{2\pi} G^{++}_{rr}(q_z,q)=\frac{T\md^2}{\qp^2(\qp^2+\md^2)}.
\end{equation}
The second line is proportional to $1/p^+$.  We show in
\Eq{Zg_minkowski} that the $Q$ structure, without the subtraction
of $G^{(0)\,rr}_{T}$, can be identified
as the two-dimensional condensate $Z_g$ defined in \Eq{defZ_g} when
expressed in Coulomb gauge;
\begin{equation}
\label{qstruct}
\int \frac{d^4 Q}{(2\pi)^4} 
  \left[  \frac{\qp^2 G^{++}_{rr}(Q)}{(q^--i\epsilon)^2} 
 + G_T^{rr}(Q)\left(1+\frac{q_z^2}{q^2}
 -\frac{2q_z\left(1-\frac{q_z^2}{q^2}\right)}{(q^--i\epsilon)}\right)
  \right] = Z_g \,.
\end{equation}
The subtraction of $G^{(0)\,rr}_{T}$
precisely removes the leading-order contribution, leaving the NLO
correction to $Z_g$, $\dZg$.  [To see this, note that
in Coulomb gauge the longitudinal spectral density
vanishes and the transverse one is proportional to $\delta(Q^2)$.]
In the UV, the bare and resummed propagator
become identical, up to suppressed $\OO(\md^2/Q^4)$ corrections, so that we can safely
integrate up to infinity; therefore $\dZg$ is finite.
We evaluate it in App.~\ref{app_zg}, finding
$\dZg=-T\md/(2\pi)$, a result originally due to Caron-Huot
\cite{CaronHuot:2008uw}.

Plugging everything into \Eq{wexpandedtotal} we obtain
	\begin{eqnarray}
		\nn  W^R(K)&=& \mathcal{B}(k) \int\frac{d^2\pp}{(2\pi)^2}\int_{\cc_A}\frac{dp^+}{2\pi}\left\{\int\frac{d^2\qp}{(2\pi)^2}\frac{2}{(p^+)^2}\left(\frac{\pp^2}{\delta E_\bp^2}-\frac{\pp^2+\bpp\cdot\bqp}{\delta E_\bp\delta E_{\bp+\bq}}\right)\frac{T\md^2}{\qp^2(\qp^2+\md^2)}\right.\\
		 &&\left.\hspace{4.7cm}+i\frac{\pp^2}{(p^+)^3\delta E_\bp^2}\frac{T\md}{2\pi}\right\}. 
				\label{wtotalqperp}
	\end{eqnarray}
We can now perform the straightforward $dp^+$ integral along $\cc_A$, leading to	
	\begin{eqnarray}
					\nn  W^R(K)&=&\mathcal{B}(k)T \int\frac{d^2\pp}{(2\pi)^2}\int\frac{d^2\qp}{(2\pi)^2}
					\frac{8\mu^+}{\pi(\pp^2+\mm)}\frac{\md^2}{\qp^2(\qp^2+\md^2)}\\ \nn&&\hspace{5cm}\times\left(\frac{\pp^2}{\pp^2+\mm}-\frac{\pp^2+\bpp\cdot\bqp}{(\bpp+\bqp)^2+\mm}\right)\\
		 &&- \mathcal{B}(k)T\int\frac{d^2\pp}{(2\pi)^2}\frac{\pp^2}{(\pp^2+\mm)^2}\frac{\md}{\pi},
					\label{wtotalcontour2}
				\end{eqnarray}
where we observe that $p^+$-independent terms give a linear divergence. We can now plug this result 
in the KMS relation \eqref{kmswightman} to obtain the corresponding Wightman amplitude 
and from that the soft contribution to the rate,
 which reads
			\begin{eqnarray}
				\nn  (2\pi)^3\ddgkv_\mathrm{soft}^{\rm diags.}&=&\Aa(k) \frac{16\mu^+}{T}
				  \int \frac{d^2\pp d^2\qp}{(2\pi)^4}
				  \frac{\md^2}{\qp^2(\qp^2+\md^2)}\\
			\nn	&&\hspace{3.2cm}\times
				 \left( \frac{\bpp}{\pp^2+\mm}
				      -\frac{\bpp+\q_\perp}{(\bpp{+}\q_\perp)^2+\mm}
				 \right)^2\\
	 &&+ \frac{\sum_s q^2_se^2 \dr \,\nfd(k)}{k}\int\frac{d^2\pp}{(2\pi)^2}\frac{\pp^2}{(\pp^2+\mm)^2}\delta \mm, 
				\label{wtotalfinal}
			\end{eqnarray}
where we have shifted the terms on the second line of \Eq{wtotalcontour2} as in 
footnote~\ref{foot_shift}.\footnote{Had we used the momentum assignments of 
Fig.~\ref{fig_diagrams}, we would have obtained the result directly in this form, see
footnotes~\ref{foot_shift_h} and \ref{foot_shift_c}.} We recall that the expression 
for the NLO correction $\delta \mm$ to the asymptotic mass is given by \Eq{minfNLO1}. 
This shows clearly how the contribution on the second line is nothing but what would 
have been obtained by substituting $\mm\to\mm+\delta\mm$ in the leading-order 
result \eqref{finalw} and expanding in $g$, as we anticipated.

The first two lines in \Eq{wtotalfinal} are equal to \Eq{divergent_bit}, which is
the soft-$p^+$ limit of the LO collinear region we have analyzed in 
Sec.~\ref{sub_coll_soft}. There we concluded that \Eq{divergent_bit} was to be subtracted from 
the rate obtained here in the soft region. Doing that removes the linear divergence and yields
			\begin{eqnarray}
			\nn	(2\pi)^3\ddgkv_\mathrm{soft}&\equiv& (2\pi)^3\ddgkv_\mathrm{soft}^{\rm diags.} -(2\pi)^3\dgk_{\rm soft}^{\rm subtr.}\\
			\nn&=&\frac{\sum_s q^2_se^2 \dr \,\nfd(k)}{k}\delta m^2_\infty\int\frac{d^2\pp}{(2\pi)^2}\frac{\pp^2}{(\pp^2+\mm)^2} .\\
			\nn&=&-\frac{\md}{\pi T}\mathcal{A}(k)\left[\ln\left(\frac{(\mu^\mathrm{NLO}_\perp)^2}{\mm}+1\right)-\frac{(\mu^\mathrm{NLO}_\perp)^2}{(\mu^\mathrm{NLO}_\perp)^2+\mm}\right]\\
			&\approx&-\frac{\md}{\pi T}\mathcal{A}(k)\left[2\ln\left(\frac{\mu^\mathrm{NLO}_\perp}{m_\infty}\right)-1\right],
			\label{softtotal}
				\end{eqnarray}
where the $d\pp$ integration has been cut off at $\mu^\mathrm{NLO}_\perp$. According to the analysis in 
Sec.~\ref{sub_overview_NLO} as summarized in Fig.~\ref{fig_nlomap}, we expect that the UV divergence				
will be removed by an IR one in the semi-collinear region, which we analyze next. Hence	$\mu^\mathrm{NLO}_\perp$ 
obeys $g^2T^2\ll(\mu^\mathrm{NLO}_\perp)^2\ll gT^2$.

The result justifies the simplified approach in Subsection
\ref{sub_soft_NLO}.  At NLO, the contribution of the soft region is
precisely the soft limit of the collinear contribution, the
leading-order soft contribution modified by the shift $\mm \rightarrow
\mm + \delta \mm$, and nothing else.

\section{The semi-collinear region }
\label{sec_semicollin}

\begin{figure}[ht]
	\begin{center}
		\includegraphics[width=14cm]{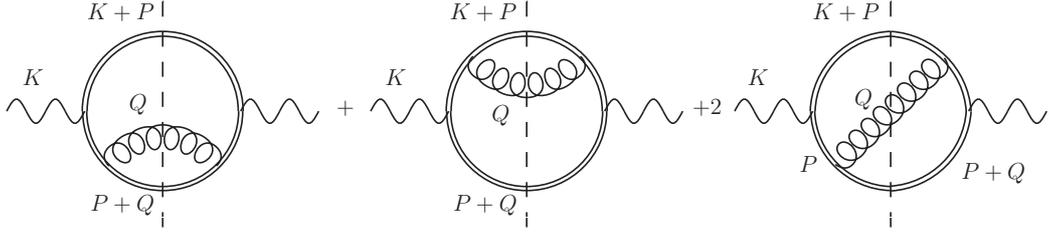}
	\end{center}
	\caption{The cuts that have to be evaluated for the semi-collinear calculation, with their corresponding
	momentum assignments.}
	\label{fig_semicuts}
\end{figure}

As we have seen in previous sections, we must treat separately the
region $Q \sim gT$ and $gT \ll \pp \ll p^+$.  Specifically we
consider $\pp^2 \sim gT^2$ while $p^+ \sim T$, which means that
the angle between $\p$ and $\k$ is small but not as small as in the
collinear region; hence we will refer to this as the
{\sl semi-collinear} region%
\footnote{With the momentum assignments in Fig.~\ref{fig_diagrams}, a semi-collinear
$P$ and a soft $Q$ imply that the momenta flowing through all fermionic lines ($P+Q$, $K+P$, $K+P+Q$)
 are semi-collinear.}.
In this case, one must
compute the diagrams shown in Fig.~\ref{fig_semicuts}, and then apply
systematically the expansion $k , p^+ \gg \pp \gg \qp,q^+$.

Actually we have already evaluated these diagrams using the collinear
expansion, since it is precisely these diagrams which give rise to the
linear-in-collisions expressions we found in
Subsec.~\ref{sub_coll_semicollin}.  In particular, \Eq{colltosemi} was derived
by making an expansion in $p^+ \gg \pp$, and it still applies, with one
proviso.  In evaluating the collision strength $\cc(\qp)$ in \Eq{C_LO},
we treated $\pp \sim \qp \sim gT$, leading to $\delta E \sim g^2 T$.
This let us neglect $\delta E$ when working out the kinematics of the
gauge bosons.  But if $\pp^2 \sim g T^2$ then $\delta E \sim gT$ and
this is no longer permissible.  In particular, when we put $(P+K)$ and
$(P+Q)$ on shell (see the cuts in Fig.~\ref{fig_semicuts}), we find
\begin{eqnarray}
\label{deltapminus}
 \delta((K+P)^2) & = & \frac{1}{2\vert
   p^++k\vert}\delta\left(p^--\frac{\pp^2}{2(p^++k)}\right), \\
\nn	\delta((P+Q)^2)&=&\frac{1}{2\vert p^++q^+\vert}\delta\left(q^--\frac{2p^+\delta E+\qp^2+2\,\bpp
\cdot\bqp-2q^+p^-}{2(p^++q^+)}\right)\\
\label{deltaqminus}
&=&\frac{\delta(q^--\delta E)}{2\vert p^+\vert}+\OO(\sqrt{g})\,, \qquad
	\delta E = \frac{k\,\pp^2}{2p^+(p^++k)},
\end{eqnarray}
where the $\OO(\sqrt{g})$ correction comes from $\bpp\cdot\bqp$ and always vanishes in the
angular integrations.  Therefore we must re-derive \Eq{C_LO} with these
somewhat different kinematics.  A straightforward computation shows that
the quantity
\begin{equation}
\label{qhat}
\frac{\hat{q}}{g^2 \crr} \equiv \frac{1}{g^2 \crr} \int \frac{d^2
  \qp}{(2\pi)^2} \,\qp^2 \,\cc(\qp)
 = \int \frac{d^4 Q}{(2\pi)^3} \delta(q^-) \qp^2 G_{rr}^{++}(Q) \,,
\end{equation}
physically interpreted as the momentum diffusion coefficient and present
in \Eq{colltosemi}, should be replaced with its finite $\delta E$
generalization,
\begin{equation}
\label{semiint}
 \Isc\equiv\int\frac{d^4Q}{(2\pi)^3}\delta(q^--\delta E)\left[	\qp^2 G^{++}_{rr}(Q) + G_T^{rr}(Q)
	\left(\left[1{+}\frac{q_z^2}{q^2}\right]\delta E^2-2q_z\delta E
 \left[1{-}\frac{q_z^2}{q^2}\right]\right)\right].
\end{equation}
This expression reverts to \Eq{qhat} in the limit $\delta E\to 0$.
Physically it represents the result of integrating over the cut
gluon line in the scattering diagrams shown in Fig.~\ref{twototwo},
treating that line as soft.  In
the case $\delta E=0$ kinematics force the gluon line to be in the
Landau cut, but for $\delta E \neq 0$ it can also be on-shell (on the
plasmon pole); therefore both processes will contribute, so
\Eq{semiint} will smoothly go over from the collinear splitting rate to
the hard scattering rate as we increase $\pp$ and hence $\delta E$.

This $\delta E$-dependent momentum diffusion coefficient can be
evaluated using Euclidean methods, see App.~\ref{app_cq}, with
the result, see \Eq{finalssgluon}, that
\begin{equation}
\Isc =  T\int\frac{d^2\qp}{(2\pi)^2}\left[
   \frac{\md^2\qp^2}{(\qp^2+\delta E^2)(\qp^2+\delta E^2+\md^2)}
   +\frac{2\delta E^2}{\qp^2+\delta E^2}\right].
\label{finalssgluon2}
\end{equation}
Unfortunately, performing the integral using Euclidean methods obscures
what part arises from the Landau cut and what part arises from the
plasmon pole.

Using \Eq{finalssgluon2} rather than \Eq{qhat} in \Eq{colltosemi}, we
find
\begin{eqnarray}
(2\pi)^3\frac{d\Gamma_\gamma}{d^3k}
    \bigg\vert_\mathrm{\sc}^{\rm diags}
 & = & \frac{2}{T} \mathcal{A}(k)\int dp^+ 
  \frac{\nfd( k+p^+) (1-\nfd(p^+))}{\nfd(k)}
  \frac{ (k+p^+)^2+(p^+)^2}{(p^+)^2 (p^++k)^2} \nn \\
\label{beforesubtr}
 & & \hspace{-2.0cm}\times\int\frac{d^2\pp}{(2\pi)^2}
  \frac{1}{\delta E^2}\int\frac{d^2\qp}{(2\pi)^2}
   \left[\frac{\md^2\qp^2}{(\qp^2+\delta E^2)
   (\qp^2+\delta E^2+\md^2)}+\frac{2\delta E^2}{\qp^2+\delta E^2}\right]. 
\end{eqnarray}
The $d^2\qp$ integration here is UV divergent, but the divergences will
be removed when we subtract the two $\OO(g)$ regions of the leading order
calculation that overlap with the semi-collinear phase space and which
have already been included in those calculations.
One such region was discussed in Sec.~\ref{sub_coll_semicollin}; when we
evaluated the collinear splitting rate, we integrated over this phase
space region but made the approximation that
$\delta E \rightarrow 0$.  Therefore we should subtract
$\hat{q}(\delta E=0)$ from $\hat{q}(\delta E)$ used above.
The other contribution we must subtract is the semi-collinear part of
the phase space of the hard $\2to2$ contribution.  The
contribution we already included when we performed the hard
$\2to2$ calculation corresponds to treating the gluon as free and
on-shell.  Therefore, we should also subtract from $\hat{q}(\delta E)$,
its value obtained by using the free gluon propagator,
$G^{rr}(Q) \rightarrow G^{(0)\,rr}(Q)=T/q^0\rho^{(0)}(Q)$, in \Eq{semiint}.
We will call this quantity $\hat{q}(\delta E)|_{\rm bare}$.  Explicitly,
we find
\begin{eqnarray}
\nn \hspace{-2ex} \Isc\Bigg\vert_{\rm bare}&\!=&\int\frac{d^4Q}{(2\pi)^3}
    \delta(q^-{-}\delta E)2G_T^{(0)rr}(Q) \delta E^2
\\
  & = & T\int\frac{d^3q}{(2\pi)^3}2 G_T^{(0)E}(0,q_z,\qp)\delta E^2\,
  2\pi\delta(q_z{+}\delta E)
= T\int\frac{d^2\qp}{(2\pi)^2}\frac{2\delta E^2}{\qp^2+\delta E^2},
\label{finalssgluonbare}
\end{eqnarray}
where, as in our treatment of \Eq{qstruct}, we have used the fact that
in Coulomb gauge the longitudinal
spectral density vanishes and the transverse one is proportional to $\delta(Q^2)$ to simplify
the integrand. We have used again the Euclidean techniques of App.~\ref{app_magic} for the evaluation of the integral, although
it can simply be evaluated in real time from the simple form of
$\rho^{(0)}_T(Q)=\sgn(q^0)\,2\pi\,\delta(Q^2)$.

Upon subtracting the two semi-collinear limits, \emph{i.e.},
 \begin{equation}
\ddgkv_{\sc} = \ddgkv_{\sc}^{\rm{diags.}} - \dgk_{\sc}^{\textrm{coll. subtr.}} - \dgk_{\sc}^{\textrm{hard subtr.}}
 \end{equation}
we have
\begin{eqnarray}
\nn	\ddgkv_{\sc}&=&\frac{2}{T} \frac{\mathcal{A}(k)}{(2\pi)^3}\int dp^+
\frac{\nfd( k+p^+) (1-\nfd(p^+))}{\nfd(k)}\frac{ (k+p^+)^2+(p^+)^2}{(p^+)^2 (p^++k)^2}\\
\nn	&&\times\int\frac{d^2\pp}{(2\pi)^2}\frac{1}{\delta E^2}\int\frac{d^2\qp}{(2\pi)^2}\left[\frac{\md^2\qp^2}{(\qp^2+\delta E^2)(\qp^2+\delta E^2+\md^2)} -\frac{\md^2}{\qp^2+\md^2}\right],\\
\label{collsubtracted}
&&
\end{eqnarray}
which is convergent in $\qp$. The region where $\qp\gg gT$ represents a very small contribution
to the integral, allowing us to extend the integration over the whole range. 

In order to evaluate the remaining integrals it is convenient
to operate a kinematical distinction. Given our momentum assignments and the fact that 
$p^+\approx p^0+q^0$, $p^++k\approx p^0+k$, we can clearly see that the region where
$p^+(p^++k)$ is positive corresponds, in terms of elementary processes, to having a semi-collinear
quark or antiquark both in the initial and final state, \emph{i.e.}, to a Compton-like (for timelike $Q$)
or bremsstrahlung-like (for spacelike $Q$) process. Conversely, a negative $p^+(p^++k)$
is associated with the pair annihilation region, with a $q\overline{q}$ pair in 
the initial state.

The integrand in \Eq{collsubtracted} is symmetric under 
$p^+ \rightarrow -k-p^+$, so the
brem/Compton region is given by $2\int_0^\infty dp^+$ and the 
annihilation region by  $2\int_{-k/2}^0 dp^+$;
\begin{equation}
\ddgkv_{\sc}= \ddgkv_{\sc}^{\rm brem./Compt.} + \ddgkv_{\sc}^{\rm pair} .
\end{equation}
We have evaluated these contributions numerically using the same  cutoff $\mu_\perp^\mathrm{NLO}$ 
for the IR
divergent $p_\perp$-integral  as in the soft region (see \Eq{softtotal}); the details are in 
Appendix.~\ref{app_sc}.
Summing up the contribution from brem/compton and pair processes, \emph{i.e.}, Eqs.~\eqref{bremcomptontotal}
and \eqref{pairtotal}, we have the full contribution from the semi-collinear region. It reads
\begin{equation}
	\label{semicollintotal}
		(2\pi)^3\frac{d\Gamma_\gamma}{d^3k}\bigg\vert_\mathrm{\sc}=-\frac{\md}{2\pi T} \mathcal{A}(k)\left[4\ln\left(\frac{\sqrt{2T\md}}{\mu_\perp^\mathrm{NLO}}\right)+C_\mathrm{brem/compton}\left(\frac{k}{T}\right)+C_\mathrm{pair}\left(\frac{k}{T}\right)\right],
\end{equation}
where the functions $C_\mathrm{brem/compton}$ and $C_\mathrm{pair}$ are fitted by \Eq{fitcbrem} and \Eq{cpairfit}, respectively.
Upon comparing this expression with the final result in the soft region, namely
\Eq{softtotal}, we notice how the dependence on $\mu_\perp^\mathrm{NLO}$
drops out of their sum.

\section{Results}
\label{sec_results}
We can now collect all contributions and write the final result for the NLO calculation.
Let us parametrize it as the sum of the leading-order result and its $\OO(g)$ correction
\begin{equation}
	\label{defnlo}
	(2\pi)^3\frac{d\Gamma_\gamma}{d^3k}\bigg\vert_\mathrm{LO+NLO}=(2\pi)^3\frac{d\Gamma_\gamma}
	{d^3k}\bigg\vert_\mathrm{LO}+(2\pi)^3\frac{d\delta\Gamma_\gamma}{d^3k},
\end{equation}
where the LO result is given by \Eq{totallo} and the $\OO(g)$ correction can 
be obtained by summing the collinear contribution, \Eq{totalcoll},  the 
soft one, \Eq{softtotal}, and the semi-collinear one, \Eq{semicollintotal}, 
yielding
\begin{eqnarray}
	\nn	(2\pi)^3\frac{d\delta\Gamma_\gamma}{d^3k}&=& \mathcal{A}(k)\left[
	\frac{\delta\mm}{\mm}\ln\left(\frac{\sqrt{2T\md}}{m_\infty}\right)
	+\frac{\delta\mm}{\mm}C_\mathrm{soft+sc}\left(\frac{k}{T}\right)\right.\\
&&\left.\hspace{1.2cm}	+\frac{\delta\mm}{\mm}C_\mathrm{coll}^{\delta m}\left(\frac{k}{T},\kappa\right)
	+\frac{g^2\ca T}{\md}C_\mathrm{coll}^{\delta\cc}\left(\frac{k}{T},\kappa\right)\right].
	\label{total!}
\end{eqnarray}
The dependence on the regulator $\mu_\perp^\mathrm{NLO}$ cancels in the sum
of the semi-collinear and soft regions, as anticipated. The function 
$C_\mathrm{soft+sc}(k/T)$ is obtained by summing the non-logarithmic
terms in the semi-collinear and soft contribution. It reads
\begin{equation}
	\label{cs+sc}
C_\mathrm{soft+sc}\left(\frac{k}{T}\right)=	\frac14\left[C_\mathrm{brem/compton}
\left(\frac{k}{T}\right)+C_\mathrm{pair}\left(\frac{k}{T}\right)\right]-\frac12.
\end{equation}
Finally, we recall that $\mathcal{A}(k)$ and $\kappa$ are given in Eqs.~\eqref{A_of_k} and
 \eqref{defkappa} and $\delta\mm/\mm=-2 \md/(\pi T)$, as given by \Eq{defdmm}.
The correction $C_{\mathrm coll}^{\delta\cc}$ is intrinsically
nonabelian, but $\delta \mm/\mm$ is nonvanishing in an Abelian theory.

\begin{figure}
	\includegraphics[width=0.495\textwidth]{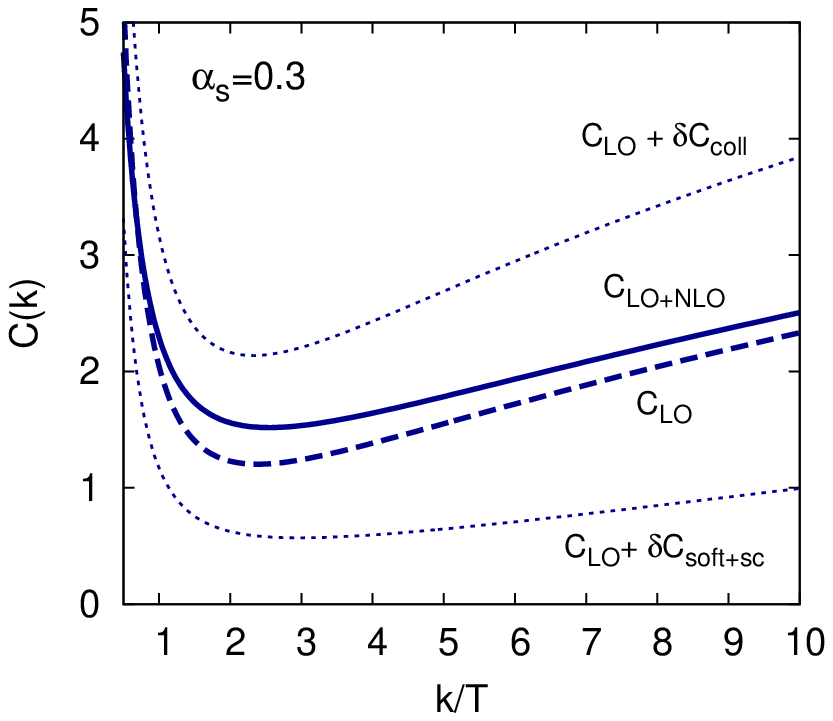}
	\includegraphics[width=0.495\textwidth]{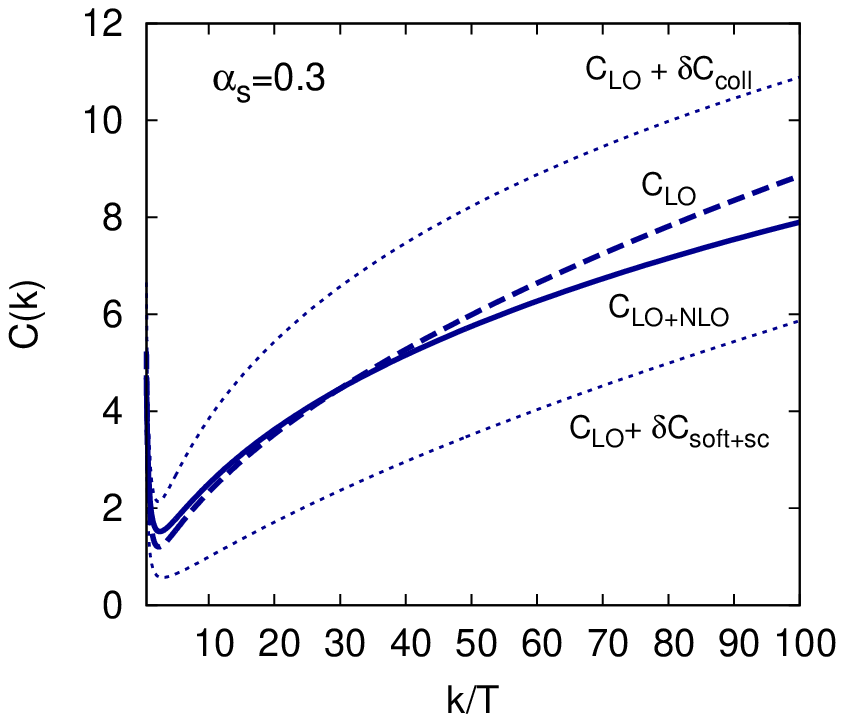}
   \caption{Left: the function, $C(k/T)$, parametrizing the photon emission rate 
      for $\nc=\nf=3$ and $\als=0.3$
      (see \Eq{gammanlo} and \Eq{A_of_k}).
         The full next-to-leading order function
         ($C_{\LO+\NLO}$) is a sum of the leading-order result ($C_{\LO}$), a 
         collinear correction ($\delta C_{\rm coll}$),  and a
         soft+semi-collinear correction ($\delta C_{\rm soft+sc}$).  
         The dashed curve  labeled
         $C_{\LO}+\delta C_{\rm coll}$ 
         shows the result when  only the collinear correction
         is included, with the analogous notation  for the  $C_{\LO} + \delta C_{\rm soft+sc}$ curve.  The difference between the dashed curves provides a 
         uncertainty estimate  for the NLO calculation.
      Right: the same as on the left but for larger $k/T$.}
	\label{plot_c_30_1}
\end{figure}

We now plot our results. Let us define
\begin{eqnarray}
	\label{defclo}
C_\mathrm{LO}\left(\frac{k}{T}\right)&\equiv&	\ln\left(\frac{T}{m_\infty}\right)
	+C_\mathrm{2\leftrightarrow2}\left(\frac{k}{T}\right)+C_\mathrm{coll}^{\mathrm{LO}}\left(\frac{k}{T},\kappa\right),\\
		\label{defcsoftsc}
	\delta C_\mathrm{soft+sc}\left(\frac{k}{T}\right)&\equiv&	\frac{\delta\mm}{\mm}\left[\ln\left(\frac{\sqrt{2T\md}}{m_\infty}\right)
		+C_\mathrm{soft+sc}\left(\frac{k}{T}\right)\right],\\
		\label{defdeltac}
	\delta C\left(\frac{k}{T}\right)&\equiv&\delta C_\mathrm{coll}\left(\frac{k}{T}\right)+	\delta C_\mathrm{soft+sc}\left(\frac{k}{T}\right),\\
	\label{defcnlo}
	C_\mathrm{LO+NLO}\left(\frac{k}{T}\right)&\equiv&C_\mathrm{LO}\left(\frac{k}{T}\right)+\delta C\left(\frac{k}{T}\right).
\end{eqnarray}
Given those definitions, it then follows that
\begin{equation}
	\label{gammanlo}
	(2\pi)^3\frac{d\Gamma_\gamma}{d^3k}\bigg\vert_\mathrm{LO}= \mathcal{A}(k)\,
	C_\mathrm{LO}\left(\frac kT\right),\qquad	(2\pi)^3\frac{d\Gamma_\gamma}
	{d^3k}\bigg\vert_\mathrm{LO+NLO}= \mathcal{A}(k)\,
	C_\mathrm{LO+NLO}\left(\frac kT\right).
\end{equation}
In Fig.~\ref{plot_c_30_1},  we start by plotting the function $C_\mathrm{LO+NLO}(k/T)$ 
for $\alpha_s=0.3$ and $\nc=\nf=3$.
In the phenomenologically interesting
momentum range, $k/T \sim 10$, the collinear 
and semi-collinear+soft corrections largely
cancel, leading to a small positive correction of order $\sim 15\%$ (Fig.~\ref{plot_c_30_1}(a)). 
At large momentum,  $k/T \simg 20$,  the LO and LO+NLO curves cross and 
the NLO correction turns negative (Fig.~\ref{plot_c_30_1}(b)).
We believe that the large cancellations we observe are rather accidental,
and one should thus consider the curves $C_\mathrm{LO}(k/T)+\delta C_\mathrm{coll}(k/T)$
and $C_\mathrm{LO}(k/T)+\delta C_\mathrm{soft+sc}(k/T)$ as  upper and lower limits respectively of an ``uncertainty estimate'' of the NLO calculation.

\begin{figure}
	\includegraphics[width=0.495\textwidth]{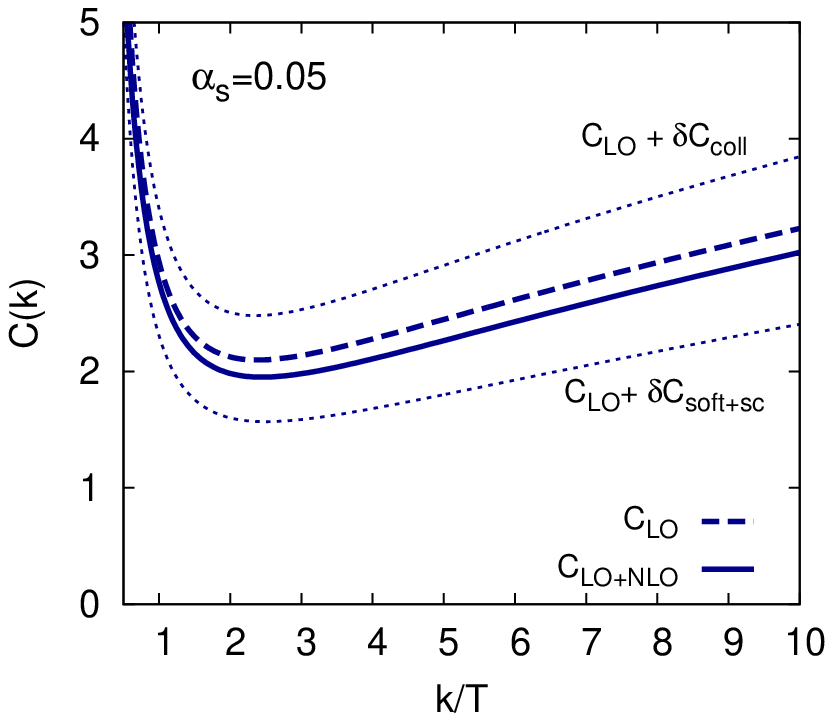}
	\includegraphics[width=0.495\textwidth]{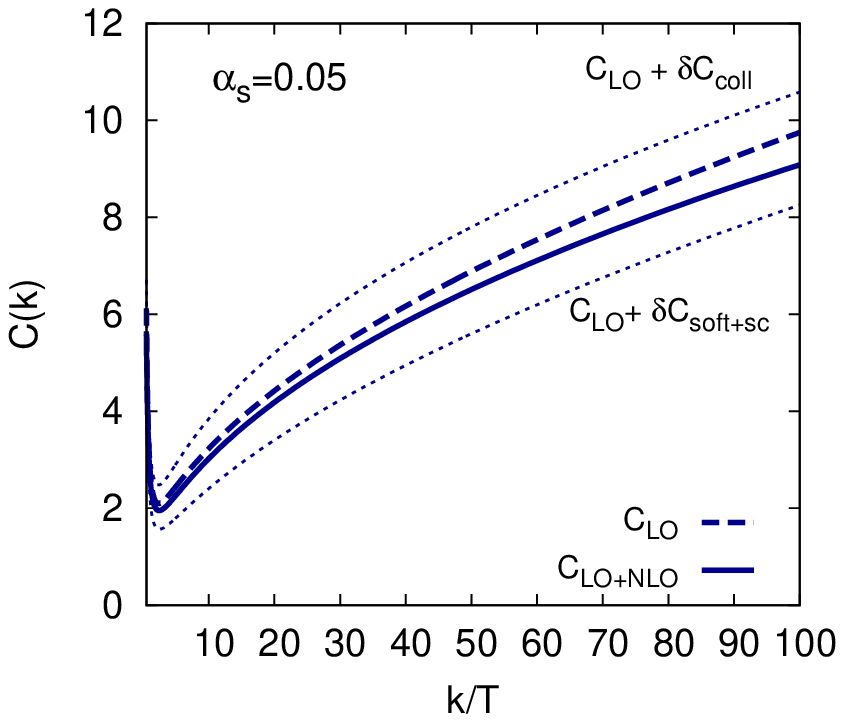}
   \caption{The functions $C(k/T)$ for $\nc=3$, $\nf=3$ as in Fig.~\ref{plot_c_30_1}, but for   $\als=0.05$.}
	\label{plot_c_5_1}
\end{figure}

In Fig.~\ref{plot_c_5_1} we plot $C_\mathrm{LO+NLO}(k/T)$ and $C_\mathrm{LO}(k/T)$
 for  $\als=0.05$, and $\nc=3$, $\nf=3$. For the smaller coupling constant the NLO correction
 is always negative and rather flat, 
 and the
magnitude of the two largely canceling contributions is also significantly smaller
than in the previous case.
\begin{figure}
		\includegraphics[width=0.495\textwidth]{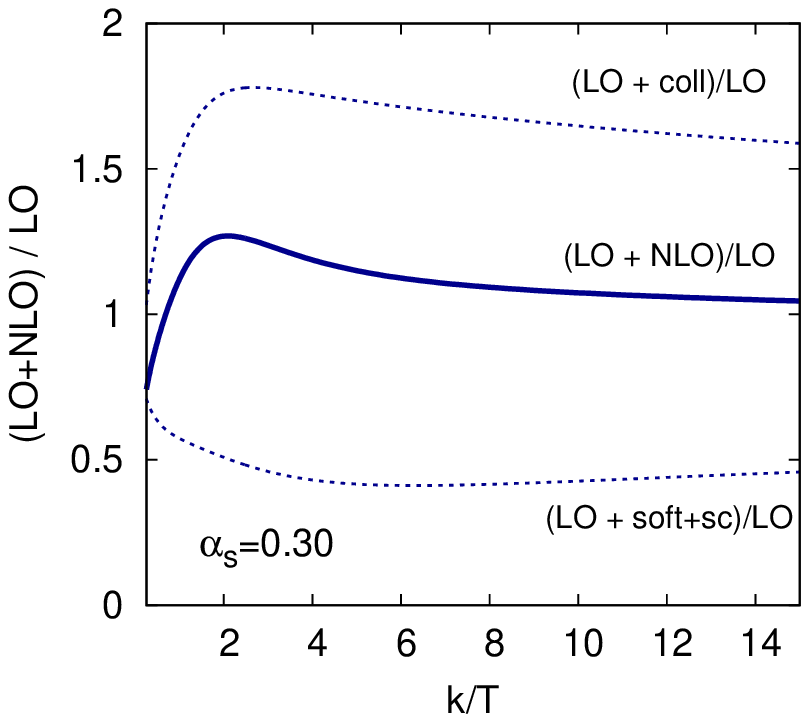}
		\includegraphics[width=0.495\textwidth]{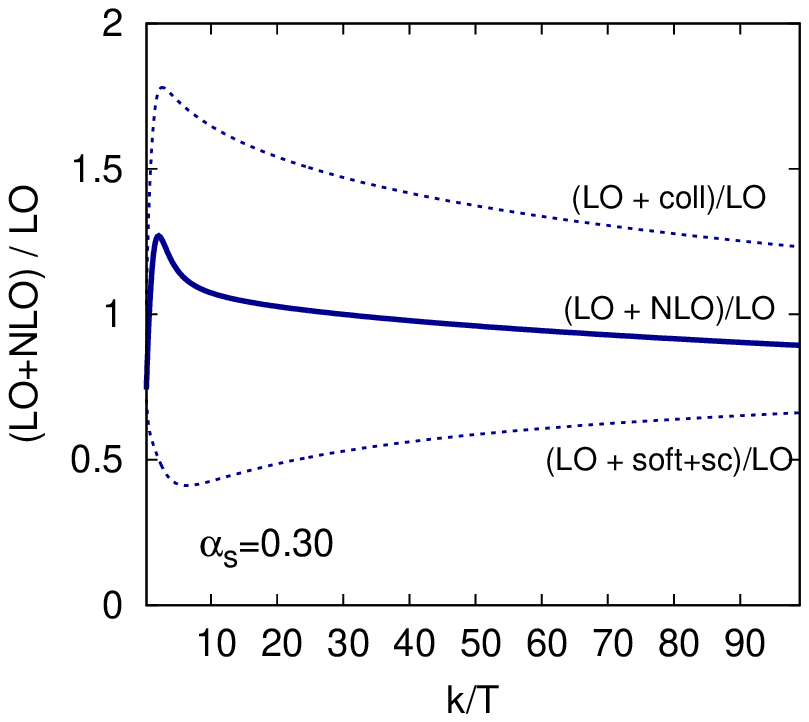}
      \caption{
         Left: the differential rate $d\Gamma_\gamma/dk$ relative to the leading
         order rate as a function of $k/T$ (or equivalently
         $C_\mathrm{LO+NLO}/C_\mathrm{LO}$).    The full next to leading order
         rate (LO+NLO)   is a sum of the leading order rate (LO), a collinear correction (coll),  and a soft+semi-collinear correction (soft+sc).  The dashed curve  labeled LO+coll shows the ratio of rates  when  only the collinear correction is included, with the analogous notation  for the  LO\,+\,soft+sc
         curve.  The difference between the dashed curves provides a 
         uncertainty estimate  for the NLO calculation. Right: the same as on the left but for larger $k/T$.
}
	\label{plot_ratio}
\end{figure}

\begin{figure}
   \includegraphics[width=0.495\textwidth]{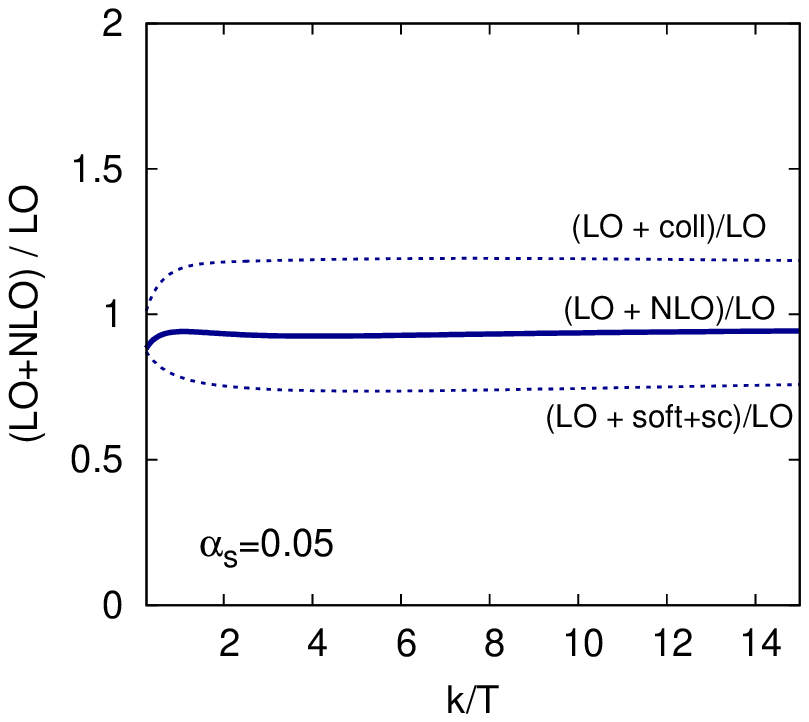}
   \includegraphics[width=0.495\textwidth]{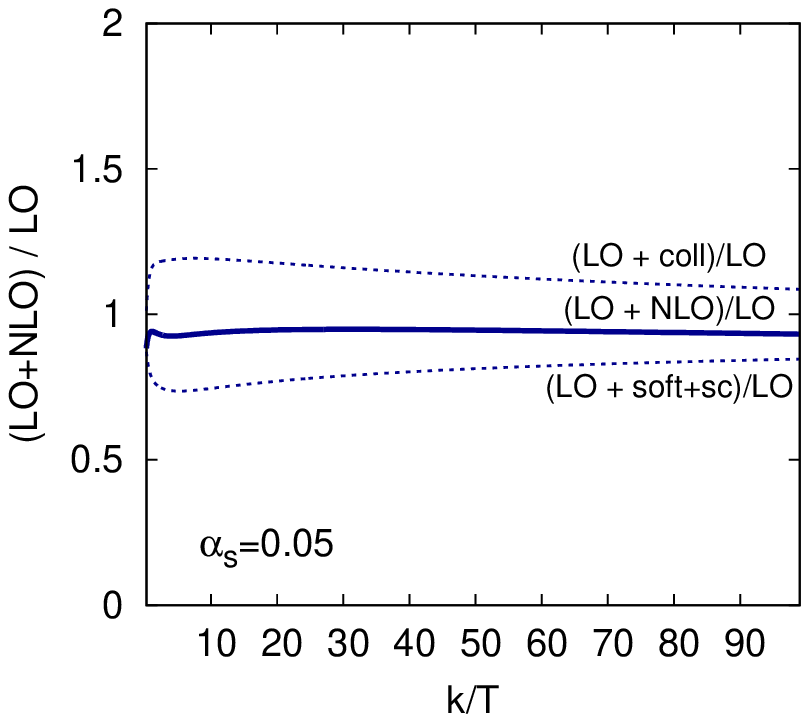}
	\caption{ The differential rate, $d\Gamma_\gamma/dk$, relative to 
   the leading order rate as described in Fig.~\ref{plot_ratio}, but for $\alpha_s=0.05$.
}\label{plot_ratio_05}
\end{figure}

In Figs.~\ref{plot_ratio} and \ref{plot_ratio_05} we plot 
the  differential photon emission rates $d\Gamma_\gamma/dk$ relative to the leading order rate, (LO+ NLO)/LO, for two different values of the coupling constant.
The reasonable,  but somewhat {\it ad hoc}, ``uncertainty
estimate'' described above can be inferred from the difference between the 
upper and lower dashed curves, which include either the collinear or the soft+semi-collinear correction, but not both.

For the largest coupling, $\als=0.3$, NLO corrections
are modest and positive, although the ``uncertainty band'' is rather large -- of order 50\% (see Fig.~\ref{plot_ratio}).
At intermediate coupling, $\als=0.15$,  the cancellation between the collinear
and semi-collinear+soft contributions is quite dramatic, causing the
LO+NLO result to be within a few percent of the LO rate (not
shown). Nevertheless, the uncertainty band remains rather large -- of
order $40\%$.
Finally, at the smallest coupling $\als=0.05$,  the (LO+NLO)/LO ratio is somewhat larger than at intermediate coupling, but with a considerably smaller uncertainty band (Fig.~\ref{plot_ratio_05}).

\section{Conclusions}
\label{sec_concl}
We have computed the photon production rate to NLO of an equilibrated, 
weakly-coupled quark-gluon plasma. 
The contributions to the LO rate can be divided into distinct kinematical regimes --- 
the hard, soft and collinear regions.
The contributions arising from the hard and the soft regions have logarithmic sensitivity
to the details of how the kinematical regions are divided. However, this dependence cancels
in the sum. At NLO the soft and collinear regions receive $\OO(g)$ corrections, and a new
``semi-collinear'' region starts to contribute here.  
We have dealt with the collinear region in Sec.~\ref{sec_coll}, with the
soft region in \ref{sec_soft}, and with the semi-collinear region in \ref{sec_semicollin}.

The collinear regime is affected by
the LPM interference of multiple scatterings through the integral
equation \eqref{fpproblem}. As we showed, computations are
most easily performed in impact parameter space and the resulting
$\OO(g)$ perturbation to the LO
result is given in Eqs.~\eqref{totalcoll}.  Furthermore two $\OO(g)$
regions of the leading-order result, the soft region and the
semi-collinear region, are identified and the asymptotic behaviors are derived
in Eqs.~\eqref{divergent_bit} and \eqref{colltosemi}. The treatment of
these $\OO(g)$ regions via the integral equation is incomplete, so we
must recompute the contributions in these regions more carefully,
subtracting off what has already been included in the collinear part of
the calculation to avoid double counting.

In the soft region, we relied on the fact that causality dictates analyticity 
properties for the $n$-point functions. Since the soft fields are lightlike separated as a result of 
the effective eikonalization of the hard fields, these analyticity properties lead to
a tremendous simplification of the calculation: the soft bosonic correlators become
the correlators of the 3D Euclidean theory, as pointed out in \cite{CaronHuot:2008ni}, 
whereas for fermions one can deform the integration contour of the non-vanishing
light-cone momentum away from the real axis towards infinity, yielding a tractable expansion in
inverse powers of that momentum.
With these technical developments the soft contribution to the NLO rate is entirely analytic
and given by \Eq{softtotal}, once the aforementioned subtraction of the collinear
limit is performed. The result is incredibly simple: the NLO correction arising from the soft sector
is the leading-order soft contribution modified by the shift $\mm \rightarrow
\mm + \delta \mm$, and nothing else. 
The contribution arising from the soft sector depends logarithmically on the 
UV regularization of the $p_\perp$-integral; this dependence cancels against
a corresponding IR divergence in the semi-collinear region.

In the semi-collinear region, the contribution from the soft gauge fields 
factorizes into a light-like separated, two-point correlator given by \Eq{semiint}.
This correlator is related to a momentum dependent transverse momentum diffusion 
coefficient experienced by a particle whose momentum obeys the semi-collinear scaling.
We use again Euclidean techniques to evaluate it, obtaining a compact analytic
result. Performing the remaining integrations (the last of which only numerically)
and upon subtracting the appropriate $\OO(g)$ regions of the leading-order result
to avoid double countings, the contribution from the semi-collinear region is 
given by \Eq{semicollintotal}, which, as expected, shows an IR log
divergence which combines with the soft contribution to make the final result
finite and cutoff-independent. 

In Sec.~\eqref{sec_results} we combine all contributions together in the full
NLO rate and plot it for different values of the coupling. The striking feature
is that the NLO correction is composed of two largely canceling contributions.
The positive one arises from the collinear region and the negative one from 
the soft and semi-collinear regions. For $\als=0.3$, 3 colors and 3 light flavors,
in the phenomenologically interesting momentum range, $k<15T$, each contribution is
a $50-75\%$ correction, but their sum is just a $10-20\%$ positive correction
(see Fig.~\ref{plot_c_30_1}(a)). At higher momenta, the NLO correction
turns negative, and the NLO+LO curve crosses the LO result (see
Fig.~\ref{plot_c_30_1}(b)).  For smaller values of the coupling,
$\als=0.05$, the NLO correction is  negative for all momenta (see
Fig.~\ref{plot_c_5_1}).

For these reasons we 
believe this cancellation to be largely accidental, and thus consider the two
separate contributions as the upper and lower bound of an uncertainty estimate 
for
the NLO calculation. In Figs.~\ref{plot_ratio} and \ref{plot_ratio_05} we plot
the (LO+NLO)/LO ratio of photo-emission rates  together with the uncertainty band for the two values of the coupling.

{}From the phenomenological point of view, the $\OO(20\%)$ correction itself in the 
relevant region for $\als=0.3$ does not alter qualitatively the current analyses 
\cite{Gale:2012xq}. On the other hand our $\OO(100\%)$ uncertainty band gives a first estimate
on the reliability of the perturbative calculation. Going to NNLO one would
also encounter UV vacuum divergences and the associated running coupling, whose
scale setting introduces another possibly large error band.

{}From the theoretical point of view, we believe that the main result of the
present work lies in the developments related to the description of soft
fields coupled to eikonalized hard fields. This progress opens new possibilities
towards the calculation of other transport coefficients, such as the shear viscosity,
at next-to-leading order. Furthermore, we believe the simple form of the NLO
soft region can be understood more transparently in terms of an effective
description of dipole propagation. Such a picture could also allow a factorization
of the non-perturbative magnetic sector. 
We plan to return to these issues elsewhere.

A clear extension of this work would be the NLO treatment of gluon radiation, 
following the generalization from photon to gluon radiation at leading order 
in \cite{Arnold:2002ja}. Other possible extensions include the calculation of the NLO
rate in $\mathcal{N}=4$ SYM. The leading-order calculation at weak coupling
was done in \cite{CaronHuot:2006te}, together with the strong-coupling result.
This could shed more light on the transition between the two regimes.

Analogously one could apply the methodologies we have developed to similar
calculations for the thermal production of light-like particles, which could be 
of relevance for cosmology and whose rates are known only to leading order. Examples are
ultrarelativistic right-handed neutrinos (see \cite{Besak:2012qm} for the LO rate), axions \cite{Graf:2010tv},
saxions \cite{Graf:2012hb}, axinos \cite{Brandenburg:2004du} and gravitinos \cite{Bolz:2000fu}.

\section*{Acknowledgments}

We would like to thank Simon Caron-Huot, Yannis Burnier, and Peter
Petreczky for useful
conversations.  We also thank the Institute for Nuclear Theory in
Seattle, where some of this work was conducted.  This work was supported
in part by the Institute for Particle Physics (Canada) and the Natural
Science and Engineering Research Council (NSERC) of Canada.
DT is supported in part by an OJI grant from the US Department of Energy
and the Sloan Foundation.

\appendix

\section{Hard Thermal Loop propagators}
\label{app_props}

In this section we detail our conventions for the HTL
propagators. Fermion propagators are most easily written in terms of
components with positive and negative chirality-to-helicity ratio. The retarded
fermion propagator reads
\begin{equation}
	\label{htlfermiondef}
	S_{R}(P)=h^+_\bp S^+_{R}(P)+h^-_\bp S^-_{R}(p)\,,
\end{equation}
where
	\begin{equation}
		\label{htlfermion}
	S^{\pm}_R(P)=\frac{i}{p^0\mp (p+\Sigma^\pm(p^0/p))}
 = \left.\frac{i}{\displaystyle p^0\mp\left[p+\frac{\mm}{2p}
  \left(1-\frac{p^0\mp p}{2p}\ln\left(\frac{p^0+p}{p^0-p}
        \right)\right)\right]}\right\vert_{p^0=p^0+i\epsilon},
\end{equation}
where the upper (lower) sign refers to the positive (negative)
chirality-to-helicity component. The  projectors are
$h^\pm_\bp\equiv(\gamma^0\mp\vec\gamma\cdot\hat p)/2$ and
$\mm=g^2\crr T^2/4$ is the fermionic asymptotic mass squared.

Gluons are described in the strict Coulomb gauge by
\begin{eqnarray}
	\label{htllong}
	G^{00}_R(Q)&=&\frac{i}{\displaystyle q^2+\md^2\left(1-\frac{q^0}{2q}\ln\frac{q^0+q+i\epsilon}{q^0-q+i\epsilon}\right)},\\
\nn G^{ij}_R(Q)&=&(\delta^{ij}-\hat q^i\hat q^j)G^T_R(Q)=\left.\frac{i(\delta^{ij}-\hat q^i\hat q^j)}{\displaystyle q_0^2-q^2-\frac{\md^2}{2}\left(\frac{q_0^2}{q^2}-\left(\frac{q_0^2}{q^2}-1\right)\frac{q^0}{2q}\ln\frac{q^0{+}q}{q^0{-}q}\right)}\right\vert_{q^0=q^0+i\epsilon}.\\
&& 	\label{htltrans}
\end{eqnarray}
The other components of the propagators in the \ra basis can be obtained 
through \Eq{raprop}.

\section{Gauge invariant condensates}
\label{condensates}

During the calculation, we encounter several condensates that can be
written as integrals of correlators separated by a spacelike or
lightlike separation:
\begin{eqnarray}
\label{defZ_g}
Z_g & \equiv & \frac{1}{\da}\left\langle v_\mu F^{\mu\rho}
   \frac{-1}{(v\cdot D)^2} v_\nu F^\nu_{\,\rho}\right\rangle \\
\label{defZ_g_xspace}
 & = & \frac{-1}{d_A}
\int_0^\infty dx^+ \, x^+ \langle \vphantom{\frac{1}{2}} v_{k\,\mu}
F_a^{\mu\nu}(x^+,0,0_\perp) U^{ab}_{\sss A}(x^+,0,0_\perp;0,0,0_\perp)
 v_{k\,\rho}{F_b^\rho}_\nu(0)\rangle , \\
\label{defZ_f}
Z_f & \equiv &
\frac{1}{2\dr}\left\langle\overline\psi\frac{\slashed{v}}{v\cdot
    D}\psi\right\rangle \\
 & = & \frac{-i}{2\dr} \int_0^\infty dx^+
 \langle \overline\psi(x^+,0,0_\perp)\,\slashed{v}_k\, U_{\sss R}(x^+,0,0_\perp;0,0,0_\perp)
   \psi(0) \rangle
\nonumber \\
\label{Isc}
\Isc&=& \int_{-\infty}^\infty dx^+\,e^{ix^+\delta E} \,
  \frac{1}{\da} \langle v^\mu_{k} {F_\mu}^{\nu}(x^+,0,0_\perp)
     U_{\sss A}(x^+,0,0_\perp;0,0,0_\perp) 
               v^\rho_{k}F_{\rho\nu}(0)\rangle, \\
\label{C_def}
\cc(\xp) & = & \lim_{x^+\rightarrow \infty}- (x^+)^{-1} 
   \log(W(x^+,\xp)), \\
\label{W_def}
\hspace{-3ex} W(x^+,\xp) & \equiv & {\rm Tr}\: \Big\langle
  \,U_{\sss R}(0,0,\xp;x^+,0,\xp)  \,U_{\sss R}(0,0,0;0,0,\xp) 
\nonumber \\ && \qquad 
   \, U_{\sss R}(x^+,0,0;0,0,0)
     \, U_{\sss R}(x^+,0,\xp;x^+,0,0)\Big\rangle \,.
\end{eqnarray}
Here coordinates are written as triples $(x^+,x^-,\xp)$ with
$x^- = (t-z)$ and $x^+ = (t+z)/2$, so that $t=z=x^+$ when $x^-=0$ and
$X\cdot P = \xp \cdot p_\perp - x^- p^+ - x^+ p^-$.
$U(x_1;x_2)$ is the Wilson line connecting the point $x_1$ to the point
$x_2$, in either the adjoint representation ($U_{\sss A}$) or
the representation of the fermions ($U_{\sss R}$),
and $W$ is the Wilson loop with a transverse segment of extent
$\xp$ and a lightlike segment of extent $x^+$ in the $(t,z)$
directions \cite{Benzke:2012sz}. For $Z_g$ and $Z_f$ we have 
employed rotational invariance and chosen the light-like vector
$v$ to be $v_k$ without loss of generality.

The condensates $Z_g$ and $Z_f$ are related to the bosonic and fermionic
hard thermal loops \cite{Braaten:1991gm}.  They describe how
fluctuations in the gauge fields ($Z_g$) and the fermionic fields
($Z_f$) can influence the propagation of a fermion moving through the
plasma at nearly the speed of light.  Therefore they determine the
dispersion correction of hard $p^+ \gg gT$ excitations; they are valid
both at leading and at
next-to-leading order \cite{CaronHuot:2008uw}.  And
$\cc(\xp)$ and $\Isc$ are related to the scattering
processes in the medium; $\cc(\xp)$ arises when treating
collinear splitting and is discussed in Sec.~\ref{sec_coll}, while
$\Isc$ arises when treating the semi-collinear regime, in \Eq{semiint}.
They are related;
\begin{equation}
\lim_{x_{\perp} \rightarrow 0} \partial^2_{\xp}
\cc(\xp) = \lim_{\delta E \rightarrow 0}
\hat{q}(\delta E) \,.
\end{equation}

In this appendix we show how these condensates are most conveniently
computed by using Euclidean methods developed in \cite{CaronHuot:2008ni}.
But first let us set up their calculation via real-time techniques,
so we can see how expressions, encountered in the main text, do indeed
correspond to these condensates.
Except for $\cc(\xp)$, we only encounter the condensates at lowest
order, where the Wilson line is set to unity and the field strengths
take their abelian form.  Working to this order,
consider first $Z_g$.  Recall that
\begin{eqnarray}
\nonumber
\langle F_{\mu\nu}(X) F_{\alpha\beta}(0) \rangle
& = & \int \frac{d^4 Q}{(2\pi)^4} \: e^{iQ\cdot X} \Big(
  Q_\mu Q_\alpha G^>_{\nu\beta} (Q)
- Q_\nu Q_\alpha G^>_{\mu\beta} (Q)
\\ && \hspace{18ex}
- Q_\mu Q_\beta  G^>_{\nu\alpha}(Q)
+ Q_\nu Q_\beta  G^>_{\mu\alpha}(Q) \Big) \,.
\end{eqnarray}
Applying this to \Eq{defZ_g_xspace} and performing the $x^-$ integral,
one finds (ignoring the difference between $G^>$ and $G_{rr}$, which is
higher order in the soft region)
\begin{equation}
\label{Zg_minkowski}
Z_g = \int\frac{d^2\qp \, dq^+ dq^-}{(2\pi)^4}
\left[ \frac{\qp^2 G^{++}_{rr}(Q)}{(q^- -i\epsilon)^2}
 - 2 \frac{q_z G_{T}^{rr}(Q)}{(q^--i\epsilon)}
         \left(1{-}\frac{q_z^2}{q^2}\right)
 + G_{T}^{rr}(Q)\left(1+\frac{q_z^2}{q^2}\right) \right] \,.
\end{equation}
This is the same as the expression encountered in the ``soft part'' of
the calculation, \Eq{qstruct}, except that there the leading-order
behavior is to be subtracted.

The calculation of $\hat{q}(\delta E)$ is analogous, except that the
$x^+$ integral produces a delta function fixing $q^-$:
\begin{equation}
\label{semiint2}
 \Isc = \int\frac{d^4Q}{(2\pi)^3}\delta(q^--\delta E)
  \left[ \qp^2 G^{++}_{rr}(Q) + G_T^{rr}(Q)
	\left(\left[1{+}\frac{q_z^2}{q^2}\right]\delta E^2-2q_z\delta E
 \left[1{-}\frac{q_z^2}{q^2}\right]\right)\right].
\end{equation}
This is identical to \Eq{semiint}.

It would be possible to compute these Minkowski-domain expressions
explicitly using the sum rule approach; but we find it simpler and more
instructive to compute them via Euclidean techniques.

\subsection{Relation to Euclidean functions:  Simple derivation}

\label{app_magic}

Caron-Huot has shown that $n$-point correlation functions, where all
fields lie on a spacelike hypersurface, can be carried out by Euclidean
techniques \cite{CaronHuot:2008ni}.  The null correlators we need can
also be computed provided that they are free of collinear singularities,
which they are.  The proof presented in Ref.~\cite{CaronHuot:2008ni} is
rather complex and technical.  Here we will present a much simpler
derivation, also due to Caron-Huot%
\footnote{S.\ Caron-Huot, oral presentation at the Institute for Nuclear
  Theory (Seattle), 29 March 2012},
which works for two point functions.  In practice this is all we need,
except for the NLO evaluation of $\cc(\xp)$.

Consider the ordering-averaged correlator of some operator (such as the
field strength), $G_{rr}(x^0,\x)$ with $\vert x^z\vert  > \vert
x^0\vert$.  (Since the separation is spacelike, operators commute, and
therefore $G_{rr}$ equals $G^<$, $G_{\sss F}$, or $G^>$.)
Write it in terms of its Fourier representation
\begin{equation}
G_{rr}(x^0,\x) = \int d\omega \int dp_z d^2 p_\perp
e^{i(x^z p^z+\bxp\cdot \bpp - \omega x^0)}\:
G_{rr}(\omega,p_z,p_\perp),
\end{equation}
and use
\begin{equation}
G_{rr}(\omega,p) = \left(\nbe(\omega)+\frac12\right) \left(G_R(\omega,p)-G_A(\omega,p)\right)=\left(\nbe(\omega)+\frac12\right) \rho(\omega,p),
\end{equation}
and define $\tilde p^z = p^z - (t/x^z) \omega$:
\begin{equation}
\label{almostthere}
G_{rr}(x^0,\x) = \int d\omega \int d\tilde{p}_z d^2 p_\perp
e^{i( x^z \tilde p^z + \bxp \cdot \bpp)}
\left(\nbe(\omega)+\frac12\right)\: 
\rho(\omega,\tilde{p}+\omega(x^0/x^z),p_\perp) \,.
\end{equation}
Now we perform the $\omega$ integration by contour methods.
The retarded function in $\rho(\omega,p)$ is related to the Euclidean function via
$G_R(\omega,p) = -i G_E(i\omega,p)$, that is, by analytic continuation.
This continuation is guaranteed not to encounter singularities in the
Green function so long as the imaginary part of the 4-momentum remains
timelike -- since then, in some frame, the continuation is purely of the
frequency.  Since $|x^0/x^z|<1$, the continuation of
$G_R(\omega,\tilde{p}-\omega(x^0/x^z),p_\perp)$ in $\omega$ will not
encounter any singularity in the upper complex $\omega$ plane. The advanced
function similarly will be free of singularities in the lower plane.
Therefore the only singularities encountered in continuing the frequency
integration are those in the statistical function $(\nbe(\omega)+1/2)$,
which has poles at $\omega = 2\pi inT$ with $n=(\ldots
-1,0,1,\ldots)$ and residue equal to $T$. Closing the contour around 
these poles, and renaming
$\tilde{p}^z$ to $p^z$, we find%
\footnote{
  The pole at $n=0$ is an artifact of the separation of $\rho$, which vanishes for $\omega=0$, into $G_R$ and $G_A$.  
The individual poles there can then be dealt with in a principal value prescription, for instance.
}
\begin{equation}
\label{simonmagic}
G_{rr}(x^0,x^z,\xp) = T \sum_{n} \int \frac{d^3 p}{(2\pi)^3}
e^{i(x^z p^z + \bxp \cdot \bpp)}
G_E(\omega_n,p_z+i\omega_n (x^0/x^z),p_\perp) \,, \quad
\omega_n = 2\pi nT \,.
\end{equation}

In any case where we need to compute the soft $gT$ contribution to such
a correlator, one may drop the nonzero Matsubara frequency
contributions; that is, we keep only the $n=0$ term in the sum.  For
this term,
$G_E(\omega_n,p_z+i\omega_n (x^0/x^z),p_\perp) = G_E(0,p_z,p_\perp)$
is the Euclidean correlation function of the 3-dimensional dimensionally
reduced (Electric QCD or EQCD) theory.  [EQCD is the 3D theory
consisting of the spatial gauge fields $A_i$ and an adjoint scalar
descended from the temporal component of the gauge field,
$\Phi=iA_0$.  The scalar is massive, $m^2_{\Phi} = \md^2$; see
for instance \cite{Braaten:1995ju}.]

\subsection{Application to Scattering}
\label{app_cq}

First we apply this method to determine $\cc(\qp)$ and
$\hat{q}(\delta E)$ at leading order.  There,
\begin{equation}
	\cc_{\rm LO}(\qp) =g^2\crr	
 \int dx^+ \int d^2\xp e^{-i\bqp\cdot\bxp} 
   G^{++}_{rr}(x^+,0,\xp)\,,
\end{equation}
where this position space expression makes clear that we are dealing
with a spacelike separation and hence \Eq{simonmagic} is applicable,
yielding
\begin{eqnarray}
\nn \cc_{\rm LO}(\qp)&=&g^2\crr T\sum_n
  \int\frac{d^3p}{(2\pi)^3}G^{++}_E(p_n^0,p_n^z,\pp)
    (2\pi)^3\delta(p_z)\delta^2(\bpp-\bqp)  \\
 & = & g^2\crr T\left(\frac{1}{\qp^2}-\frac{1}{\qp^2+\md^2}\right)
+\mbox{hard ($n\neq 0$) contributions}\,.
\label{simonagz}
\end{eqnarray}
The $1/\qp^2$ and $1/(\qp^2+\md^2)$ terms are the contributions from the
$A_z A_z$ correlator and the $A_0 A_0$ correlator respectively.  The
nonzero Matsubara frequency contributions are suppressed, for
$\qp \sim gT$, by a power of $g^2$ and may be neglected.
This result was first found by Aurenche Gelis and
Zaraket \cite{Aurenche:2002pd}, by a rather more complicated sum rule
procedure, which involved canceling $1/(\qp^2+\md^2/3)$ poles.
The current procedure, originally due to Caron-Huot
\cite{CaronHuot:2008ni}, avoids this complication.

With this Euclidean framework the NLO (1-loop) corrections also become
tractable as a computation within dimensionally-reduced EQCD
\cite{CaronHuot:2008ni}.  We will return to this result in the next
appendix.

Next consider $\hat{q}(\delta E)$, \Eq{Isc}.  We work to lowest order,
replacing the Wilson line with the identity and keeping only the
two-point correlator of the $A$-fields in the field strengths.
We may again apply
\Eq{simonmagic} and to find the infrared contribution we may keep only
the $n=0$ term.  This corresponds to the replacements
$G_{rr}\to G_E$, $\int dq^0/(2\pi)\to T$ and
$q^0\to 2\pi i n T \to 0$, $q^+ \to q_z$.  Writing the gauge field
correlator in terms of its momentum space representation, we then obtain
\begin{eqnarray}
\nn \Isc & = & T\int\frac{d^3q}{(2\pi)^3} \left[ 
  \qp^2 G^{++}_{E}(0,q_z,\qp) + G_T^{E}(0,q_z,\qp)
  \left(\left[1{+}\frac{q_z^2}{q^2}\right]\delta E^2
  -2q_z\delta E\left[1{-}\frac{q_z^2}{q^2}\right]\right)\right] \\
 & & \hspace{2.2cm} \nn \times 2\pi\delta(q_z+\delta E) \\
 & = & T\int\frac{d^2\qp}{(2\pi)^2}\left[
   \frac{\md^2\qp^2}{(\qp^2+\delta E^2)(\qp^2+\delta E^2+\md^2)}
   +\frac{2\delta E^2}{\qp^2+\delta E^2}\right].
\label{finalssgluon}
\end{eqnarray}

\subsection{Application to $\dZg$}
\label{app_zg}

Let us now apply the Euclidean formalism to compute
$Z_g$ at NLO.  Our starting point will be the position space expression
of \Eq{defZ_g_xspace}.  Since the positions involved are lightlike
separated, we may apply \Eq{simonmagic}.

The leading order contribution to $Z_g$ arises from hard $p^+\sim T$
excitations; it therefore arises from a range of $n$ values and is not
easily established by Euclidean methods.  However the leading order
value is easy to determine by conventional methods, and equals
\begin{equation}
Z_g^{\rm LO} = 2\int \frac{d^3 p}{(2\pi)^3 p} \nbe(p) = \frac{T^2}{6}.
\label{Zg_LO}
\end{equation}
For $n\neq 0$ the correlation functions receive $\OO(g^2)$ loop
corrections, so any corrections from $n\neq 0$ are $\OO(g^2)$ and
therefore beyond our current precision goal.  To find the $\OO(g)$ NLO
corrections, we need to compute the contribution from $n=0$ modes and
subtract the leading-order, unresummed-theory value.  A
straightforward evaluation of \Eq{defZ_g_xspace}, replacing $x^+$ with
$x_z$ and using 3-dimensional Euclidean correlation functions, gives
\begin{equation}
	\label{zgsimon}
\dZg = T\!\int\!\frac{d^3q}{(2\pi)^3}\!
  \left[ \frac{\qp^2 G_E^{++}(0,q_z,\qp)}{(q_z+i\epsilon)^2} 
  + G_T^{E}(0,q_z,\qp) \left(3 - \frac{q_z^2}{q^2}\right)
  -\mbox{ same, free propagators}
 \right],
\end{equation}
where the $i\epsilon$ prescription in $1/(q_z+i\epsilon)^2$ arises from
the boundary conditions on the $z$ integration, $\int_0^\infty z d z$.
Now the bare and interacting $A_i$ correlators are the same at leading
order, but the temporal mode $A_0$ develops a (Debye) mass;
$\langle A_0 A_0(q) = \frac{-1}{q^2+\md^2}$ whereas the bare value is
$-1/q^2$.  (The minus sign is because there is an $i$ difference between
the Minkowski and Euclidean $A^0$ field, so $A^+ = A^z +i A^0$.)
Therefore only the $G_E^{00}$ part of the first term is not canceled by
the relevant free version.  Its subtracted contribution is then
\begin{equation}
\dZg = T\int\frac{d^3q}{(2\pi)^3} \frac{\qp^2}{(q_z+i\epsilon)^2}
\left( \frac{-1}{q^2+\md^2} + \frac{1}{q^2} \right)
 = - \frac{T\md}{2\pi}.
\end{equation}
The correction is $\OO(g)$ relative to \Eq{Zg_LO} and negative,
representing a reduction in the thermal mass due to the screening
of infrared gauge modes.

No such $\OO(g)$ correction arises for $Z_f$ because the correlator
directly involves fermionic fields which do not have zero modes.  To
$\OO(g)$ accuracy,
\begin{equation}
Z_f = 2\int \frac{d^3 p}{(2\pi)^3 p} \nfd(p) = \frac{T^2}{12} \,.
\end{equation}

\subsection{Thermal mass at NLO}
\label{app_mass}

We now apply these results for $Z_g$ and $Z_f$ to the fermionic
effective mass.
The thermal dispersion relation for a particle approaches $p_0^2=p^2+\mm$ for 
$p^0\approx p\gg gT$, where $\mm$ is the asymptotic mass.
The asymptotic mass is given  by the real 
part of the fermion self-energy. In more detail we have \cite{Blaizot:2005wr}
\begin{equation}
	\label{defminfdiagram}
	\mm =2p\,\mathrm{Re}\,\Sigma_R^+(p^0=p)\,,
        \qquad\Sigma_R^+(P)\equiv\frac{1}{2}\Tr{h^+_p\Sigma_R(P)}.
\end{equation}  

\begin{figure}[ht]
\begin{center}
\includegraphics[width=7cm]{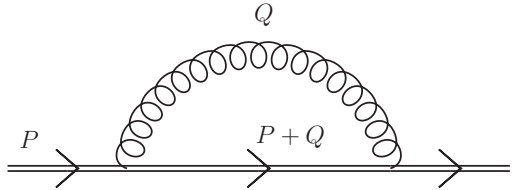}
\end{center}
\caption{The diagram contributing to the asymptotic mass at
leading and next-to-leading order.}
	\label{fig_minfty}
\end{figure}

To the order of interest, one diagram contributes to the retarded
self-energy, see Fig.~\ref{fig_minfty}.  There are two \ra
assignments; either the fermion is retarded and the gluon is $rr$, or
the gluon is retarded and the fermion is $rr$.  The retarded line, which
in the NLO calculation always carries a large momentum, enforces the
eikonality; the operator correlation in $Z_g$ ($Z_f$) corresponds to the
cut gauge boson (fermion) line.  One easily finds that
\cite{CaronHuot:2008uw}
\begin{equation}
	\label{minf}
	\mm = g^2 \crr( Z_g + Z_f )\,.
\end{equation}
We just found that, at NLO, the condensates read \cite{CaronHuot:2008uw}
\begin{eqnarray}
\label{Z_g}
Z_g & = & \ZgLO+\dZg, \quad  
      \ZgLO =\frac{T^2}{6}, \quad \dZg =  - \frac{T \md}{2\pi} \,,
\\
\label{Z_f}
Z_f & = & 
    Z_f^{LO} + \delta Z_f, \quad Z_f^{LO} =  \frac{T^2}{12}, \quad \delta Z_f = 0\,,
\end{eqnarray}
which, together with the well known result for the leading order Debye mass
\begin{equation}
	\label{mD}
	\md^2=\frac{g^2}{3}(\ca+T_R \nf)T^2\,,
\end{equation}
yields the fermionic $\mm$ at NLO in QCD,%
\footnote{
In \cite{Blaizot:2005wr} the asymptotic mass was computed
numerically at NLO in the large-$\nf$ non-abelian theory. 
The authors found a strong momentum dependence of their NLO result, in sharp contrast 
with \Eq{minfNLO1}, which however reproduces their average shift. It is 
our understanding that, in the context of a strict $g$ expansion, the 
result of Caron-Huot \cite{CaronHuot:2008uw} is correct for fermion
momenta much larger than $gT$. 
}
\begin{equation}
\label{minfNLO1}
m^2_{\infty,\rm NLO} = \mm +\delta \mm= g^2 \crr \left( \frac{T^2}{4}
  - \frac{g T^2}{2\pi}
  \sqrt{\frac{2N_c + \nf}{6}} \right) .
\end{equation}

\section{NLO collision kernel}
\label{sub_nlo_cq}

In the previous appendix we presented the definition of the differential
collision rate $\cc(\xp)$ and found its leading order
expression in transverse momentum space, $\cc(\qp)$.  Its NLO expression
in $\qp$ space has also been found \cite{CaronHuot:2008uw}; writing
$\cc_{\textrm{NLO}} = \cc + \delta \cc$, explicitly
\begin{eqnarray}
\delta\cc(\qp)& \!=\! & (g^4 T^2 \crr \ca) \left[
 - \frac{\md{+}2\frac{q_\perp^2-\md^2}{q_\perp} 
    \arctan\frac{q_\perp}{\md}}{4\pi (q_\perp^2{+}\md^2)^2}
 + \frac{\md-\frac{q_\perp^2{+}4\md^2}{2q_\perp} 
    \arctan\frac{q_\perp}{2\md}}{8\pi q_\perp^4} 
\right. \nonumber \\ \nn&& \hspace{3cm} - \frac{
    \arctan\frac {q_\perp}{\md}}{2\pi q_\perp(q_\perp^2{+}\md^2)}
 + \frac{\arctan\frac{q_\perp}{2\md}}{2\pi q_\perp^3}
 + \frac{7}{32 q_\perp^3}\\
 && \hspace{3cm}\left.+\frac{\md}{4\pi(q_\perp^2{+}\md^2)}
     \left( \frac{3}{q_\perp^2{+}4\md^2} - \frac{2}{q_\perp^2{+}\md^2}
     -\frac{1}{q_\perp^2} \right) \!\right].
\label{C_NLO}
\end{eqnarray}
In order to use the collision kernel in \Eq{bspace}, we need to Fourier transform this
expression to find $\cc'(b)$, preferably analytically but at minimum via
accurate numerical integration.

The integral at leading order can be performed by taking partial
fractions and considering $e^{i\b \cdot \q_\perp}/(q_\perp^2+\md^2)$:
\begin{eqnarray}
& & \int_{-\infty}^{\infty} \frac{dq_1}{2\pi} e^{ib q_1}
\int_{-\infty}^{\infty} \frac{dq_2}{2\pi} \frac{1}{q_2^2 + q_1^2+\md^2}
= \int_{-\infty}^{\infty} \frac{dq_1}{2\pi} e^{ib q_1}
   \frac{1}{2\sqrt{q_1^2+\md^2}}
\nonumber \\
& = & \frac{1}{2\pi} \int_{\md}^{\infty} \frac{dx}{\sqrt{x^2-\md^2}} e^{-bx}
= \frac{K_0(b\md)}{2\pi}\,,
\end{eqnarray}
where, after performing the trivial $q_2$ integration, we
changed contours to pick up the discontinuity along the cut of
the $\sqrt{q^2+\md^2}$ function; $x\equiv{\rm Im}\, q$.
Ultimately we need the same expression with $\md \rightarrow 0$, and with
$b \rightarrow 0$ for each (finite and zero $\md$) case;
\begin{eqnarray}
\cc'(b) & \equiv & \int \frac{d^2\qp}{(2\pi)^2}
     \Big( 1 - e^{i\b \cdot \q_\perp} \Big) \cc(\qp)
\cc_{\rm LO}'(b)
\nonumber \\
   & = & g^2 T \crr \lim_{\epsilon \rightarrow 0} \frac{1}{2\pi}
\left( K_0(\epsilon b \epsilon \md) - K_0(b \epsilon \md) 
     - K_0(\epsilon b \md) + K_0(b\md) \right) \,,
\nonumber \\
\frac{\cc_{\rm LO}'(b)}{g^2 T \crr}
 & = & \frac{1}{2\pi} \left( K_0(b\md) + \gamma_{\rm E}
  +\ln(b\md/2) \right) \,.
\end{eqnarray}

We follow the same strategy for the terms in the NLO correction
\eqref{C_NLO}.  We first perform the $q_2$ integration (the direction
orthogonal to $\b$).
This can generally be done by deforming
the contour to pick up all poles and cuts.  In every case the resulting
integral can be done analytically.  Then we
perform the $q_1$ integration.  In some cases this can be done
analytically, in other cases the $b=0$ case is analytic but the finite
$b$ case involves an integral along a cut.  Finally there are cases
where the difference, $1-e^{-bx}$ must be integrated along a cut.
Some of these integrals remain numerical, but all converge exponentially
and are small for large $b$.

Without going into detail, the result (writing the terms in the same
order as they appear in \Eq{C_NLO}) is
\begin{eqnarray}
&&
\nn\frac{\md\, \delta \cc'(b)}{g^4 T^2 \crr \ca}
\\ \nn
 & = &
- \frac{1}{8\pi^2} \left[ \frac{b\md K_1(b\md) - 1 +4 - 4 e^{-b\md}}{2}
+ \int_1^\infty dz \left( e^{-b\md} - e^{-b\md z} \right)
         \frac{\ln\frac{z^2}{z^2 - 1}}{(z^2-1)^{3/2}} \right]
\nonumber \\
& &{} - \frac{1}{96\pi^2} \int_0^\infty \frac{dz}{z} (1-e^{-b\md z})
      \left( 1 - \frac{(z^2-4)^{3/2} \theta(z-2)}{z^3} \right)
\nonumber \\
\nn& & - \frac{1}{32} + \frac{1}{8\pi^2} \int_1^\infty \!\! dz
        \frac{e^{-b\md z} \ln \frac{z^2}{z^2-1}}{\sqrt{z^2-1}}
{} + \frac{1}{8\pi^2} \int_0^\infty \frac{dz}{z} \left( 1-e^{-b\md z}\right)
     \left( 1 - \frac{\theta(z{-}2)\sqrt{z^2{-}4} }{z} \right)
\nonumber \\
& & {} + \frac{7 b\md}{64 \pi}
    + \frac{1}{8\pi^2} \Big[K_0(2 b\md) - 2 K_0(b\md) + \ln\frac{4}{\md b} 
- \gamma_{\rm E} + b\md K_1(b\md) - 1 \Big] \,. 
\label{nasty}
\end{eqnarray}
We have been unable to perform these integrals analytically.  However,
they are quite straightforward numerically, and they also have a simple
behavior in the large $b\md$ limit:
\begin{equation}
		\frac{\md\, \delta \cc'(b)}{g^4 T^2 \crr \ca} 
	\xrightarrow[b\md\gg 1]{}
	  \frac{7b\md}{64\pi}
 + \frac{-9\pi^2 -58 + 72 \ln(2) -3 ( \ln(b\md)+\gamma_{\rm E} )}
   {288\pi^2}
 + \OO(\exp(-b\md)).
\end{equation}
This has the interpretation as the total scattering rate, IR regulated at
a scale $q\sim 1/b$. While the leading order total scattering rate
has a logarithmic divergence, the NLO has a linear divergence.
Note that the corrections to this linear + constant + log behavior are
exponentially suppressed.

The small $b\md$ behavior is
\begin{eqnarray}
\frac{\md\, \delta \cc'(b)}{g^4 T^2 \crr \ca} 
	\xrightarrow[b\md\gg 1]{} 
\frac{-b\md}{32\pi} + \frac{\hat{q}_{\rm nlo} b^2 \md^2}{4} + {\cal O}(b^3) \,,
\quad \hat{q}_{\rm nlo} \!\!=\!\! & \frac{3\pi^2 + 10 - 4 \ln(2)}{32\pi^2} \,.
\end{eqnarray}
Here $\hat{q}_{\rm nlo}$ is the NLO correction to the momentum
broadening rate.  To see this, expand \Eq{eq:C(b)} in small $b$ and
angle average,
$\exp(i\b\cdot \q_\perp) \sim 1-(\b\cdot \q_\perp)^2/2 \sim 1-b^2 q^2/4$.
The integral $\int d^2 q \; q^2 \cc(q)$ is what we usually mean by
$\hat{q}$; hence $\hat{q}_{\rm nlo}$ is the numerical coefficient on the
NLO contribution to $\hat{q}$.
We have thus verified that the
expansion performed on the expression in \Eq{nasty} reproduces
the expected linear-in-$b$ and $\hat{q}$ behavior found by Caron-Huot \cite{CaronHuot:2008ni}.

Note that the small $b$ behavior is actually negative.  Our understanding
is that this is actually correct.  Basically, small $b$ corresponds to large
momentum transfers, a limit where our treatment of exchange momenta as
$gT \ll T$ breaks down. In \cite{CaronHuot:2008ni}, Caron-Huot showed
how an opposite term arises in the soft limit of the hard contribution to
$\hat{q}$, leading to a cancellation. Had we used dimensional regularization
to perform the integration of $b^2q^2 \cc_\mathrm{NLO}(q)/4$, the linear
term would simply have vanished.

\section{Evaluation of the semi-collinear integrations }

In this Appendix we evaluate numerically the integrals 
appearing in \Eq{collsubtracted}.
In order to match properly with the
UV divergence in the soft region, which was regulated by a cutoff
$\mu_\perp^\mathrm{NLO}$ ($gT\ll\mu_\perp^\mathrm{NLO}\ll \sqrt{g}T$)
in the transverse momentum $\pp$, we regulate the IR region with the 
same transverse cutoff. As we mentioned, we split the calculation
into a bremsstrahlung/Compton contribution ($p^+>0$) and an annihilation contribution
($-k/2<p^+<0$), where we used the symmetry under $p^+\to-p^+-k$.
We start with the former.

\label{app_sc}
\subsection{The bremsstrahlung/Compton contribution: $\Gamma_{\sc}^{\rm brem/Compt.}$}
As we shall show, it is  technically convenient to first
introduce an intermediate regulator $\mu^+$ for the $dp^+$ integration, 
with $gT\ll \mu^+\ll T$, hence $\mu^+\gg \mu_\perp^\mathrm{NLO}$. 
In practice we divide the phase space in two regions, \emph{i.e.}
\begin{enumerate}
	\item First we take $p^+>\mu^+$, $\pp>\mu_\perp^\mathrm{NLO}$.
	\item We then consider the slice $0<p^+<\mu^+$, $\pp>\mu_\perp^\mathrm{NLO}$,
	where only the IR asymptotic behavior in $p^+$ needs to be considered.
\end{enumerate}
The dependence on $\mu^+$ cancels in the sum of the two regions.

The $\pp$ and $\qp$ integrations
on the second line of \Eq{collsubtracted} yield
\begin{eqnarray}
\nn	&&\frac{1}{2\pi}\int_{\mu_\perp^\mathrm{NLO}}^\infty\frac{d\pp\,\pp}{\delta E^2}\int\frac{d^2\qp}{(2\pi)^2}
	\left[\frac{\md^2\qp^2}{(\qp^2+\delta E^2)(\qp^2+\delta E^2+\md^2)} 
	-\frac{\md^2}{\qp^2+\md^2}\right]\\
	\label{perpintegrals}&&=-\frac{\md}{4\pi k}\vert p^+(k+p^+)\vert+\order{(\mu_\perp^\mathrm{NLO})^2},
\end{eqnarray}
where we have used the fact that $p^+\gg \mu_\perp^\mathrm{NLO}$. The first term on the 
second line is the result one would obtain with vanishing cutoff. As we mentioned in the
previous section, matching regions
for momenta of the order of the cutoffs are equally described by the regions on
either side of it and the dependence on the cutoff has to vanish at all orders. Power-law
terms can then be neglected and we can just plug the first term in \Eq{collsubtracted},
yielding
\begin{equation}
		(2\pi)^3\frac{d\Gamma_\gamma}{d^3k}\bigg\vert_\mathrm{brem/compton}^{(1)}=-\frac{\md}{\pi T} \mathcal{A}(k)\int_{\mu^+}^\infty dp^+
\frac{\nfd( k+p^+) (1-\nfd(p^+))}{\nfd(k)}\frac{ (k+p^+)^2+(p^+)^2}{k\,p^+\,(p^++k)}.
\label{perpintegrated}
\end{equation}
This expression is IR log divergent. The logarithm can be extracted by adding 
and subtracting $\theta(T-p^+)/(2p^+)$ under the integral sign, so that
\begin{equation}
		-\frac{\md}{\pi T} \mathcal{A}(k)\int_{\mu^+}^T dp^+
\frac{1}{2p^+}=		-\frac{\md}{2\pi T} \mathcal{A}(k)\ln\frac{T}{\mu^+},
\label{muzregulate}
\end{equation}
and
\begin{equation}
		(2\pi)^3\frac{d\Gamma_\gamma}{d^3k}\bigg\vert_\mathrm{brem/compton}^{(1)}=-\frac{\md}{2\pi T} \mathcal{A}(k)\left[\ln\left(\frac{T}{\mu^+}\right)+C_\mathrm{brem/compton}\left(\frac{k}{T}\right)\right].
\label{oneregulated}
\end{equation}
$C_\mathrm{brem/compton}(k/T)$ is defined as
\begin{eqnarray}
\nn	C_\mathrm{brem/compton}\left(\frac{k}{T}\right)&=&2\int_{0}^\infty dp^+\left[
\frac{\nfd( k+p^+) (1-\nfd(p^+))}{\nfd(k)}\frac{ (k+p^+)^2+(p^+)^2}{k\,p^+\,(p^++k)}\right.\\
\label{defcbrem}
&&\hspace{6.8cm}\left.-\frac{\theta(T-p^+)}{2p^+}\right],
\end{eqnarray}
where we stretch the integral to $0$ since it is now finite, the difference being 
given by negligible positive powers of $\mu^+/T$.
For further convenience, we parametrize $C_\mathrm{brem/compton}(k/T)$ with an accuracy of
 $2\%$ or better as
\begin{equation}
	\label{fitcbrem}
  C_\mathrm{brem/compton}(x)= \frac{4 - \ln(4) - \frac{\pi^2}{6}}{x} \sum_{n=1}^4 d_n e^{-n x}  
  +     \left( -0.12563 +  \frac{\ln(4)}{x} + \frac{\pi^2}{6}  \frac{1}{x^2 + x} \right),
\end{equation}
with
\begin{equation}
   [d_1\ldots d_4] = [ 2.29467534576455, -3.0183977101591, 1.2374580732449, 0.486264291149683  ].
\end{equation}
Let us now turn to region 2. By imposing the cutoffs and expanding for $p^+\ll k$ we have
\begin{eqnarray}
\nn	(2\pi)^3\frac{d\Gamma_\gamma}{d^3k}\bigg\vert_\mathrm{brem/compton}^{(2)}&=&\frac{4}{T} \mathcal{A}(k)\int_0^{\mu^+} dp^+\int_{\mu^\mathrm{NLO}_\perp}^\infty\frac{d\pp}{(2\pi)}\frac{ \pp}{2(p^+)^2\delta E^2 }\\
\nn	&&\times\int\frac{d^2\qp}{(2\pi)^2}\left[\frac{\md^2\qp^2}{(\qp^2+\delta E^2)(\qp^2+\delta E^2+\md^2)} -\frac{\md^2}{\qp^2+\md^2}\right],\\
\label{region2brem}
&&
\end{eqnarray}
where now $\delta E\approx\pp^2/(2p^+)$. By changing integration variable from $dp^+$ to $d\delta E$ we have
\begin{eqnarray}
\nn	(2\pi)^3\frac{d\Gamma_\gamma}{d^3k}\bigg\vert_\mathrm{brem/compton}^{(2)}&=&\frac{4}{T} \mathcal{A}(k)\int_{\mu^\mathrm{NLO}_\perp}^\infty\frac{d\pp}{(2\pi)}\int_{\pp^2/(2\mu^+)}^{\infty} d\delta E 
\frac{1}{\pp\delta E^2}\\
\nn	&&\times\int\frac{d^2\qp}{(2\pi)^2}\left[\frac{\md^2\qp^2}{(\qp^2+\delta E^2)(\qp^2+\delta E^2+\md^2)} -\frac{\md^2}{\qp^2+\md^2}\right],\\
\label{deltaeint}
&&
\end{eqnarray}
which is easier to integrate. Upon expanding the result in $\mu^+\gg\mu_\perp^\mathrm{NLO}$ we have
\begin{equation}
	\label{final2region}
		(2\pi)^3\frac{d\Gamma_\gamma}{d^3k}\bigg\vert_\mathrm{brem/compton}^{(2)}=-\frac{\md}{2\pi T} 
\mathcal{A}(k)\ln\frac{2\md\mu^+}{(\mu_\perp^\mathrm{NLO})^2}+\order{\frac{(\mu_\perp^\mathrm{NLO})^2\mathcal{A}(k)}{\mu^+ T}}.
\end{equation}
Let us remark that the $\OO((\mu_\perp^\mathrm{NLO})^2\mathcal{A}(k))/(\mu^+ T))$ term
obtained here cancels exactly with the one that would be obtained by keeping the suppressed 
$\OO((\mu_\perp^\mathrm{NLO})^2)$ term in
\Eq{perpintegrals}, confirming the correctness of our two-region matching procedure 
and the aforementioned cancellation of power-law terms in the cutoffs.

Hence, the contribution from the brem/Compton region is
\begin{equation}
		(2\pi)^3\ddgkv_{\sc}^{\rm brem./Compt.}=-\frac{\md}{2\pi T} \mathcal{A}(k)\left[\ln\left(\frac{2T\md}{(\mu_\perp^\mathrm{NLO})^2}\right)+C_\mathrm{brem/compton}\left(\frac{k}{T}\right)\right].
\label{bremcomptontotal}
\end{equation}

\subsection{The pair annihilation contribution: $\Gamma_{\sc}^{\rm pair} $}
We employ the same two-region strategy for this process too. In principle we would have
IR divergences in $p^+$ at both endpoints $0$ and $-k$. However, using the symmetry
at $-k/2$ we discussed before, we can restrict the integration to $(-k/2,0)$ and worry
about a single IR divergence.  In practice, in region 1,
plugging \Eq{perpintegrated} in \Eq{collsubtracted} and keeping only
the $\mu_\perp^\mathrm{NLO}$-independent term, we have
\begin{equation}
	(2\pi)^3\frac{d\Gamma_\gamma}{d^3k}\bigg\vert_\mathrm{pair}^{(1)}=\frac{\md}{2\pi T} \mathcal{A}(k)\int_{-k+\mu^+}^{-\mu^+} dp^+
\frac{\nfd( k+p^+) (1-\nfd(p^+))}{\nfd(k)}\frac{ (k+p^+)^2+(p^+)^2}{k\,p^+\,(p^++k)}.
\end{equation}
Upon changing the sign of the integration variable and restricting the integration to $(0,k/2)$
we have
\begin{equation}
	(2\pi)^3\frac{d\Gamma_\gamma}{d^3k}\bigg\vert_\mathrm{pair}^{(1)}=-\frac{\md}{\pi T} \mathcal{A}(k)\int^{k/2}_{\mu^+} dp^+
\frac{\nfd( k-p^+) \,\nfd(p^+)}{\nfd(k)}\frac{ (k-p^+)^2+(p^+)^2}{k\,p^+\,(k-p^+)}.
\end{equation}
The IR logarithm can be extracted by adding and subtracting $1/(2p^+)$ under the 
integral sign, \emph{i.e.},
\begin{equation}
		-\frac{\md}{\pi T} \mathcal{A}(k)\int_{\mu^+}^{k/2} dp^+
\frac{1}{2p^+}=		-\frac{\md}{2\pi T} \mathcal{A}(k)\ln\frac{k}{2\mu^+}.
\label{muzregulatepair}
\end{equation}
With a slight rearrangement we can write the result from region 1 as
\begin{equation}
		(2\pi)^3\frac{d\Gamma_\gamma}{d^3k}\bigg\vert_\mathrm{pair}^{(1)}=-\frac{\md}{2\pi T} \mathcal{A}(k)\left[\ln\left(\frac{T}{\mu^+}\right)+C_\mathrm{pair}\left(\frac{k}{T}\right)\right],
\label{oneregulatedpair}
\end{equation}
with
\begin{equation}
	\label{defcpair}
	C_\mathrm{pair}\left(\frac{k}{T}\right)=\ln\frac{k}{2T}+2\int_{0}^{k/2} dp^+\left[
\frac{\nfd( k-p^+) \,\nfd(p^+)}{\nfd(k)}\frac{ (k-p^+)^2+(p^+)^2}{k\,p^+\,(k-p^+)}-\frac{1}{2p^+}\right],
\end{equation}
We can fit $C_\mathrm{pair}$ with a very good accuracy as
\begin{equation}
\label{cpairfit}
C_{\rm pair}(k/T) =  - \gamma_E + {\rm Ei}(-k/2T) + 
I_{\rm pair}(k/T),
\end{equation}
with the function 
\begin{equation}
I_\mathrm{pair}(x) \simeq  2\ln(g(x)) + b_1 + \frac{b_2}{g(x)} + 
\frac{b_3}{g(x)^2} + \left(\sum_{n=0}^3 a_n x^n\right) e^{-x},
\end{equation}
where $g(x) =  e^{-x/2} + x$ and
\begin{equation}
b_1              = -1.29715   \,, 
\quad b_2              = 1.38486 \,,    
\quad b_3              = -1.58046   \,, 
\end{equation}
and
\begin{equation}
a_0              = 1.2014        \,,
\quad a_1              = -0.303456  \,,    
\quad a_2              = 0.00446236  \,,    
\quad a_3              = -0.0451118  \,.     
\end{equation}
Region 2, upon exploiting again the symmetry with respect to $k/2$, turns out to be 
identical to its brem/compton counterpart, yielding again a log that removes the
$\mu^+$ dependence. The final contribution from the pair processes is thus
\begin{equation}
		(2\pi)^3\frac{d\Gamma_\gamma}{d^3k}\bigg\vert_{\sc}^{\rm pair}=-\frac{\md}{2\pi T} \mathcal{A}(k)\left[\ln\left(\frac{2T\md}{(\mu_\perp^\mathrm{NLO})^2}\right)+C_\mathrm{pair}\left(\frac{k}{T}\right)\right].
\label{pairtotal}
\end{equation}

\section{The contribution from HTL vertices in the soft region}
\label{app_htl}
In this section we analyze the contribution from the HTL vertices within the 
framework of light-cone fermionic sum rules we have introduced before. The analysis 
performed in Section \ref{sub_soft_NLO} relied heavily on analyticity arguments; 
relations such as the KMS relation were employed to rewrite propagators or amplitudes 
in terms of fully retarded and fully advanced functions. This is particularly 
advantageous in the current analysis of the contribution of the HTL vertices, 
since, as we shall show, only the fully retarded/advanced vertices are needed, 
i.e. only those with one external $a$ line, which correspond to the analytic 
continuation of the Euclidean Hard Thermal Loops.

As we mentioned, a full treatment of the HTL effective theory within the context of the \ra
basis was carried out
 in \cite{CaronHuot:2007nw} in the gauge sector only. There it was 
observed that hard loops with two external $a$ lines are enhanced by 
a Bose factor of $T/p^0\sim 1/g$, $P$ being the external momentum, 
with respect to the ones with one external $a$ line, which are instead 
the fully retarded functions obtained by analytic continuation of the 
Euclidean amplitudes. However, when attaching propagators to these loops, 
the enhanced ones can only be connected by $ra$ propagators, which 
scale like $1/(g^2T^2)$ (see Table~\ref{tab_pc}), whereas a $rr$ propagator 
can be attached to the standard, fully retarded ones. The $rr$ propagator 
has a relative $T/p^0\sim1/g$ enhancement, thereby making the connected
amplitudes of the same size.

An opposite behavior is observed when the analysis of 
\cite{CaronHuot:2007nw} is generalized to include fermions, as we 
have done. Consider for simplicity the HTL self-energy: the $ra$ 
and $ar$ amplitudes, which are the fully retarded and fully advanced 
amplitudes, corresponding to analytic continuation of Euclidean loops,
 scale like $gT$. The 
$aa$ amplitude scales instead like $g^2T$, the suppression being due 
to Pauli-blocking; this can easily be seen by noting that the $aa$ 
amplitude is linked by the KMS relation to the difference of the 
$ra$ and $ar$ self-energies times $(1/2-\nfd(p^0))\sim p^0/(4T)\sim g$.

\begin{figure}[ht]
	\begin{center}
		\includegraphics[width=14cm]{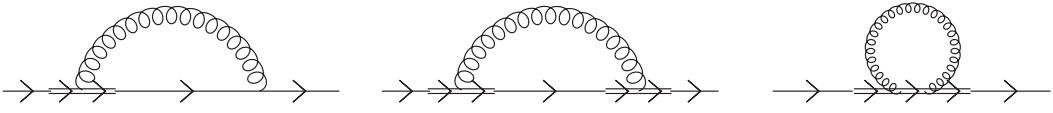}
	\end{center}
	\caption{The relevant diagrams and assignments for the soft, retarded self-energy 
	with HTL vertices. The gluon is always $rr$. The notation for the HTL vertices 
	follows the one in \cite{CaronHuot:2007nw}: the double line is the eikonalized 
	hard mode flowing in the loop.}
	\label{fig_self}
\end{figure}

Going to the effective quark-gluon vertex, one obtains that the 
$raa$ amplitudes are all Pauli-blocked by a factor of $g$ with 
respect to the fully retarded $rra$, irrespective of which particle is assigned 
the $r$ label, whereas the $aaa$ vertex scales like the $rra$, in agreement with
the results of \cite{Fueki:2001nm}. This would then 
give a more complicated set of power-counting rules than in the pure gauge theory. 
For the problem at hand, however, a limited number of hard loops is required: the 
requirements that the gluon be $rr$, in order to gain a $1/g$ enhancement, and 
that the soft quarks be $ra$ or $ar$, in order not to be Pauli-blocked, imply that 
the only needed HTL vertices are of the simple $rra$ and $rrra$ type. For what concerns 
the cat eye diagram one just needs to replace the bare vertex on the soft quark line 
in Fig.~\ref{fig_cateye} with the HTL vertex having the same \ra assignments. Regarding 
 the soft-soft self-energy and tadpole diagrams, the three possibilities are depicted 
in Fig.~\ref{fig_self}.
 They are:
\begin{enumerate}
	\item A single vertex connecting the gluon to the soft line is replaced with 
	its fully retarded (or fully advanced) HTL counterpart, whereas the other 
	vertex is kept in its bare form. One of the two possibilities is shown on 
	the left in Fig.~\ref{fig_self}.
	\item Both vertices are replaced by HTL vertices, as in the middle diagram 
	of Fig.~\ref{fig_self}.
	\item Only the two-quark,two-gluon $rrra$ HTL vertex is relevant for the tadpole, 
	as shown on the 	right in  Fig.~\ref{fig_self} 
\end{enumerate} A detailed inspection shows that no other assignment can 
contribute at the same order in $g$.

At the practical level, the quark-gluon effective vertex is obtained with this 
simple replacement
\begin{equation}
	\label{htlvertex}
	\gamma^\nu\to\gamma^\nu+ \frac{\mm}{2}
        \int\frac{d\Omega_l}{4\pi}\frac{\slashed{v}_l
	 v_l^\nu}{(v_l\cdot P+i\epsilon)(v_l\cdot(P+Q)+i\epsilon)},
\end{equation}
in Eqs.~\eqref{defws} and \eqref{defwc}. $v_l=(1,\mathbf{v}_l)$ is a lightlike 
four-vector, corresponding to the hard momentum in the loop (hence the label $l$), 
whose direction we integrate on. \\
The contribution of the tadpole to the retarded, soft self-energy reads instead
\begin{eqnarray}
\nn	\Sigma_{R\, \mathrm{tad}}(P)&=&
\frac{g^2\crr \mm}{4}\int\frac{d^4Q}{(2\pi)^4}\int\frac{d\Omega_l}{4\pi}
\left[G_{\mu\nu}(Q)\right]_{rr}v_l^\mu v_l^\nu\,\frac{ \slashed{v}_l }{(v_l\cdot 
P+i\epsilon)^2}\left(\frac{1}{v_l\cdot(P-Q)+i\epsilon}\right.\\
&&\left.\hspace{7.8cm}+\frac{1}{v_l\cdot(P+Q)+i\epsilon}\right).
\label{tad}
\end{eqnarray}
This expression contains a symmetry factor of $1/2$.

As we already remarked, the analytic structure in $p^+$ of the cat eye and 
self-energy insertion diagrams, which was crucial for the derivation of the 
sum rule, remains unchanged with the addition of the HTL vertices, so that we 
can still deform the integration contours at $\vert p^+\vert\gg gT$. In order to prove that the HTL 
vertices are irrelevant in the sum rule context, we then need to show that the 
amplitudes that include them go to zero on the arcs  faster than $1/p^+$. 
Although the HTL vertices introduce two more powers of $p^+$ at the denominator, the 
fact that $P$ is a spacelike vector and that $P+Q$ might also be spacelike causes 
possible enhancements in the $d\Omega_l$ angular integration. We thus proceed by 
performing the traces obtained by inserting the HTL vertices in the amplitudes $W_c$ 
and $W_s$ through the rules we have just introduced. One then obtains a basis of 
angular integrals of the kind
\begin{equation}
\label{angustruct}
\int\frac{d\Omega_l}{4\pi}\frac{f(v_l)}{(v_l\cdot P+i\epsilon)(v_l\cdot(P+Q)+i\epsilon)},
\end{equation}
where $f(v_l)$ is a set of functions obtained by contracting $v^\nu_l$ with all other 
4-vectors available, \emph{i.e.}, 
\begin{equation}
v_l\cdot U =1,\quad v_l\cdot P,\quad v_l\cdot (P+Q),\quad v_l\cdot v_k,\quad v_l\cdot v_{l^\prime},
\end{equation}
and products thereof. $U=(1,0,0,0)$ is the four-velocity of the plasma in its rest 
frame and $v_{l^\prime}$ is the hard loop velocity of the second HTL vertex, which 
arises in the evaluation of the middle diagram in Fig.~\ref{fig_self}. The 
contribution of the tadpole can also be related to this basis by differentiation 
with respect to $p^0$ and $q^0$.\\
The angular integrations are known in the literature and can be read from 
\cite{Frenkel:1989br,Ayala:2001mb}. Upon inserting the results back in the 
amplitudes $W_s$ and $W_c$ and expanding  for large $p^+$ one obtains that all 
contributions containing one or two HTL vertices behave on the arcs 
as $1/(p^+)^2$ or smaller, and therefore do not contribute.

\bibliographystyle{JHEP}
\bibliography{photonbib}

\providecommand{\href}[2]{#2}\begingroup\raggedright\begin{thebibliography}{10}

\bibitem{Adare:2008ab}
{\bf PHENIX} Collaboration, A.~Adare {\em et.~al.}, {\it {Enhanced production
  of direct photons in Au+Au collisions at $\sqrt{s_{NN}}=200$ GeV and
  implications for the initial temperature}},  {\em Phys.Rev.Lett.} {\bf 104}
  (2010) 132301, [\href{http://xxx.lanl.gov/abs/0804.4168}{{\tt
  arXiv:0804.4168}}].

\bibitem{Adare:2011zr}
{\bf PHENIX} Collaboration, A.~Adare {\em et.~al.}, {\it {Observation of
  direct-photon collective flow in $\sqrt{s_{NN}}=200$ GeV Au+Au collisions}},
  \href{http://xxx.lanl.gov/abs/1105.4126}{{\tt arXiv:1105.4126}}.

\bibitem{Afanasiev:2012dg}
{\bf PHENIX} Collaboration, S.~Afanasiev {\em et.~al.}, {\it {Measurement of
  Direct Photons in Au+Au Collisions at $\sqrt{s_{NN}} = 200$ GeV}},
  \href{http://xxx.lanl.gov/abs/1205.5759}{{\tt arXiv:1205.5759}}.

\bibitem{Lee:2012cd}
{\bf CMS} Collaboration, Y.-J. Lee, {\it {Measurement of isolated photon
  production in pp and PbPb collisions at $\sqrt{s_{NN}} =2.76$ TeV with CMS}},
   \href{http://xxx.lanl.gov/abs/1208.6156}{{\tt arXiv:1208.6156}}.

\bibitem{delaCruz:2012ru}
{\bf CMS} Collaboration, B.~de~la Cruz, {\it {W, Z and photon production in
  CMS}},  \href{http://xxx.lanl.gov/abs/1208.4927}{{\tt arXiv:1208.4927}}.

\bibitem{Milov:2012pd}
{\bf ATLAS} Collaboration, A.~Milov, {\it {Measurement of the $W$, $Z$ and
  photon production in lead-lead collisions at $\sqrt{s_{NN}} = 2.76$ TeV with
  the ATLAS detector}},  \href{http://xxx.lanl.gov/abs/1209.0088}{{\tt
  arXiv:1209.0088}}.

\bibitem{Steinberg:2012tv}
{\bf ATLAS} Collaboration, P.~Steinberg, {\it {Measurement of high $p_T$
  isolated prompt photons in lead-lead collisions at $\sqrt{s_{NN}}=2.76$ TeV
  with the ATLAS detector at the LHC}},
  \href{http://xxx.lanl.gov/abs/1209.4910}{{\tt arXiv:1209.4910}}.

\bibitem{Gordon:1993qc}
L.~Gordon and W.~Vogelsang, {\it {Polarized and unpolarized prompt photon
  production beyond the leading order}},  {\em Phys.Rev.} {\bf D48} (1993)
  3136--3159.

\bibitem{Fries:2002kt}
R.~J. Fries, B.~M{\"u}ller, and D.~K. Srivastava, {\it {High-energy photons
  from passage of jets through quark gluon plasma}},  {\em Phys.Rev.Lett.} {\bf
  90} (2003) 132301, [\href{http://xxx.lanl.gov/abs/nucl-th/0208001}{{\tt
  nucl-th/0208001}}].

\bibitem{Zakharov:2004bi}
B.~Zakharov, {\it {Induced photon emission from quark jets in ultrarelativistic
  heavy-ion collisions}},  {\em JETP Lett.} {\bf 80} (2004) 1--6,
  [\href{http://xxx.lanl.gov/abs/hep-ph/0405101}{{\tt hep-ph/0405101}}].

\bibitem{Kapusta:1991qp}
J.~I. Kapusta, P.~Lichard, and D.~Seibert, {\it {High-energy photons from quark
  - gluon plasma versus hot hadronic gas}},  {\em Phys.Rev.} {\bf D44} (1991)
  2774--2788.

\bibitem{Baier:1991em}
R.~Baier, H.~Nakkagawa, A.~Niegawa, and K.~Redlich, {\it {Production rate of
  hard thermal photons and screening of quark mass singularity}},  {\em
  Z.Phys.} {\bf C53} (1992) 433--438.

\bibitem{Aurenche:1998nw}
P.~Aurenche, F.~Gelis, R.~Kobes, and H.~Zaraket, {\it {Bremsstrahlung and
  photon production in thermal QCD}},  {\em Phys.Rev.} {\bf D58} (1998) 085003,
  [\href{http://xxx.lanl.gov/abs/hep-ph/9804224}{{\tt hep-ph/9804224}}].

\bibitem{Arnold:2001ba}
P.~B. Arnold, G.~D. Moore, and L.~G. Yaffe, {\it {Photon emission from
  ultrarelativistic plasmas}},  {\em JHEP} {\bf 0111} (2001) 057,
  [\href{http://xxx.lanl.gov/abs/hep-ph/0109064}{{\tt hep-ph/0109064}}].

\bibitem{Arnold:2001ms}
P.~B. Arnold, G.~D. Moore, and L.~G. Yaffe, {\it {Photon emission from quark
  gluon plasma: Complete leading order results}},  {\em JHEP} {\bf 0112} (2001)
  009, [\href{http://xxx.lanl.gov/abs/hep-ph/0111107}{{\tt hep-ph/0111107}}].

\bibitem{Arnold:1994eb}
P.~B. Arnold and C.-x. Zhai, {\it {The Three loop free energy for high
  temperature QED and QCD with fermions}},  {\em Phys.Rev.} {\bf D51} (1995)
  1906--1918, [\href{http://xxx.lanl.gov/abs/hep-ph/9410360}{{\tt
  hep-ph/9410360}}].

\bibitem{Arnold:1994ps}
P.~B. Arnold and C.-X. Zhai, {\it {The Three loop free energy for pure gauge
  QCD}},  {\em Phys.Rev.} {\bf D50} (1994) 7603--7623,
  [\href{http://xxx.lanl.gov/abs/hep-ph/9408276}{{\tt hep-ph/9408276}}].

\bibitem{Braaten:1995jr}
E.~Braaten and A.~Nieto, {\it {Free energy of QCD at high temperature}},  {\em
  Phys.Rev.} {\bf D53} (1996) 3421--3437,
  [\href{http://xxx.lanl.gov/abs/hep-ph/9510408}{{\tt hep-ph/9510408}}].

\bibitem{Kajantie:2002wa}
K.~Kajantie, M.~Laine, K.~Rummukainen, and Y.~Schr{\"o}der, {\it {The Pressure
  of hot QCD up to g6 ln(1/g)}},  {\em Phys.Rev.} {\bf D67} (2003) 105008,
  [\href{http://xxx.lanl.gov/abs/hep-ph/0211321}{{\tt hep-ph/0211321}}].

\bibitem{CaronHuot:2008uh}
S.~Caron-Huot and G.~D. Moore, {\it {Heavy quark diffusion in QCD and N=4 SYM
  at next-to-leading order}},  {\em JHEP} {\bf 0802} (2008) 081,
  [\href{http://xxx.lanl.gov/abs/0801.2173}{{\tt arXiv:0801.2173}}].

\bibitem{CaronHuot:2008ni}
S.~Caron-Huot, {\it {O(g) plasma effects in jet quenching}},  {\em Phys.Rev.}
  {\bf D79} (2009) 065039, [\href{http://xxx.lanl.gov/abs/0811.1603}{{\tt
  arXiv:0811.1603}}].

\bibitem{CaronHuot:2008uw}
S.~Caron-Huot, {\it {On supersymmetry at finite temperature}},  {\em Phys.Rev.}
  {\bf D79} (2009) 125002, [\href{http://xxx.lanl.gov/abs/0808.0155}{{\tt
  arXiv:0808.0155}}].

\bibitem{Braaten:1989mz}
E.~Braaten and R.~D. Pisarski, {\it {Soft Amplitudes in Hot Gauge Theories: A
  General Analysis}},  {\em Nucl.Phys.} {\bf B337} (1990) 569.

\bibitem{Aurenche:1999tq}
P.~Aurenche, F.~Gelis, and H.~Zaraket, {\it {KLN theorem, magnetic mass, and
  thermal photon production}},  {\em Phys.Rev.} {\bf D61} (2000) 116001,
  [\href{http://xxx.lanl.gov/abs/hep-ph/9911367}{{\tt hep-ph/9911367}}].

\bibitem{Aurenche:2000gf}
P.~Aurenche, F.~Gelis, and H.~Zaraket, {\it {Landau-Pomeranchuk-Migdal effect
  in thermal field theory}},  {\em Phys.Rev.} {\bf D62} (2000) 096012,
  [\href{http://xxx.lanl.gov/abs/hep-ph/0003326}{{\tt hep-ph/0003326}}].

\bibitem{Aurenche:2002pd}
P.~Aurenche, F.~Gelis, and H.~Zaraket, {\it {A Simple sum rule for the thermal
  gluon spectral function and applications}},  {\em JHEP} {\bf 0205} (2002)
  043, [\href{http://xxx.lanl.gov/abs/hep-ph/0204146}{{\tt hep-ph/0204146}}].

\bibitem{Aurenche:2002wq}
P.~Aurenche, F.~Gelis, G.~Moore, and H.~Zaraket, {\it
  {Landau-Pomeranchuk-Migdal resummation for dilepton production}},  {\em JHEP}
  {\bf 0212} (2002) 006, [\href{http://xxx.lanl.gov/abs/hep-ph/0211036}{{\tt
  hep-ph/0211036}}].

\bibitem{CaronHuot:2007nw}
S.~Caron-Huot, {\it {Hard thermal loops in the real-time formalism}},  {\em
  JHEP} {\bf 0904} (2009) 004, [\href{http://xxx.lanl.gov/abs/0710.5726}{{\tt
  arXiv:0710.5726}}].

\bibitem{Besak:2012qm}
D.~Besak and D.~B{\"{o}}deker, {\it {Thermal production of ultrarelativistic
  right-handed neutrinos: Complete leading-order results}},  {\em JCAP} {\bf
  1203} (2012) 029, [\href{http://xxx.lanl.gov/abs/1202.1288}{{\tt
  arXiv:1202.1288}}].

\bibitem{Gale:2012xq}
C.~Gale, {\it {Electromagnetic radiation in heavy ion collisions: Progress and
  puzzles}},  \href{http://xxx.lanl.gov/abs/1208.2289}{{\tt arXiv:1208.2289}}.

\bibitem{Arnold:2002ja}
P.~B. Arnold, G.~D. Moore, and L.~G. Yaffe, {\it {Photon and gluon emission in
  relativistic plasmas}},  {\em JHEP} {\bf 0206} (2002) 030,
  [\href{http://xxx.lanl.gov/abs/hep-ph/0204343}{{\tt hep-ph/0204343}}].

\bibitem{CaronHuot:2006te}
S.~Caron-Huot, P.~Kovtun, G.~D. Moore, A.~Starinets, and L.~G. Yaffe, {\it
  {Photon and dilepton production in supersymmetric Yang-Mills plasma}},  {\em
  JHEP} {\bf 0612} (2006) 015,
  [\href{http://xxx.lanl.gov/abs/hep-th/0607237}{{\tt hep-th/0607237}}].

\bibitem{Graf:2010tv}
P.~Graf and F.~D. Steffen, {\it {Thermal axion production in the primordial
  quark-gluon plasma}},  {\em Phys.Rev.} {\bf D83} (2011) 075011,
  [\href{http://xxx.lanl.gov/abs/1008.4528}{{\tt arXiv:1008.4528}}].

\bibitem{Graf:2012hb}
P.~Graf and F.~D. Steffen, {\it {Axions and saxions from the primordial
  supersymmetric plasma and extra radiation signatures}},
  \href{http://xxx.lanl.gov/abs/1208.2951}{{\tt arXiv:1208.2951}}.

\bibitem{Brandenburg:2004du}
A.~Brandenburg and F.~D. Steffen, {\it {Axino dark matter from thermal
  production}},  {\em JCAP} {\bf 0408} (2004) 008,
  [\href{http://xxx.lanl.gov/abs/hep-ph/0405158}{{\tt hep-ph/0405158}}].

\bibitem{Bolz:2000fu}
M.~Bolz, A.~Brandenburg, and W.~Buchmuller, {\it {Thermal production of
  gravitinos}},  {\em Nucl.Phys.} {\bf B606} (2001) 518--544,
  [\href{http://xxx.lanl.gov/abs/hep-ph/0012052}{{\tt hep-ph/0012052}}].

\bibitem{Benzke:2012sz}
M.~Benzke, N.~Brambilla, M.~A. Escobedo, and A.~Vairo, {\it {Gauge invariant
  definition of the jet quenching parameter}},
  \href{http://xxx.lanl.gov/abs/1208.4253}{{\tt arXiv:1208.4253}}.

\bibitem{Braaten:1991gm}
E.~Braaten and R.~D. Pisarski, {\it {Simple effective Lagrangian for hard
  thermal loops}},  {\em Phys. Rev.} {\bf D45} (1992) 1827--1830.

\bibitem{Braaten:1995ju}
E.~Braaten and A.~Nieto, {\it {On the convergence of perturbative QCD at high
  temperature}},  {\em Phys.Rev.Lett.} {\bf 76} (1996) 1417--1420,
  [\href{http://xxx.lanl.gov/abs/hep-ph/9508406}{{\tt hep-ph/9508406}}].

\bibitem{Blaizot:2005wr}
J.-P. Blaizot, A.~Ipp, A.~Rebhan, and U.~Reinosa, {\it {Asymptotic thermal
  quark masses and the entropy of QCD in the large-N(f) limit}},  {\em
  Phys.Rev.} {\bf D72} (2005) 125005,
  [\href{http://xxx.lanl.gov/abs/hep-ph/0509052}{{\tt hep-ph/0509052}}].

\bibitem{Fueki:2001nm}
Y.~Fueki, H.~Nakkagawa, H.~Yokota, and K.~Yoshida, {\it {N point vertex
  functions, Ward-Takahashi identities and Dyson-Schwinger equations in thermal
  QCD / QED in the real time hard thermal loop approximation}},  {\em
  Prog.Theor.Phys.} {\bf 107} (2002) 759--784,
  [\href{http://xxx.lanl.gov/abs/hep-ph/0111275}{{\tt hep-ph/0111275}}].

\bibitem{Frenkel:1989br}
J.~Frenkel and J.~Taylor, {\it {High Temperature Limit of Thermal QCD}},  {\em
  Nucl.Phys.} {\bf B334} (1990) 199.

\bibitem{Ayala:2001mb}
A.~Ayala and A.~Bashir, {\it {Longitudinal and transverse fermion boson vertex
  in QED at finite temperature in the HTL approximation}},  {\em Phys.Rev.}
  {\bf D64} (2001) 025015, [\href{http://xxx.lanl.gov/abs/hep-ph/0103030}{{\tt
  hep-ph/0103030}}].

\end{thebibliography}\endgroup
\end{document}